\documentclass[manuscript]{acmart}

\usepackage{acro}
\usepackage{arydshln}
\usepackage{xspace}
\DeclareAcronym{IV}{short = IV, long = independent variable}
\DeclareAcronym{DV}{short = DV, long = dependent variable}
\DeclareAcronym{UI}{short = UI, long = user interface}
\DeclareAcronym{TLX}{short = NASA-TLX, long = NASA Task Load Index}
\DeclareAcronym{RTLX}{short = raw TLX, long =raw NASA-Task Load Index}
\DeclareAcronym{ER}{short = ER, long = error rate}
\DeclareAcronym{TCT}{short = TCT, long = Task Completion Time}
\DeclareAcronym{HCI}{short = HCI, long = Human-Computer Interaction}
\DeclareAcronym{HFE}{short = HFE, long = Human Factors and Ergonomics}
\DeclareAcronym{cuDNN}{short = cuDNN, long =  CUDA Deep Neural Network library}
\DeclareAcronym{RMSE}{short = RMSE, long = root mean squared error}
\DeclareAcronym{HMD}{short = HMD, long = head-mounted display}
\DeclareAcronym{RF}{short = RF, long = Random Forest}
\DeclareAcronym{GP}{short = GP, long = Gaussian process, long-plural = Gaussian processes}
\DeclareAcronym{KNN}{short = \textit{k}NN, long = \textit{k}-nearest neighbor}
\DeclareAcronym{NN}{short = NN, long = Neural Network}
\DeclareAcronym{DNN}{short = DNN, long =  Deep Neural Network}
\DeclareAcronym{CNN}{short = CNN, long = Convolutional Neural Network}
\DeclareAcronym{FCL}{short = FCL, long = fully connected layer}
\DeclareAcronym{BoD}{short = BoD, long = Back-of-Device}
\DeclareAcronym{VR}{short = VR, long = Virtual Reality}
\DeclareAcronym{AR}{short = AR, long = Augmented Reality}
\DeclareAcronym{MR}{short = MR, long = Mixed Reality}
\DeclareAcronym{FOV}{short = FoV, long = field of view}
\DeclareAcronym{RW}{short = RW, long = real world}
\DeclareAcronym{IFRC}{short = IFRC, long = index finger ray cast}
\DeclareAcronym{FARC}{short = FARC, long = forearm ray cast}
\DeclareAcronym{EFRC}{short = EFRC, long = eye-finger ray cast}
\DeclareAcronym{HRC}{short = HRC, long = head ray cast}
\DeclareAcronym{DOF}{short = DOF, long = degree-of-freedom, long-plural-form = degrees-of-freedom}
\DeclareAcronym{6DOF}{short = 6DOF, long = six-degree-of-freedom}
\DeclareAcronym{3DOF}{short = 3DOF, long = three-degree-of-freedom}
\DeclareAcronym{LOOCV}{short = LOOCV, long = leave-one-out cross-validation}
\DeclareAcronym{LOPOCV}{short = LOPOCV, long = leave-one-participant-out cross-validation}
\DeclareAcronym{CV}{short = CV, long = cross-validation}
\DeclareAcronym{RM}{short = RM, long = repeated measure}
\DeclareAcronym{ANOVA}{short = ANOVA, long = analysis of variance}
\DeclareAcronym{RMANOVA}{short = RM-ANOVA, long = repeated measures analysis of variance}
\DeclareAcronym{MANOVA}{short = MANOVA, long = multivariate analysis of variance}
\DeclareAcronym{AGATe}{short = AGATe, long = AGreement Analysis Toolkit}
\DeclareAcronym{GHoST}{short = GHoST, long = Gesture Heatmap Toolkit Gesture Heatmaps Toolkit}
\DeclareAcronym{GREAT}{short = GREAT, long = Gesture Relative Accuracy Toolkit}
\DeclareAcronym{GRT}{short = GRT, long = Gesture Recognition Toolkit}
\DeclareAcronym{DTW}{short = DTW, long = Dynamic Time Warping}
\DeclareAcronym{LHRD}{short = LHRD, long = large high resolution display}
\DeclareAcronym{GEQ}{short = GEQ, long = Game Experience Questionnaire}
\DeclareAcronym{SPGQ}{short = SPGQ, long = Social Presence Gaming Questionnaire}
\DeclareAcronym{SUS}{short = SUS, long = Slater-Usoh-Steed questionnaire}
\DeclareAcronym{IPQ}{short = IPQ, long = igroup presence questionnaire}
\DeclareAcronym{PQ}{short = WS, long = Witmer and Singer presence questionnaire}
\DeclareAcronym{ITQ}{short = ITQ, long = immmersive tendency questionnaire}
\DeclareAcronym{BIP}{short = BIP, long = break-in-presence}
\DeclareAcronym{VE}{short = VE, long = virtual environment}
\DeclareAcronym{TPI}{short = TPI, long = temple presence inventory}
\DeclareAcronym{ITC}{short = ITC-SOPI, long = ITC-Sense of presence inventory}
\DeclareAcronym{IMU}{short = IMU, long = inertial measurement unit}
\DeclareAcronym{GUI}{short = GUI, long = graphical user interface}
\DeclareAcronym{TAM}{short = TAM, long = Technology Acceptance Model}
\DeclareAcronym{UTAUT}{short = UTAUT, long = Unified Theory of Acceptance and Use of Technology}
\DeclareAcronym{AI}{short = AI, long = Artificial Intelligence}
\DeclareAcronym{DALI}{short = DALI, long = Driving Activity Load Index}
\DeclareAcronym{BLMM}{short = BLMM, long = Bayesian linear mixed model} 
\usepackage{svg}
\usepackage[caption=false]{subfig}
\usepackage{todonotes}
\usepackage{multicol}
\usepackage{multirow}
\usepackage{booktabs, array}

\usepackage{tabularx}
\usepackage{lipsum}
\usepackage{xcolor}
\newcommand{\mychange}[1]{\textcolor{black}{#1}}

\AtBeginDocument{%
  \providecommand\BibTeX{{%
    \normalfont B\kern-0.5em{\scshape i\kern-0.25em b}\kern-0.8em\TeX}}}

\newcommand{\lastaccess}{2022-04-10}

\setcopyright{acmlicensed}
\acmJournal{TOCHI}
\acmYear{2022} \acmVolume{1} \acmNumber{1} \acmArticle{1} \acmMonth{1} \acmPrice{15.00}\acmDOI{10.1145/3529225}

\graphicspath{ {./figures/} }
\newcolumntype{P}[1]{>{\centering\arraybackslash}p{#1}}
\begin{document}

\title[The Placebo Effect of AI in HCI]{The Placebo Effect of Artificial Intelligence in Human-Computer Interaction}

\author{Thomas Kosch}
\authornote{Contributed equally to this research.}
\affiliation{%
	\institution{Utrecht University}
	\city{Utrecht}
	\country{the Netherlands}}
\email{t.a.kosch@uu.nl}

\author{Robin Welsch}
\authornotemark[1]
\affiliation{%
	\institution{LMU Munich}
	\city{Munich}
	\country{Germany}}
\email{robin.welsch@ifi.lmu.de}

\author{Lewis Chuang}
\affiliation{%
	\institution{Chemnitz University of Technology }
	\city{Chemnitz}
	\country{Germany}}
\email{clew@hrz.tu-chemnitz.de}

\author{Albrecht Schmidt}
\affiliation{%
	\institution{LMU Munich}
	\city{Munich}
	\country{Germany}}
\email{albrecht.schmidt@ifi.lmu.de}

\renewcommand{\shortauthors}{T. Kosch et al.}

\renewcommand{\transparent}{Error-based Adaptation\xspace}

\newcommand{\nontransparent}{Physiology-based Adaptation\xspace}

\begin{abstract}
In medicine, patients can obtain real benefits from a sham treatment. These benefits are known as the placebo effect. We report two experiments (Experiment I: N=369; Experiment II: N=100) \mychange{demonstrating} a placebo effect in adaptive interfaces. Participants were asked to solve word puzzles while being supported by no system or an adaptive AI interface. All participants experienced the same word puzzle difficulty and had no support from an AI throughout the experiments. Our results showed that the belief of receiving adaptive AI support increases expectations regarding the participant's own task \mychange{performance, sustained after interaction}. These expectations were positively correlated to performance, as indicated by the number of solved word puzzles. We integrate our findings into technological acceptance theories and discuss implications for the future assessment of AI-based user interfaces and novel technologies. We argue that system descriptions can elicit placebo effects \mychange{through user expectations} biasing the results of user-centered studies.
\end{abstract}



\maketitle

\begin{figure}
    \centering
    \centering
    \subfloat[][]{
        \includegraphics[height=0.225\columnwidth]{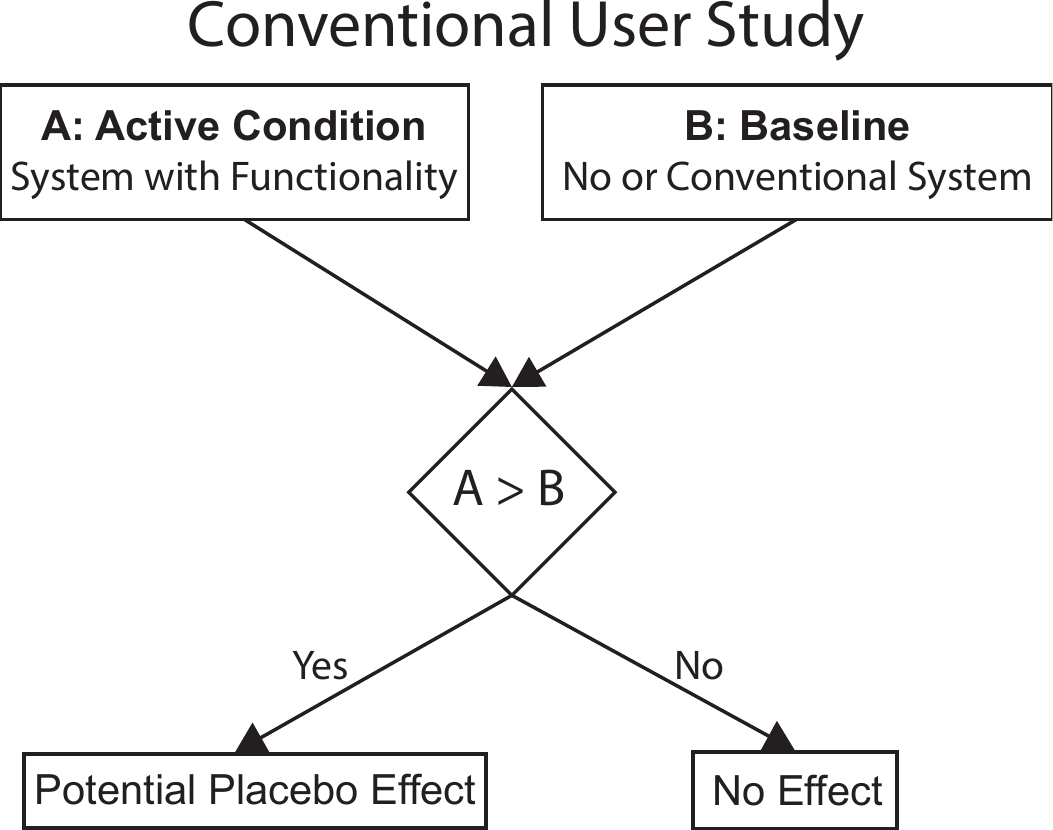}
        \label{fig:cus}}
    \subfloat[][]{
        \includegraphics[height=0.225\columnwidth]{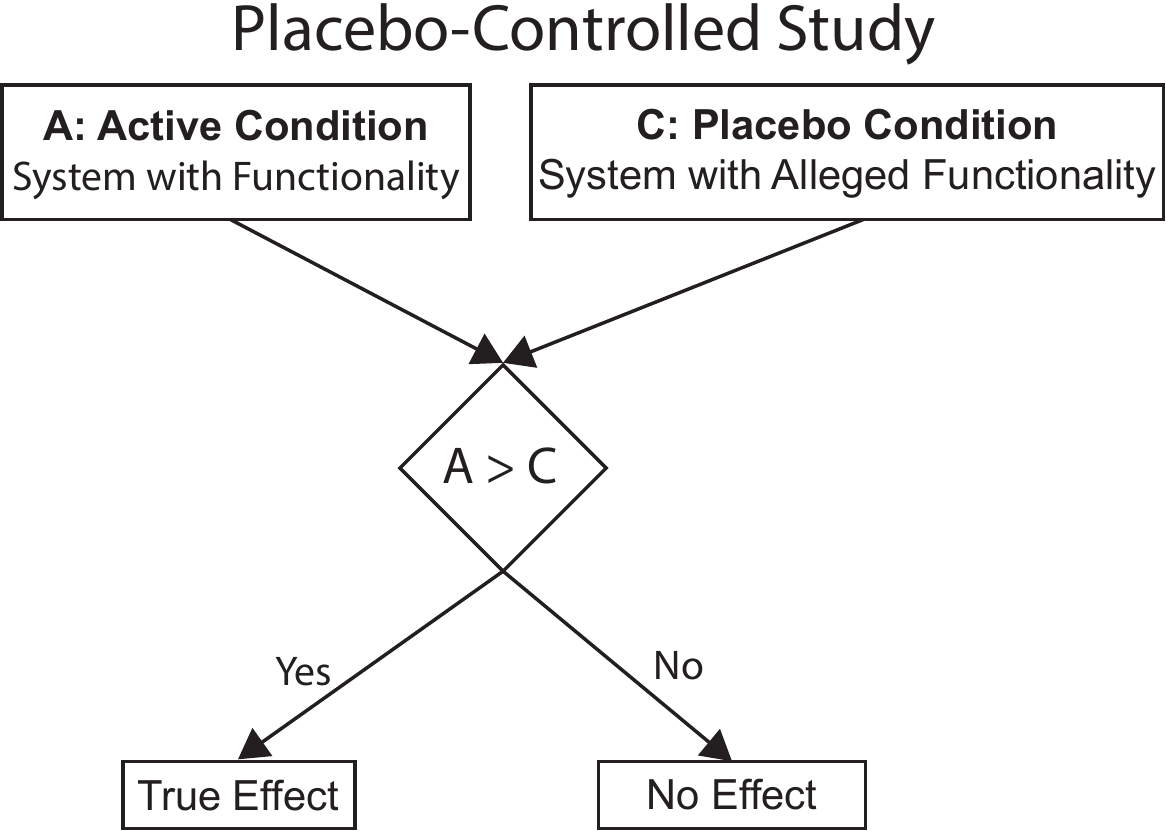}
        \label{fig:pcs}}
    \subfloat[][]{
        \includegraphics[height=0.225\columnwidth]{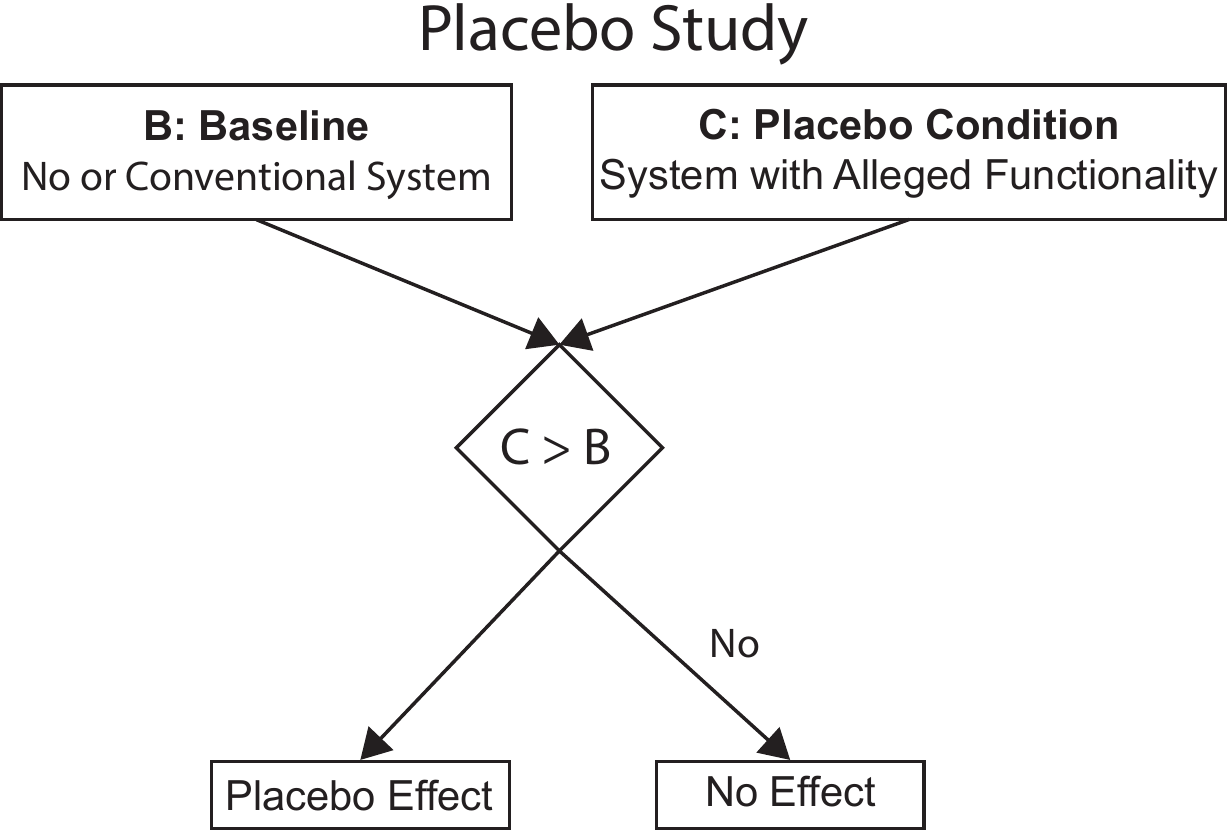}
        \label{fig:ps}}
    \caption{Flow charts comparing conventional and placebo-controlled studies. \textbf{(a):} Conventional user studies comparing novel systems with a baseline do not take the potential presence of a placebo into account, which may be responsible for the perceived superiority of the novel system. \textbf{(b):} Instead, comparing a novel system with a system that pretends an alleged novel functionality does reveal if the actual novel system has an effect. \textbf{(c):} A placebo effect in the study measures is present if the alleged novel system is considered superior to a conventional baseline.}
    \label{fig:teaser}
\end{figure}

\section{Introduction}
A medical placebo; a sham treatment (e.g., a sugar pill without active ingredients), can improve a patient's subjective condition\mychange{~\cite{finniss2010biological,powerless2001}} without treating the patient with an active substance or specific procedure. Such a placebo can alleviate pain~\cite{lasagna1954study} or support the treatment of ailments~\cite{beecher1955powerful, kaptchuk1998powerful}, thereby providing an effective medical treatment without an illness-specific mechanism of effect. The only critical determinant of the placebo effect is the patient's expectation in the placebo's efficacy, \mychange{leading to a positive evaluation after treatment~\cite{stewart2004placebo, montgomery1996mechanisms}}. 

Regardless of its usefulness to treat ailments, the placebo effect obscures the evaluation of novel medical interventions. A novel cure could be evaluated positively only due to the expectation of improvement. Therefore, medical studies employ placebo-control when evaluating a novel treatment. Treatments are deemed effective only if their effect is larger than a placebo-control. The improvement caused by a specific treatment has to exceed the improvement brought about by expectations alone. This placebo-control is widely implemented in other scientific fields that evaluate human responses, such as psychological treatment~\cite{Boot2013Psychplac}, sport science~\cite{ojanen1994can}, and even visualization research~\cite{correll2020actually}. However, it is not the norm when evaluating the effectiveness of novel human-machine interfaces.

What would the placebo effect in \ac{HCI} look like? Novel technological systems are evaluated by presenting systems to users and measuring their responses (see Figure \ref{fig:cus}). A placebo effect in HCI would consist of a system to be judged as more usable or to return better task performance \mychange{after interaction because participants have heightened expectations in the system's capabilities prior to the interaction}. In medicine, expectations of improvement are controlled for by giving a non-active substance to participants who are led to believe that the substance is active and will have an effect. The participants will not know whether they have been given the active treatment (e.g. a painkiller) or an inactive substance (e.g. a sugar pill). In \ac{HCI}, participants often recognize the system or technique and can tell apart the experimental conditions, e.g. a novel AI-system against a baseline system (see Figure \ref{fig:cus}). Instructions might even explicitly describe the novelty of the evaluated systems and, hence, set participants' expectations. \mychange{Thus, a  placebo effect in \ac{HCI} would consist of a participant's favorable evaluation in terms of effectiveness after the interaction. This includes an improvement of task performance in the absence of an active system (e.g., thinking an AI would support users even when \textit{No Adaptivity} is given, see Figure \ref{fig:ps})}. The current work investigates to what extent descriptions of systems can alter subjective evaluations and task performance in user studies.

Previous HCI studies have demonstrated that a placebo can improve user experience. In games, fake power-up elements that make no difference to gameplay~\cite{Denisova2019}, and sham descriptions of \ac{AI} \mychange{adaptation} increase the self-reported game immersion~\cite{denisova2015placebo}. In social media, providing control settings for prioritizing items in one's news feed can result in higher subjective ratings of user satisfaction, even when these control settings do not influence anything~\cite{vaccaro2018illusion}. Sham design elements thus affect the pleasantness of the experience with interactive technologies. We are unaware, however, of any evidence for a placebo effect in \ac{HCI} \mychange{on personal effectiveness after interaction or objective task performance, similar to placebos in medical trials.}

The current work investigates if a sham treatment (i.e., a placebo), as operationalized by system descriptions, \mychange{can influence subjective and objective metrics of effectiveness when engaging with a system. Drawing from placebo-theory in psychology and medicine \cite{Boot2013Psychplac,stewart2004placebo,beecher1955powerful,lasagna1954study},} we systematically list \ac{HCI} studies controlling for placebo effects. We then provide an \mychange{literature search on the consideration of placebo effects in \ac{HCI}. Motivated by the classification of the search} results (see Figure \ref{fig:teaser}), we ran two experiments manipulating system descriptions \cite{10.1145/3290607.3312787} aligning with placebo-theory \cite{finniss2010biological,Boot2013Psychplac,stewart2004placebo} (see Figure \ref{fig:ps}). 

We provided participants with different descriptions of an \mychange{adaptive AI} that supports the completion of a word puzzle task following the example of previous studies~\cite{Denisova2019, vaccaro2018illusion, denisova2015placebo}. We compare the participants' evaluations and task performance to control conditions. We realized this by instructing participants that a novel AI-based system would analyze their webcam image and choose word puzzles (see Figure \ref{fig:wordpuzzle}), varying in difficulty to optimize their task performance. In reality, all participants underwent the same procedure --- \textit{No Adaptivity} was always provided where no adaptation took place. Note that different definitions of AI exist (for a range of potential definitions and conceptualizations, see \cite{helm2020machine, ottenbreit2021, li2020}). Here, we see AI as a machine learning-based algorithm that is capable of processing and integrating complex data (e.g. a video feed) to make decisions. Input and output resemble human decision-making (e.g. seeing arousal in the face of a participant and giving an easier item). In our experiments, only the implementation of expectations by our system descriptions and not the concept of \mychange{adaptation} was critical.  

In two online experiments, we investigated the expectancy-placebo effect of system descriptions for a between-subjects (i.e. Experiment I) and within-subjects (i.e. Experiment II) design. In Experiment I, we tested two versions of an \mychange{adaptive} AI system against a \textit{No Adaptivity}-condition. Thus, a third of participants received a description of a \textit{\nontransparent}, AI-based assistance system adjusting the word puzzle difficulty using the video feed of the participant webcam. Another third received a description of an \textit{\transparent}, where the word puzzle difficulty was adapted based on task errors. Here, erroneous responses would simplify the following questions and correct responses increase the difficulty. The final third of participants were told that they would not receive any support, hence receiving \textit{No Adaptivity} at all. While the \textit{\transparent} condition presents a control condition of an AI system description in which participants could validate the adaptive system, the \textit{\nontransparent} presents a placebo condition --- a description of an AI-based system that has been employed in several \ac{HCI} studies \cite{Denisova2019, vaccaro2018illusion, denisova2015placebo}. Experiment II compared a \textit{\nontransparent} to a \textit{No Adaptivity} condition in a more controlled within-subjects study. 

\mychange{Experiment I found increased user expectations for \textit{\nontransparent} and \textit{\transparent} system descriptions. However, we did not find a significant effect between the system descriptions and the error rate, showing that the objective performance was not impacted using different system narratives. To investigate further if individual expectations are responsible for this, Experiment II used a within-subject design to control for potential subjective differences. Here, we utilized the \textit{No Adaptivity} and \textit{\nontransparent} condition, showing a placebo effect on the subjective performance ratings when using \textit{\nontransparent}. Again, no statistical significant effect was found for the error rate, suggesting that placebo effects of AI systems do not affect objective measures but are instead impacting subjective measures.}

\section*{Contribution Statement}
\mychange{This paper makes two contributions: Inspired by a literature search on placebo elements in HCI studies, we demonstrate a placebo effect in HCI in two experiments:} (1) In a word puzzle task, we present participants with three different system descriptions suggesting improvements in the subjective performance ratings after interaction. (2) We then discuss the implications of a placebo effect in evaluation studies of AI and other systems and show how user expectations can be assessed and controlled for between- and within-subject experimental designs.

\section{Related Work}
First, we present standard measures that assess the usability and acceptance of novel user interfaces. Next, we describe research on adaptive AI-based interfaces. Finally, we provide a systemic literature survey summarizing research in the domain of placebo effects in \ac{HCI}.

\subsection{User Expectations}
User expectations concerning satisfaction are particularly relevant when evaluating novel technology. If interaction matches the user's prior expectations, user interfaces are typically considered more favorably~\cite{thong2006effects, venkatesh2016unified,davis1985technology} (e.g., an adaptive system supporting task completion) in terms of user satisfaction. 
Disconfirmed expectations (e.g., when an innovative technology makes task completion harder or does not add support), can produce adverse effects concerning user satisfaction and can be the primary factor in the rejection of a novel technology~\cite{Liu2011FB}. This has been shown in survey-based methods \cite{thong2006effects, venkatesh2016unified,davis1985technology} and also experimentally \cite{bentley2000biasing,hartmann2008framing,oliver1977effect}. 
In a gaming context, Michalco et al.~\cite{doi:10.1080/10447318.2015.1065696} primed users with positive, neutral, or negative game ratings to investigate how knowledge about ratings influences the user's assessment of the game. Their results show that a positive prime can foster more positive ratings after playing the game if user expectations are confirmed, but also that dis-confirmation of expectation leads to even stronger contrasting effects on ratings, i.e., positively primed. Still, unenjoyable games are rated to be more negative. Therefore, providing users with prior knowledge on the satisfaction of others can significantly influence user evaluation. \mychange{In line with this, Hollis et al. \cite{hollis2018} and Costa et al. \cite{costa2016} showed that providing users with made-up positive feedback on their physiological state (e.g. electrodermal activity or heart rate) can influence their subjective emotional state \cite{hollis2018} and can be used to reduce feelings of anxiety \cite{costa2016}.} \mychange{The focus of these studies was on user satisfaction with the interactive technology or well-being in response to fake feedback. In our study, we implemented expectations with regard to the effectiveness of the system that if sustained after interaction can be considered placebo effects.} Although user satisfaction and effectiveness are correlated~\cite{hornbaek2007meta}, it is unclear to which extent user expectations are manipulated.

\mychange{Finally, Spiel et al. investigated the priming of two difficulty adaptations of a TETRIS game, where the speed of falling pieces is actively adapted based on the player’s performance and eye movements. The participants were divided into two groups, where one group was informed about the difficulty adaptations while the other was not. The authors found significant differences in the brick fall speed between both groups, implying that prior framing of the participants can result in changing playing performances. While the authors used a functional adaptation throughout their conditions, the question of how a suggested but non-functional placebo affects user expectations remains open.}

\subsection{Usability and Assessment of Adaptive User Interfaces}

How can we infer if one technology is better than another? In user studies, subjective measures for acceptance and usability such as the \ac{TAM}~\cite{davis1985technology, 10.2307/249008, doi:10.1287/mnsc.46.2.186.11926}, \ac{UTAUT}~\cite{indrati2014comparation}, or the \ac{TLX}~\cite{HART1988139} for the measurement of workload, are widely used metrics in the field of \ac{HCI}~\cite{doi:10.1177/154193120605000909}. However, such subjective measures are prone to subjective biases. They require the participant's ability to think back and evaluate the interaction with the UI, can be susceptible to socially desirable responding (e.g., see \cite{natesan2016cognitive,kwak2019measuring} or confirmation-bias \cite{nickerson1998confirmation}.  Objective measures such as task completion time, error rates or physiological measures (e.g. heart rate), are less affected by such biases \cite{trewin2015usage}. The adaptation of UIs has become an integral part of interactive systems to satisfy the user's goals, widen the user base, and extend the system's lifespan~\cite{browne2016adaptive}. User-centric adaptation can span from the content or the interface format and the different user input modalities. For example, one can adapt letter presentation in an E-reader to alpha oscillations; an indicator of workload in the electroencephalogram \cite{koschrsvp}, reduce the number of interaction possibilities with the number of errors made by the user \cite{hudlicka2002assessment}, or adapt the interface to the mood of the user \cite{10.1145/2668056.2668058}. \mychange{Further examples include the user's context~\cite{10.1145/1216295.1216357} and their interaction performance as well as their physiology~\cite{10.1145/2808228} serving as implicit measurements for user interface adaptations. Many user interfaces base their adaptations on \ac{AI}-methods. The availability of large-scale user data allowed recent advances in machine learning and neural networks to make robust predictions about user-related issues during interaction. This includes adaptations based on the user's perceived workload~\cite{koschrsvp, 10.1145/3173574.3174010, kosch2019your}, emotions~\cite{10.1145/3472749.3474775, 10.1145/3399715.3399928}, and interaction difficulties due to the complex visualization of UI elements~\cite{10.1145/3229093}.} AI-based adaptive interfaces optimize usability by linking the system design to user input~\cite{HORNBAEK200679}, where the overall goal of user interface adaptations is the improvement of the interaction efficiency and perceived usability~\cite{HORNBAEK200679}. However, previous placebo research~\cite{Duarte12, strait2014reliability, vaccaro2018illusion, Paepcke2010} indicated that the sole description of a non-functional adaptive system can positively increase user satisfaction without system functionality. \mychange{Garcia et al. \cite{garcia2021} showed that adding an animation that supports a user's mental model can increase performance ratings of a system, even if it performs objectively in the same manner or even worse than the non-animated system.} Therefore, descriptions of systems can be regarded as a potential placebo, if aside from perceived effectiveness the task performance is susceptible to such manipulations \cite{Duarte12, 10.1145/1133265.1133306}. 

\mychange{Although previous work successfully demonstrated placebo effects in difficulty-adapting video games, the systematic investigation of placebo effects on user expectations has received little attention in \ac{HCI} research. We argue that unintentionally manipulating the user's expectancy, which is a necessary condition for placebo effects, is likely to manipulate the results of \ac{HCI} studies on a larger scale. This placebo effect is potentially biasing conclusions in novel and high-expectancy areas such as interaction with AI. We address this research gap by presenting two experiments investigating the impact of user expectations and placebo effects on user experience measures and user performance.}

\subsection{A Placebo Effect in User-Centered Studies}
\mychange{We performed a systemic literature search to identify previous work concerning placebo effects in \ac{HCI}.} We used the regular expression (\textit{``placebo''} AND \textit{``placebic''}) to search for relevant publications in the ACM Digital Library that investigated placebo effects in HCI studies. This includes publications from more than 50 journals and proceedings of more than 170 conferences. Our initial query resulted in 420 publications that we further narrowed down by excluding those situated in the medical field, which briefly mentioned placebos as a medical or psychological science method. The resulting data set contained 27 publications (see Table \ref{tab:studies-table}). 

\begin{table}[]
\caption{HCI studies retrieved from the literature search. We categorize the studies into \textbf{(a)} studies that employed a placebo-control, \textbf{(b)} studies that specifically investigate the existence of a placebo effect, and \textbf{(c)} studies that conclude with a need for placebo-controls.}
\label{tab:studies-table}
\begin{tabular}{lll}
\toprule
\textbf{Category}        & \textbf{N} & \textbf{Studies} \\
\midrule
Placebo-Control          & 4          &   \cite{10.1145/3027063.3053164,10.1145/1952222.1952262,10.1145/2856767.2856785,10.1145/3178876.3186162}               \\
Placebo Study            & 12         &   \cite{10.1145/3379503.3403554,Duarte12,duarte2013cake,denisova2015placebo,Denisova2019,vaccaro2018illusion,10.1145/3099023.3099080,farahat2013,10.1145/3379503.3403554,10.1145/3064663.3064712}               \\ \cdashline{1-2}
Acknowledge Placebo Effects as a Confound & 11         &   \cite{10.1145/2020408.2020535,10.1145/3306618.3314281,10.1145/2674396.2674435,10.1145/2370216.2370455,10.1145/3374135.3385274,lin2020,10.1145/2724660.2724668,10.1145/1281700.1281710,10.1145/2559978,10.1145/3116595.3116623,10.1145/3359203,10.1145/3301275.3302322}\\
\bottomrule
\end{tabular}
\end{table}
We identified three types of \ac{HCI} user studies that account for the potential influence of placebo effects (see Figure~\ref{fig:cus}). First (N = 12), some studies entertained the possibility of placebo effects within their or other studies (see Figure~\ref{fig:cus}). Second (N = 4), some studies employed placebo-controls (see Figure~\ref{fig:ps}), comparing user evaluations of a sham-treatment/system against a functional system, e.g. a non-functional brain-computer interface against a functional brain-computer interface \cite{Duarte12}. Third, and most importantly, some studies (N = 11) compared a sham treatment against a baseline condition. These studies were designed specifically to provide evidence of a placebo effect in HCI studies. They employed system descriptions to implement a placebo, describing a function of a system that was not implemented. In line with practices in Psychology \cite{Boot2013Psychplac}, and Medicine \cite{finniss2010biological}, we will use the term ``placebo-studies'' to refer to studies that employ a sham-treatment and compare it to a baseline condition, and ``placebo-controlled studies'', studies that compare a sham-treatment against a functional system. 

Placebo studies in HCI suggest that user evaluations of technology can be biased by knowledge and expectations about a given system. Vaccaro et al.~\cite{vaccaro2018illusion} demonstrated that when giving users an illusory control with buttons supposedly intelligently regulating social media content, these elements increase feelings of control and trust in a system. Rutten and Geerts~\cite{rutten2020better} demonstrated this for the evaluation of novel systems, which is a prominent case in \ac{HCI}. In an experimental design, they compared two haptic touch devices. They could show that ratings of attractiveness and pleasure post-interaction can largely be attributed to the perceived novelty of a system. Therefore, evaluation can be strongly affected by descriptions of the system. 

Also, more specific evaluation dimensions such as immersion and heart rate seem to be affected by knowledge or belief about a system. Denisova et al.~\cite{DENISOVA201956,Denisova2019} told players that a game adapted the difficulty to their performance, even though no adaptation took place. The players receiving this sham system description judged the game to be more immersive. In a similar experiment, Duarte et al.~\cite{Duarte12, duarte2013cake} found that players' heart rates increased in response to sham game elements that allegedly empowered the game character. Paepcke and Takayama~\cite{Paepcke2010} implemented a similar placebo manipulation in a study of two robot interfaces. They described an interface with differing levels of functionality; one with a lot of features, the other with only a few. However, the participants were using the same interface. The participants preferred interaction with the feature-rich robot as compared to the robot with only minor features. In sum, these studies indicate that descriptions of a system can be considered placebo in user-centered evaluations of usability. As technology is typically evaluated in terms of usability and interaction efficiency, \mychange{we operationalize a placebo effect in terms of different system descriptions that affect our measured metrics. Changes in metrics could be due to an interaction with the alleged AI system, including the percentage of correct answers or subjective ratings of the user's performance after interaction with the AI.}

\section{Evaluating Placebo Effects of AI in User Studies}
We conducted two experiments to evaluate the placebo effect in Human-AI interfaces. Previous research shows that placebos can be induced by social learning \cite{kirsch1999specifying} (e.g. observing someone else improving), classical conditioning (e.g. associating an inert pill with relaxation) \cite{flaten1999caffeine} or verbal information (e.g. expert instructions) \cite{stewart2004placebo}. Motivated by the literature survey, we sought to induce a placebo effect by manipulating system descriptions. Both experiments were conducted remotely and were started by launching a web app in a browser. Participants were asked to solve word puzzles with varying difficulties. In the first experiment, we explored how informing participants about experiencing an alleged \textit{Difficulty Adaptivity} can change the user's {expectations, subjective performance ratings and objective task performance. In the second experiment, we tested the placebo effect by comparing subjective performance ratings and objective performance in a control condition and a \textit{\nontransparent} condition within-subjects. Regardless of the assigned adaptivity, the task difficulty remained constant in all studies, i.e. no changes to the task difficulty took place at any time. Understanding whether expert instructions can prompt placebo effects would enable researchers to control or correct for their influence in future evaluations.

Our studies investigated the following research question and hypotheses:

\begin{itemize}
    \item[] \textbf{RQ:} Do system descriptions confound the evaluation of adaptive systems?
\end{itemize}
We investigated the following hypotheses to answer this research question:
\begin{itemize}
    \item[] \textbf{H1:} User expectations can be manipulated by a description of an adaptive AI system.
    \item[] \textbf{H2:} User performance \mychange{and performance ratings} can be manipulated by a description of an adaptive system.
    \item[] \textbf{H3:} Workload decreases when participants believe they are being supported by an adaptive AI system.
\end{itemize}

\subsection{General Study Approach}

\mychange{We performed two experiments to investigate the hypotheses and answer the research questions. In Experiment I, we examined the possibility of a placebo effect using a between-subject design: Participants were informed about all possible adaptation scenarios (e.g., \textit{No Adaptivity} vs. \textit{\nontransparent}) beforehand and were then told which of these conditions they would be assigned to. The between-subject design was to ensure that probing the functionality of one system did not interfere with evaluating the other. In Experiment II we ran a within-subject experiment. This would allow users to experience a non-functional assistance system against an allegedly functional system by each participant, hence increasing statistical power and experimental control. In detail, participants could compare their performance for the \textit{\nontransparent} and the \textit{No Adaptivity}. Here, we could investigate if a placebo effect of \ac{AI} emerged even when the baseline and placebo-control conditions were known.}

\section{Experiment I: Exploring the Placebo Effect of AI}
This experiment presents an experimental between-subject design to explore the possibility of a placebo effect in Human-AI interfaces. We describe our experimental design below.

\begin{table}[]
    \centering
    \caption{Experimental conditions and their assigned properties. \textit{No Adaptivity} and \textit{\transparent} convey a transparent functionality while maintaining a low complexity. In contrast, the \textit{\nontransparent} provides a low transparency and high complexity.}
    \label{tab:conditions}
    \begin{tabular}{llll}
        \toprule
         &  \textbf{Transparency} & \textbf{Complexity} & \textbf{Interactivity}\\
         \midrule
         \textit{No Adaptivity} & High & Low & Low \\
        \textit{\transparent} & High & Low & High \\
         \textit{\nontransparent} & Low & High & High \\
         \bottomrule
    \end{tabular}
\end{table}

\subsection{Experimental Design and System Description}
The experiment was conducted remotely using a web app. Participants were instructed that they had been selected for one out of three \textit{Adaptivities}, and that their task was to solve word puzzles of varying task difficulty compromising our manipulation of the system description. They were advised that the difficulty of the word puzzles would be adapted according to the assigned \textit{Adaptivity}. Each participant processed the same word puzzles in a randomized order without any difficulty adjustments. \mychange{The participants were informed about the presence of all \textit{Adaptivities} before the experiment. We have intentionally decided to inform the participants about all available conditions before asking them for their anticipated task performance and expectations using the assigned Adaptivity. Hence, the participants can compare their expectations, provided by the description of the assigned \textit{Adaptivity}, against the other two \textit{Adaptivities}.}

The participants were informed about the presence of all conditions before starting the experiment, as is standard practice in medical studies. However, we also disclosed the assigned adaptivity to the participants since we were investigating the AI placebo effect of knowing an alleged working treatment (i.e., \textit{\transparent} or \textit{\nontransparent}), compared to the baseline (\textit{No Adaptivity}). The system descriptions were shown to all participants before assigning them randomly to one. We summarize the system descriptions below. Table \ref{tab:conditions} shows an overview of the conditions and their assigned properties manipulating the participants' expectations, i.e. a non-blinded between-subjects design with the three following groups:

\subsubsection*{\textbf{No Adaptivity:}}
The participants were told that no assistant was selected to adjust the difficulty of the word puzzles. The task difficulty was randomized and was not influenced by the user's performance or physiological responses. Therefore, no changes in task difficulty took place. This condition served as a control.

\subsubsection*{\textbf{\transparent:}}
Adapting the task difficulty according to the task performance is an established paradigm~\cite{10.1145/2800835.2807942, 10.1145/3025453.3025721} enabling the user to self-validate the functionality of the task difficulty adaptation. Participants were told the task difficulty would be adjusted according to the number of errors they made, so  that correct answers would increase the task difficulty and a streak of wrong answers would trigger decreased difficulty. This condition presents a control condition concerning validation of system functionality by the user, i.e., the user could check whether an error resulted in the algorithm choosing an easier item in the subsequent trial. In contrast to the \textit{\nontransparent}, the \textit{\transparent} allows participants to self-validate their results, i.e., the algorithm that reflects the item choice of the system is specified.

\subsubsection*{\textbf{\nontransparent:}}
Participants were informed about the presence of a physiology-based adaptation of an \ac{AI}  changing the word puzzle difficulty by analyzing the heart rate~\cite{6078233, doi:10.1161/01.HYP.22.4.479} and emotions via facial expressions~\cite{ekman2003unmasking} (i.e., an AI-based adaptation). A scientifically valid explanation was provided in the system descriptions to the participants to convince them of the efficacy of the physiologically-based approach. The participants were told that changes in heart rate~\cite{LUQUECASADO201683, doi:10.1142/S0219519413500383} and perceived emotions~\cite{prakash2015role} are correlated to cognitive demand and stress-levels that can be captured using a webcam, and that an \ac{AI} (i.e., neural network) aggregates those two metrics to calculate the currently perceived cognitive workload and stress level. The difficulty of the word puzzles was then supposedly adapted according to these measures to optimize user performance. The participant's webcam was activated, and the camera picture could be viewed and validated by the participant during the experiment to ensure that its functionality was being utilized. No data from the camera was recorded or transmitted. Participants were also assured that the processing of the video stream took place on the device itself. The exact algorithm that reflected the word puzzle difficulty was not specified.

\mychange{Both the \textit{\nontransparent} and the \textit{\transparent} are AI interventions. While the \textit{\nontransparent} is described as a complex non-linear and non-traceable algorithm, the \textit{\transparent} is traceable and verifiable by the participant. Previous research showed that the mental model of a system is crucial for manipulating performance ratings~\cite{garcia2021}. Therefore, we compare both descriptions of theoretically simple verifiable algorithms (i.e., \textit{\transparent}) to a more complex but supposedly more intelligent system (i.e., \textit{\nontransparent}}).}

\subsubsection{Task Description and Apparatus}
The experiment was remotely conducted via a web app in the participant's web browser. Each interaction took place in the browser, including the experimental instructions and the system descriptions. The participant's demographics, consent, task performance, and self-report measures were logged exclusively on the server. The participants were then randomly selected for one of the three aforementioned conditions after explaining each of them. 

\begin{figure*}
    \centering
    \subfloat[][]{
        \includegraphics[width=0.48\columnwidth]{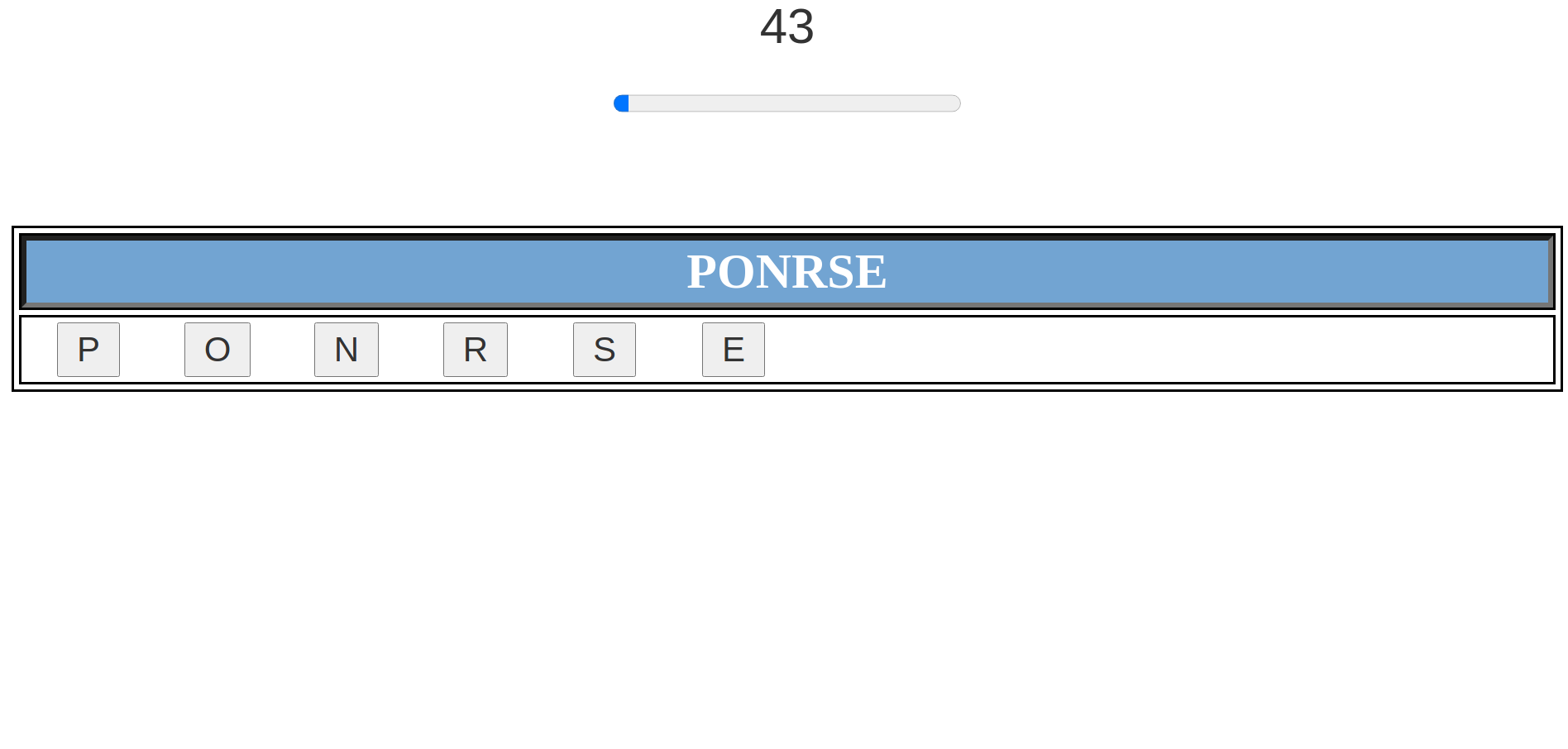}
        \label{fig:task-performance}}
    \subfloat[][]{
        \includegraphics[width=0.48\columnwidth]{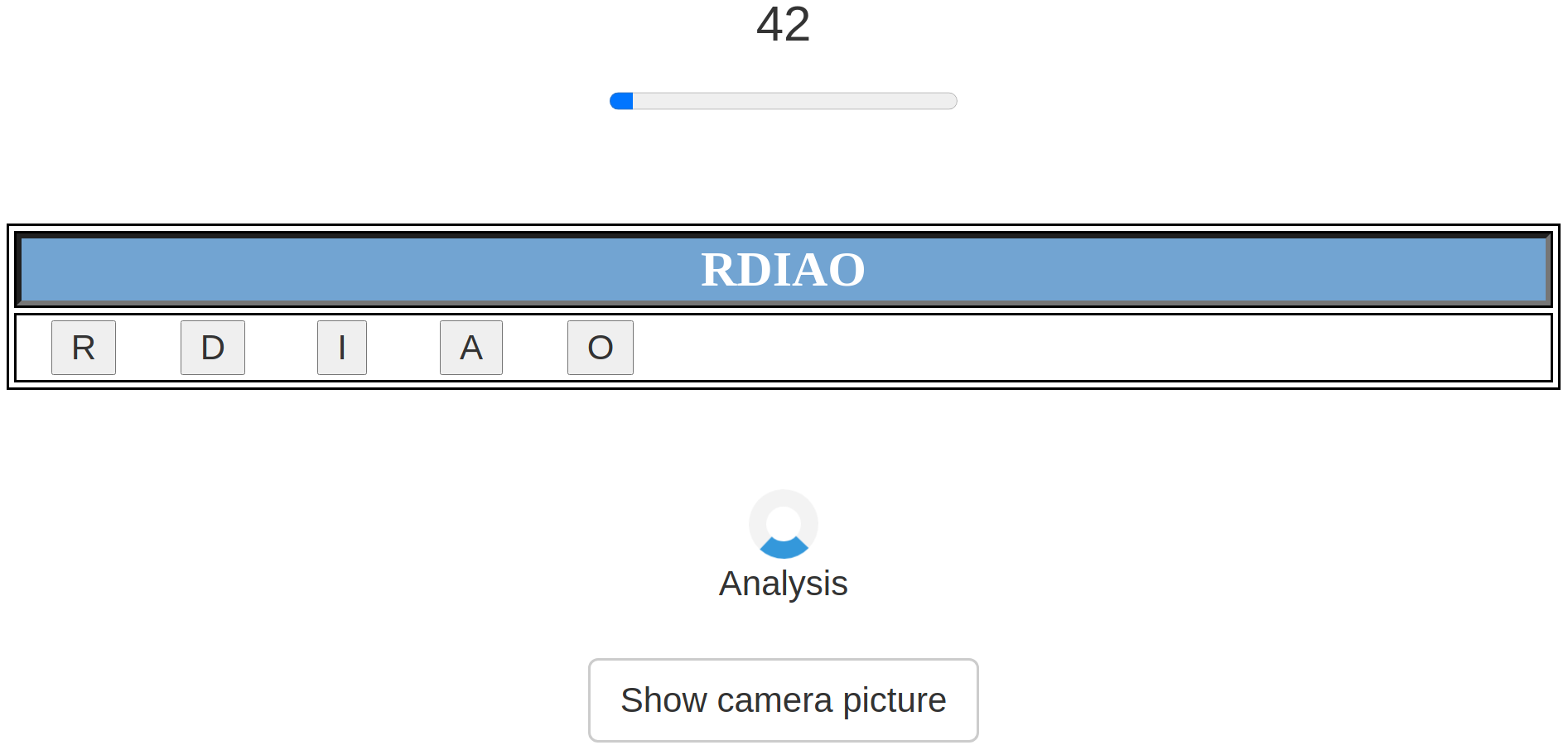}
        \label{fig:camera}}

    \caption{Participants were randomly assigned to one of the three \textit{Adaptivities}. Each word puzzle had a time limit of 45 seconds, upon which the experiment continued with the next word puzzle. \textbf{(a):} Participants either learned that the task difficulty was not adapted at all or was adapted according to the number of correct and wrong answers in a staircase model. The correct word was ``PERSON''. \textbf{(b):} We explained that the difficulty was adapted using an \ac{AI}, which uses the camera image to deduce workload and stress-levels. An animation pretended a constant analysis where the camera feed could be viewed. The correct word was ``RADIO''.}
    \label{fig:wordpuzzle}
\end{figure*}

The experiment presented participants with word puzzles to solve. Each puzzle consisted of a randomized sequence of characters, which they had to select in the correct order to spell an English word. Their selection was incrementally displayed in a text field above this random sequence. Upon selection, characters were removed and thus could not be selected again. After choosing all available characters the puzzle ended, and the next word puzzle was presented. A visible time limit of 45 seconds was set for each trial. The participant's responses were automatically recorded when the last character was selected or after the 45 seconds, whichever came first. The \textit{Adaptivity} condition was displayed throughout the experiment. This word puzzle task allowed us to modulate the task difficulty noticeably for participants (i.e., through the number of presented characters) while not revealing the correct answer immediately. Figure~\ref{fig:wordpuzzle} depicts an example of the word puzzles for each condition.

Overall, participants solved 20 different word puzzles during the experiment. All participants received the same word puzzles, presented in random word order and a randomized arrangement of characters. \mychange{Hence, the participants solved 20 word puzzles in every condition. The word pool in Experiment I consisted of the same 20 words.} Greater task difficulty was defined by increased word lengths, where the word length was normally distributed (i.e., between three and eight characters). Words were selected from the Oxford 5000\footnote{\url{www.oxfordlearnersdictionaries.com/wordlists/oxford3000-5000} - last access \lastaccess} list, which contains the essential English basic words. \mychange{The rules of the word puzzle were well-defined and had an unambiguous solution. The game rules were explained to the participants while the game complexity was varied through the number of displayed characters (i.e., higher number of characters lead to a higher task complexity) and the familiarity rating of the words. Hence, the game difficulty could be quickly varied to suggest difficulty adaptations to the player.}

\subsubsection{Measures}
After informing the participants about the \textit{Adaptivity} they had been assigned to, we asked them about their expected performance of the \textit{Adaptivity} on a seven-point Likert scale: ``You are in the group where the word puzzles are adjusted by your task performance. Do you think you will do better compared to the other groups?''\footnote{1 = I will perform much worse; 7 = I will perform much better.}. This quantified the effect of our system description manipulation~\cite{de1996placebo}. When solving the word puzzles, we measured the required time in seconds and the number of errors for each answer. An error was logged whenever a wrong word was chosen by selecting the last remaining character, or when the time limit of 45 seconds was up. We employed the raw \ac{TLX}~\cite{HART1988139, doi:10.1177/154193120605000909} questionnaire after the experiment to measure the perceived workload.

\subsubsection{Procedure}
\begin{figure}
    \centering
    \includegraphics[width=0.95\columnwidth]{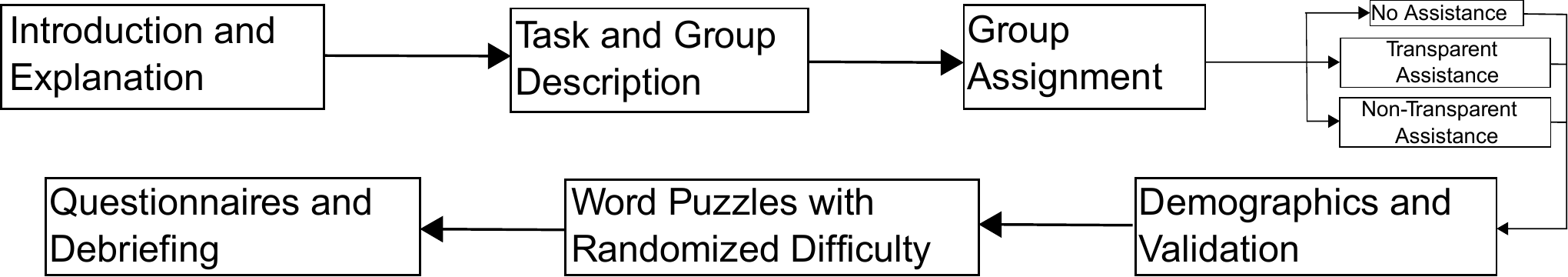}
    \caption{The study procedure of Experiment I.}
    \label{fig:study_procedure_exp1}
\end{figure}

The course of the study was explained to the participants by visiting a web page that described the evaluation of an intelligent online assistant. Afterwards, participants provided consent in accordance with the declaration of Helsinki. Then we explained the three adaptivities. This was followed by demographic questionnaires, including the participant's self-estimated technical competence and belief in improved performance over the other two adaptivities, which was probed on a seven-point Likert scale, respectively. Next, a validation of the participant's understanding of the placebo stories was conducted. Participants were asked to fill out three multiple-choice questions that tested their understanding of the intention of the study, modulation of task difficulty, and the available adaptivities (see Appendix). This information was later used to exclude participants who misconceived the study details. Before the quiz, participants were randomly assigned to one of the three adaptivities. Following that, they were asked to solve the 20 word puzzles and to fill in a \ac{TLX} questionnaire~\cite{HART1988139, doi:10.1177/154193120605000909}. The experiment took approximately 15 minutes. The participants were debriefed regarding the deception after the experiment. They could then choose if they wanted to withdraw their data or still provide them for further processing. The experiment procedure is depicted in Figure~\ref{fig:study_procedure_exp1}. The German Society for Psychology and the local institutional ethics board provided ethical approval for the study\footnote{Ethical assessment ID: EK-MIS-2020-023}.

\subsection{Participants}
\mychange{We used Prolific\footnote{\url{https://prolific.co}} to recruit participants for our study. In Experiment I, we set the filter criteria to native speakers from the US and the UK. We used the same filter criteria for Experiment II as in Experiment I in addition to excluding participants from Experiment I. Participants were compensated with nine Pounds per hour. We removed participants who did not pass the validation questions or who completed the word puzzle in a very short time. Subsequently, participants providing invalid data were removed from the analysis and did not receive compensation.} Three hundred forty nine participants completed our experiment successfully. The data was cleaned before analysis. Participants who did not answer all three validation questions correctly, achieved less than 10\% of correct answers in the word puzzles (i.e., 2 answers or fewer), took less than 100 seconds to complete all puzzles, or did not answer the word puzzles or questionnaires completely were excluded from the data analysis. Participants were compensated with three British pounds for their participation. Of the 369 participants, 75 (21.48\%) did not carefully read the instructions as they could not answer all three questions regarding the system descriptions. Further, 16 participants (4.58\%) did not comply with our instructions and solved less than 2 of the 20 word puzzles, three of whom (0.86\%) did run through in a reasonable time frame, but they were therefore still excluded from our statistical analysis. This left 255 (69.11\%) participants after preprocessing the data. One hundred sixty six participants were female, 80 were male, four identified with another gender, and one participant preferred not to disclose their gender. On average, participants were of medium age (\textit{M} = 32.13; \textit{SD} = 12.41), and living in English-speaking countries. Participants rated their technical competency with novel technologies on a seven-point Likert scale, all reporting a medium to good proficiency with technological systems (\textit{M} = 4.61; \textit{SD} = 1.30). Eighty four participants were briefed with \textit{No Adaptivity}, 93 received the \textit{\transparent} and 78 obtained the \textit{\nontransparent} narrative\footnote{Sample size was determined by simulation of a between-subjects ANOVA on percent correct scores ($\mu$1 = .60; $\mu$2 = .65 $\mu$3 = .70.; \textit{SD} = .1; $\alpha$ = .05; N = 300, 1-$\beta$=100\%, number of simulations = 2000), which yielded a minimal detectable effect size of partial a eta =.06 and thus small effects are detectable in our experimental design.}.

\subsection{Analysis}
\mychange{In the following, we analyze differences between the three groups regarding their subjective and objective performance indices. \mychange{We will test for the effects of our manipulation by comparing each adaptive condition to the \textit{No Adaptivity} condition Mann-Whitney U-tests or Wilcoxon-rank test for paired samples. We will test user expectations before and after interaction with the system in a rank-aligned repeated-measures \ac{ANOVA} \cite{kay2016package}. Follow-up analysis is based on Bonferroni-corrected ART-C contrasts \cite{Elkin21} Correlations for Likert-type scales (ordinal) are estimated using Spearman-rank correlations.} The \textit{No Adaptivity} always compromises the baseline for all measurements. NULL-hypotheses relevant for our main research questions are evaluated using Bayes factors computed using the BayesFactor package \cite{morey2015package} or JASP \cite{wagenmakers2018bayesian}\footnote{Note that we have tried different weakly informative priors. However, none of them have changed the results substantially, so we stuck with the default priors of the BayesFactor-Package.}.}

\subsection{Subjective Evaluation}
We analyzed the self-assessed subjective task performance before and after the word puzzle quiz, contrasting each assistance condition and the control-condition as a between-subjects factor. If system descriptions change expectations and the evaluation of effectiveness, we would expect mean differences in performance expectancy in favor of the \textit{\nontransparent} and \textit{\transparent} conditions as compared to the \textit{No Adaptivity}-condition to carry to performance ratings after the interaction. 

\begin{figure*}
    \centering
    \subfloat[][]{
        \includegraphics[width=0.5\columnwidth]{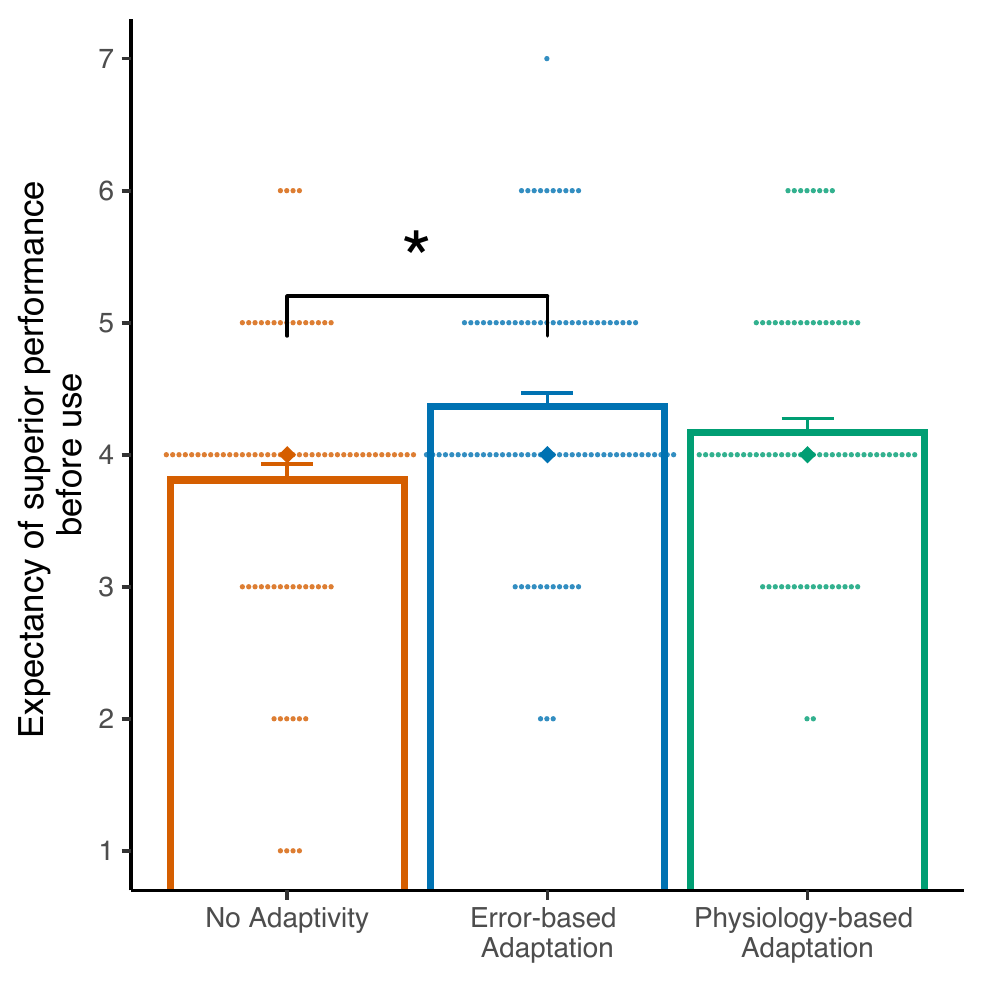}
        \label{fig:expect_pre_group}}
    \subfloat[][]{
        \includegraphics[width=0.5\columnwidth]{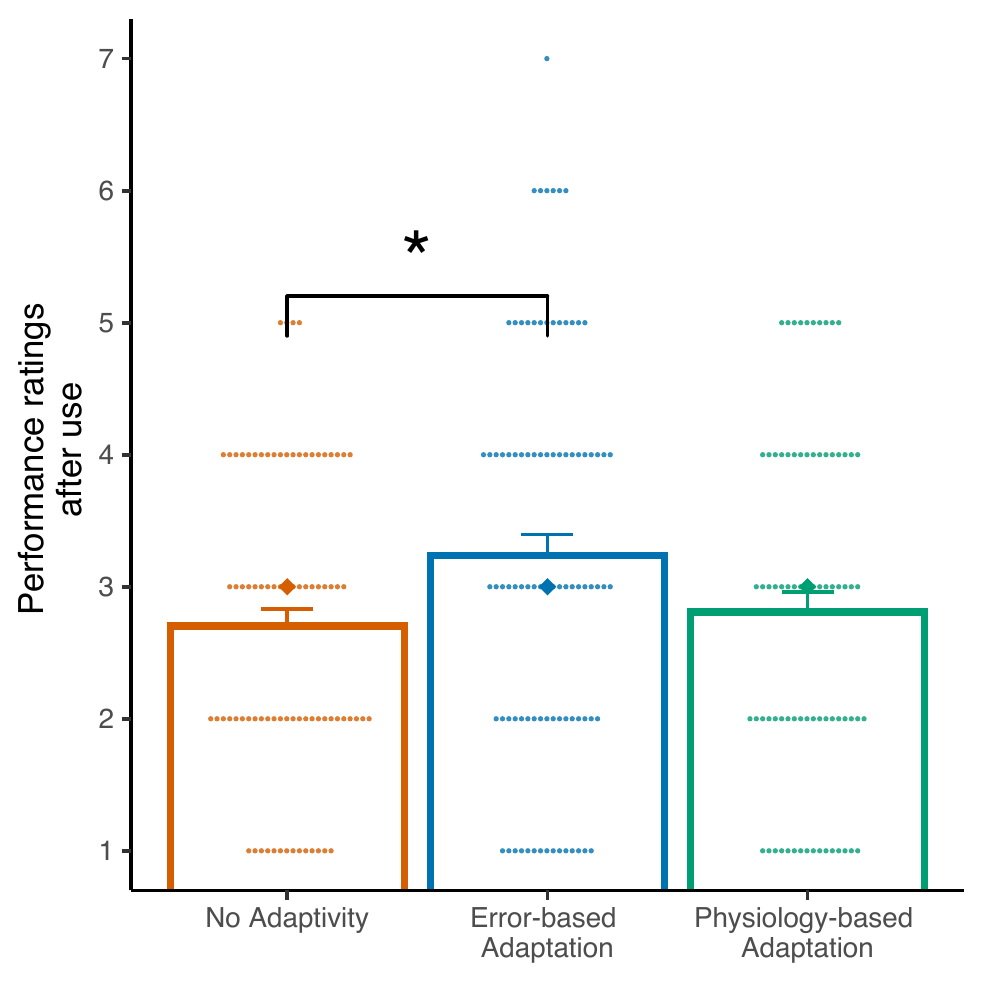}
        \label{fig:expect_post_group}}
     \caption{\textbf{(a):} Averaged measured expectations regarding the anticipated task performance of the assigned \textit{Adaptivity} \textbf{before} the experiment. \textbf{(b):} Averaged measured expectations regarding the perceived performance \textbf{after} the experiment. The error bars depict the standard error of the mean. Brackets indicate significant differences. Dots are individual data points. \mychange{The rhombus indicates the median value.}}
    \label{fig:expect_pre_group_corrplot}
\end{figure*}

In accordance with our hypothesis, the narrative of the adaptivity did affect the users' subjective estimates of task performance prior to solving the word puzzles for the \textit{\transparent}. Participants in the \textit{\transparent}, \textit{U} = 6434.50, \textit{p} = .001, \textit{Z} = -3.24, \textit{r} = 0.24, expected their performance to be superior to the \textit{No Adaptivity} modality (i.e. the control group). Figure~\ref{fig:expect_pre_group_corrplot} shows this change in performance expectations.\mychange{ This was not significant for the  \textit{\nontransparent}, \textit{U} = 6364.50, \textit{p} = .086, \textit{Z} =-1.72, \textit{r} = 0.135. \footnote{Here, using the Bayesian equivalent of a Mann-Whitney U-Test, \cite{wagenmakers2018bayesian}, we can show that that this is not due to the NULL-Hypothesis (no difference between ranks) being true, BF$01$ = 1.20, but rather due to insufficient amount of information.} Note that this pattern was mirrored in performance ratings after interaction with the system.  In the \textit{\transparent}, \textit{U} = 6714.50, \textit{p} = .022, \textit{Z} = -2.29, \textit{r} = 0.172 but not in the \textit{\nontransparent}, \textit{U} = 6479.50, \textit{p} = .673, \textit{Z} = 0.42, \textit{r} = 0.03. \footnote{The Null-hypothesis, is about 5 times more likely than the alternative hypothesis, BF01=5.989}, when compared to \textit{No Adaptivity} (see Figure~\ref{fig:expect_post_group})}. 

\mychange{We computed a repeated-measures rank-aligned \ac{ANOVA} (between = \textit{Adaptivities}, within= \textit{Time point (before vs. after UI interaction)} using the ArTool that revealed a significant effect within the \textit{Adaptivities},  \textit{F}(2, 252) = 5.36, \textit{p} < .001.  We also found a significant main effect of \textit{Time point (before vs. after UI interaction)}, \textit{F}(1, 252) = 179.50, \textit{p} < .001. Participants judged their performance to be better before solving the word puzzles (\textit{M} = 4.12; \textit{SD} = 1.04) as compared to after completing the word puzzles (\textit{M} = 2.93; \textit{SD} = 1.37). The interaction of the factors in the \ac{ANOVA} was not significant, \textit{F}(2, 252) = 0.54, \textit{p} = .584. \footnote{To verify if it is justified to conclude that expectation readjustment is uniform across \textit{Adaptivity} levels, which amounts to rejecting the NULL-hypothesis, we carried out an additional Bayesian analysis. We performed a parametric repeated-measures Bayesian ANOVA \cite{wagenmakers2018bayesian} on performance ratings for each time-point. Note that this parametric analysis could be considered more liberal. The Bayes factor comparing the model with and without interaction effect suggested that the data were 14.78 : 1 in favor of the simple effect model, making it 15 times more likely to be true as compared to the alternative hypothesis of differing means for each \textit{Adaptivity} over time given the data. Thus, the data are in agreement with user expectation decrease being uniform.} Bonferroni-adjusted post-hoc contrasts on ranks across \textit{Adaptivities} could show that our manipulation resulted in an expectation of superior task performance, averaged across Time Point only for the \transparent and the control-condition, \textit{t}(252) = -3.16, \textit{p} = .005 but no difference in ranks between the \nontransparent and the control-condition, \textit{t}(252) = 0.85, \textit{p} >.99,  or for the comparison of \nontransparent and \transparent, \textit{t}(252) = -2.23, \textit{p} =.080. We can conclude that performance ratings before and after interaction differed for the \textit{\transparent} from ratings the \textit{No Adaptivity}-group but not for the \nontransparent.} 

Both assistants were believed to be specifically adaptive after interaction. Participants responded to the question ``The word puzzles have adapted well to the task difficulty'', rated on a seven-point Likert scale (1 = Do not agree at all; 7 = Absolutely agree), with a relatively more agreement as compared to the \textit{No Adaptivity} condition, (\textit{M} = 3.54; \textit{SD} = 1.43), \textit{\transparent}: (\textit{M} = 4.30; \textit{SD} = 1.18), \mychange{\textit{U} = 6253.00, \textit{p} < .001, \textit{Z} = -3.69, \textit{r} = 0.27, and \textit{\nontransparent}: (\textit{M} = 4.00; \textit{SD} = 1.37), \textit{U} = 7004.00, \textit{p} = .026, \textit{Z} = -2.20, \textit{r} = 0.17}. Therefore, making perceived adaptation in the \textit{\transparent}, where participants could validate functionality similar to the \textit{\nontransparent} condition, we follow that it is highly unlikely that subjects' validation of system functionality had any impact on performance expectations and thus effectiveness. 

\subsection{Number of Errors}
 Task performance was operationalized as the number of correct responses participants made during the word puzzle task (i.e., percent of correct answers). Participants solved the tasks with an average accuracy of 69.90\% (\textit{SD} = 14.88\%) and agreed on a 7-point Likert scale (\textit{M} = 3.53; \textit{SD} = 1.19) that the word puzzles were easy to solve. Thus, objective and subjective difficulty of the word puzzles were considered easy to medium, allowing for an ideal level of variation. We further investigated if manipulation of system description changed task performance. None of the conditions differed significantly (\textit{\nontransparent}: \textit{M} = 67.63\% \textit{SD} = 14.61\%, \textit{t}(159.86) = 1.28, \textit{p} = .203, \textit{d} = 0.20; \textit{\transparent}: \textit{M} = 71.14\% \textit{SD} = 14.68\% , \textit{t}(171.54) = 0.23, \textit{p} = .821, \textit{d} = 0.03) as compared to the control-condition (\textit{M} = 70.63\%, \textit{SD} = 15.28\%). 

The expectancy of task performance (before interaction) was substantially correlated \mychange{(spearman-correlations)} with the relative number of correct responses during interaction, \textit{r} = .25, \textit{p} < .001. This correlation was not uniform across groups, \textit{\nontransparent}: \textit{r} = .40, \textit{p} < .001, \textit{p} < .001 , \textit{\transparent}: \textit{r} = .27, \textit{p} = .008, control-condition: \textit{r} = .12, \textit{p} = .289. Thus, we concluded that if performance expectations can influence actual task performance, then this is most likely for the \textit{\nontransparent}.\mychange{ We found a similar pattern for the performance ratings after the interaction, \textit{r} = .40, \textit{p} < .001, which were largely uniform across \textit{Adaptivities}, .39 > \textit{r} < .44. The higher they rated their own performance, the higher their individual percentage was correct. Therefore, self-assessment of performance was valid across condition.} 

\subsection{Workload}
There was no effect on workload measured by the raw \ac{TLX} for the placebo conditions (\textit{\nontransparent}: \textit{M} = 65.91; \textit{SD} = 13.68), \textit{t}(156.70) = -0.21, \textit{p} = .831, \textit{d} = -0.03; \textit{\transparent}: \textit{M}= 64.26; \textit{SD} = 14.91),  \textit{t}(165.74) = 0.47, \textit{p} = .640, \textit{d} = 0.07, control; \textit{M} = 65.39; {SD} = 17.07) . \mychange{This is likely because subjective performance expectations (before UI use) were unrelated, \textit{r(spearman)} = -.01, \textit{p} = .818. However, performance ratings (after UI use) were  negatively related to the perceived workload, \textit{r(spearman)} = -.24, \textit{p} < .001.}  Task performance (percent correct) was also strongly negatively related to workload, \textit{r}(253) = -.35, \textit{p} < .001. Participants who reported low workload performed significantly better. Note that this relation was largely uniform across groups, -.30 > \textit{r} < -.40,' and thus did not warrant a more fine grained analysis. 

\begin{table}
\caption{Means (\textit{M}) and standard deviations (\textit{SD}) of technology acceptance scales as a function of groups. }
\label{tab:UTAUTWELLS}
\resizebox{\columnwidth}{!}{
\begin{tabular}{rrrrrrr}
\toprule
 &
  \multicolumn{1}{l}{\begin{tabular}[c]{@{}l@{}} \textbf{\transparent}
  \end{tabular}} &
  \multicolumn{1}{l}{\textbf{\nontransparent}}

&
  \multicolumn{1}{l}{} &
  \multicolumn{1}{l}{} &
  \multicolumn{1}{l}{} &
  \multicolumn{1}{l}{} \\
 
 &
  \multicolumn{1}{r}{\textit{M (SD)}} &
  \multicolumn{1}{r}{\textit{M (SD)}} &
  \multicolumn{1}{l}{\textit{df}} &
  \multicolumn{1}{l}{\textit{t}} &
  \multicolumn{1}{l}{\textit{p}} &
  \multicolumn{1}{l}{\textit{d}} \\
  \cline{2-7}
Performance Expectancy  & 14.80 (4.96) & 12.17 (6.08) & 148.41 & 3.06  & \textbf{.003} & 0.48  \\
Effort Expectancy       & 16.62 (4.92) & 15.14 (6.57) & 140.48 & 1.64  & .102 & 0.26  \\
Hedonic Motivation      & 11.42 (4.15)  & 10.40 (4.74) & 154.46 & 1.48  & .140 & 0.23  \\
Personal Innovativeness & 13.35 (4.29)  & 12.47 (4.68) & 158.18 & 1.27  & .205 & 0.20  \\
Novelty                 & 13.02 (3.54)  & 12.47 (5.00) & 135.21 & 0.81  & .419 & 0.13  \\
Attitude                & 12.97 (3.51)  & 11.63 (4.89) & 136.45 & 2.02  & \textbf{.045} & 0.32   \\
Overall Reward          & 12.23 (3.80)  & 11.00 (4.90) & 143.66 & 1.80  & .074 & 0.28  \\
Behavioral Intention   & 9.96 (3.93)   & 8.79 (4.62)  & 151.91 & 1.75  & .082 & 0.27  \\
Overall Risk            & 10.04 (4.10)   & 10.33 (4.55) & 156.60 & -0.43 & .665 & -0.07\\
\bottomrule
\end{tabular}
}
\end{table}

\subsection{Specificity of Perceived Adaptiveness}
We calculated a sum score for the perceived adaptiveness regarding the individual stress-level and adaptiveness based on prior performance as presented in previous research~\cite{denisova2015placebo} (see Appendix). The score could range from five to 35. High scores here indicate a high level of perceived adaptiveness based on either the physiological stress-level or performance. Adaptiveness based on task performance did differ between the \textit{No Adaptivity} (\textit{M} = 17.36; \textit{SD} = 7.49) and \textit{\transparent} (\textit{M} = 23.52; \textit{SD} = 6.23), \textit{t}(162.13) = -5.91, \textit{p} < .001, \textit{d} = -0.90. This difference was also significant when comparing the \textit{No Adaptivity} to the \textit{\nontransparent} (\textit{M} = 21.27; \textit{SD} = 6.71), \textit{t}(159.81) = -3.51, \textit{p} < .001, \textit{d} = -0.55. However, still the \textit{\transparent} was rated to be more adaptive than the  \textit{\nontransparent}, \textit{t}(159.06) = -2.25, \textit{p} = .026, \textit{d} = -0.35.  A different pattern of results was found when comparing \textit{No Adaptivity} (\textit{M} = 15.13; \textit{SD} = 6.44) with the adaptive conditions on the scale measuring perceived adaptiveness based on stress-levels, \textit{\nontransparent}: (\textit{M} = 19.09; \textit{SD} = 6.06), \textit{t}(159.97) = -4.03, \textit{p} < .001, \textit{d} = -0.63, \textit{\transparent} (\textit{M} = 16.57; \textit{SD} = 5.74), \textit{t}(167.13) = 1.56, \textit{p} = .120, \textit{d} = 0.24. Mean adaptiveness based on stress-levels differed significantly between the \textit{\nontransparent} and the \textit{\transparent} \textit{t}(160.41) = -2.78, \textit{p} = .006, \textit{d} = -0.43. Therefore, both adaptivities were perceived to adapt to performance, however only the \textit{\nontransparent} was perceived to adapt to the user's stress. We can therefore conclude that the illusion of adaptiveness is specific to the conditions even after system evaluation, and that users did not validate the adaptive system.

\subsection{Technology Acceptance}
For both adaptive conditions, we measured a range of scales regarding technology acceptance on a seven-point Likert scale. All of the scales compromised sum scores. The scales' performance expectancy (e.g., ``I find the assistant useful in my daily life''), effort expectancy (e.g., ``I find the assistant easy to use'') and hedonic motivation (e.g., ``Using the assistant is fun'') were adapted from the \ac{UTAUT}~\cite{venkatesh2016unified}. Each scale had four items rated from 1 (Do not agree at all) to 7 (Absolutely Agree). Therefore sum-scores could have a range of four to 28. Personal Innovativeness (e.g., ``I like to experiment with new information technologies''), Novelty (e.g., ``Using the assistant is new and refreshing''), Attitude (e.g., ``Overall, it is wise to use the assistant''), Overall reward (e.g., ``Using the assistant will be rewarding''), behavioral intention (e.g., ``I intend to use the assistant to adapt the task difficulty'') and overall risk (e.g., ``Using the assistant would be risky'') were adapted from Wells et al.~\cite{wells2010effect}. The scales had three items, each rated from 1 (Do not agree at all) to 7 (Absolutely Agree). Therefore, sum scores could have a range of three to 21. 

Most of the scales differentiated between the conditions (see Table~\ref{tab:UTAUTWELLS}). Only performance expectancy was slightly increased in the \textit{\transparent} condition compared to the \textit{\nontransparent} which largely resembles the decline in performance expectations (see Figure~\ref{fig:expect_pre_group}). The attitude scale followed a similar trend. We therefore chose not to include the UTAUT in our second study. 

\begin{figure*}
    \centering
    \includegraphics[width=.5\columnwidth]{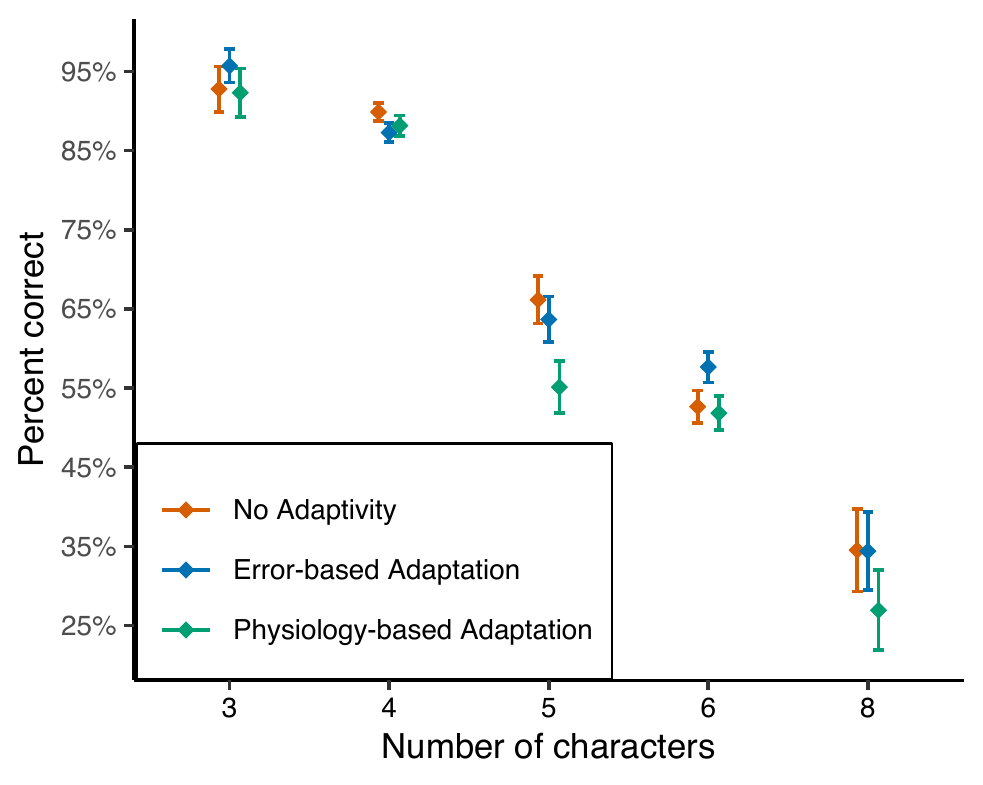}
     \caption{Average percentage correct for the assigned \textit{Adaptivity}. Error bars denote +/- 1 standard error of the mean.}
    \label{fig:nchar_s1}
\end{figure*}

\subsection{Length of Word Puzzles}
We found a potential problem area regarding word length for participants' performance within Experiment I when exploring our data set; namely, that variation between groups is small for easy, three character words and for more difficult words of eight characters, but may differ more when encountering items of medium difficulty of five characters, see Figure~\ref{fig:nchar_s1}. As this variation could have decreased our ability to detect effects, we chose to use a validated set of word puzzles in the English language with five characters for Experiment II, where we conducted a within-subjects study.

\subsection{Discussion}

Our results show that the description of an adaptivity changes the perceived system efficiency of the user. \mychange{Task performance expectations were increased before use when comparing the \textit{\transparent} and \textit{No Adaptivity}-condition (H1). This difference was not found for the \textit{\nontransparent}-condition. In consequence the placebo effect estimates of higher performance after interaction only surfaced for the \textit{\transparent}.} 

Participants perceived the adaptiveness to be specific to their respective groups. Thus, participants in the \textit{\nontransparent} reported that the word puzzles were adapted to their perceived stress-level, and participants in the group that believed they were receiving \textit{\transparent} were strongly convinced that the word puzzles were adapted to fit their performance.\mychange{ We conclude that the possibility of validating the adaptive algorithm in the \textit{\transparent}-condition did not seem to hinder observing an effect on performance ratings. }

Still, belief in the superiority of an adaptive system differed between individuals, and therefore, may need to be systematically controlled to obtain a measurable behavioral effect. \mychange{No effect was found for the \nontransparent condition, the number of errors (H2) and perceived workload (H3) between the groups, likely due to the substantial variance in performance expectation across subjects and the associated low statistical power.} This variance, however, was predictive of the number of errors in the task, especially for the \textit{\nontransparent}-condition. Therefore, we chose to replicate and extend our findings in a within-subjects experimental study to control for the individual perception and variance of the \textit{Adaptivities}.

\section{Experiment II: A Placebo Effect When Comparing Interfaces}
We revealed that an \textit{\transparent} system adaptivity produces placebo effects that impact subjective performance expectations and performance ratings. After interaction with the system, participants still believed in its adaptive function and could distinguish between \mychange{\textit{\nontransparent}} and the \textit{\transparent} condition. We did not find an effect on objective measures regarding task performances, such as percentage of correct answers, possibly due to varying difficulty levels in the puzzles. However, there was a strong correlation of expectancy and task performance for the \textit{\nontransparent}-condition; thus, if user expectations can be manipulated more consistently, \mychange{we expect a placebo effect of the \textit{\nontransparent} on objective task performance and performance ratings.}  Now that we have established the possible size of a placebo effect in a user study with the \textit{\transparent} condition, we will compare the \textit{\nontransparent} adaptive system to a control condition in a within-subject design. \mychange{Hence, we compare the \textit{No Adaptivity} with the \textit{\nontransparent}. We removed the \textit{\transparent} due to its transparency and simplicity. The \textit{\transparent} can be verified regarding its correctness, hence mitigating a potential placebo effect.} Thus, every participant encountered both the \textit{No Adaptivity} and the \textit{\nontransparent} condition in a counterbalanced order. We recruited 100 participants who had not participated in Experiment I, where half of the participants (N = 50) started with the \textit{\nontransparent} followed by the \textit{No Adaptivity} condition. The other half started with \textit{No Adaptivity} followed by the \textit{\nontransparent}.

\begin{figure}
    \centering
    \includegraphics[width=0.95\columnwidth]{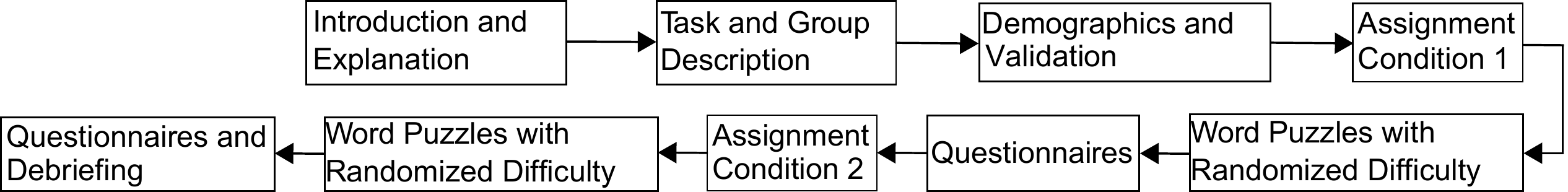}
    \caption{Study procedure of Experiment II.}
    \label{fig:study_procedure_exp2}
\end{figure}

\subsection{Method}
We used a similar web app as described in Experiment I. The web app was modified to implement a within-subjects design where participants were confronted with two conditions, see Figure \ref{fig:study_procedure_exp2}. The participants were instructed with a cover story to solve word puzzles using the \textit{No Adaptivity} and \textit{\nontransparent} condition as described in Experiment I. \mychange{In other words, participants solved word puzzles using each \textit{Adaptivity} in a counterbalanced study design.} 

Each condition contained 40 word puzzles with words that were selected from Gilhooly and Johnson~\cite{doi:10.1080/14640747808400654}. Participants were asked to solve word puzzles of varying difficulty with the use of the two different adaptivities, while items were presented randomly throughout both conditions. \mychange{The participants solved 40 words per adaptivity, resulting in 80 word puzzles for the experiment. The words and their character arrangement were randomized for each condition.} The description of the two adaptivities, that is \textit{No Adaptivity} and \textit{\nontransparent} remained the same as in Experiment I.

We added a ``calibration phase'' to the procedure of Experiment II to familiarize participants with the word puzzles. Before starting with the assigned adaptivity condition, participants were instructed to solve five word puzzles to get used to the task. Before starting with the \textit{\nontransparent} condition, participants were instructed to solve five word puzzles to calibrate an artificial intelligence driving the assistance system. We added a numeric measure of performance expectancy. Before each quiz, we asked participants how many word puzzles they were estimating they could solve correctly. Here, participants could indicate if they expected between zero to 40 correct answers in the upcoming quiz. This provided us with an estimate of the participant's expectancy regarding the anticipated efficiency while solving the task. After each adaptivity, participants filled in a NASA-TLX questionnaire.  As we had established acceptance and usability concerning our system descriptions in Experiment I, we did not collect data on the UTAUT or specificity of adaptiveness in Experiment II. Participants were debriefed after the experiment and could choose if they wanted to withdraw their data or still provide them for processing. The overall experiment took approximately 40 minutes. The experiment procedure is depicted in Figure~\ref{fig:study_procedure_exp2}.

\subsection{Results}

\subsubsection{Participants}
We recruited 100 participants\footnote{Sample size was determined analytically of paired \textit{t}-test on percentage of correct scores (\textit{SD$diff$} = .1;  $\alpha$ = .05; N = 100, 1-$\beta$=80\%), which yielded a minimal detectable effect size of $\mu$ = .028 thus rather small effects were detectable with our experimental design and sample given the statistical model.} through Prolific using the same selection parameters as in Experiment I, adding that they must not have been participants in Experiment I to the recruitment query. Data cleaning was done resembling the previous study: 19\% (19 of 100) participants who did not answer all three validation questions correctly were removed. We also removed 10 instances where participants completed less than 10\% (4 of 40 puzzles) one participant who only completed half the study. No suspiciously short experimental runs were found in this study. This procedure left 75 participants in our sample to analyze. 40 participants reported to be female, 34 reported participants to be of male gender and one participant did not disclose their gender.   

\begin{figure*}
    \centering
    \includegraphics[width=.5\columnwidth]{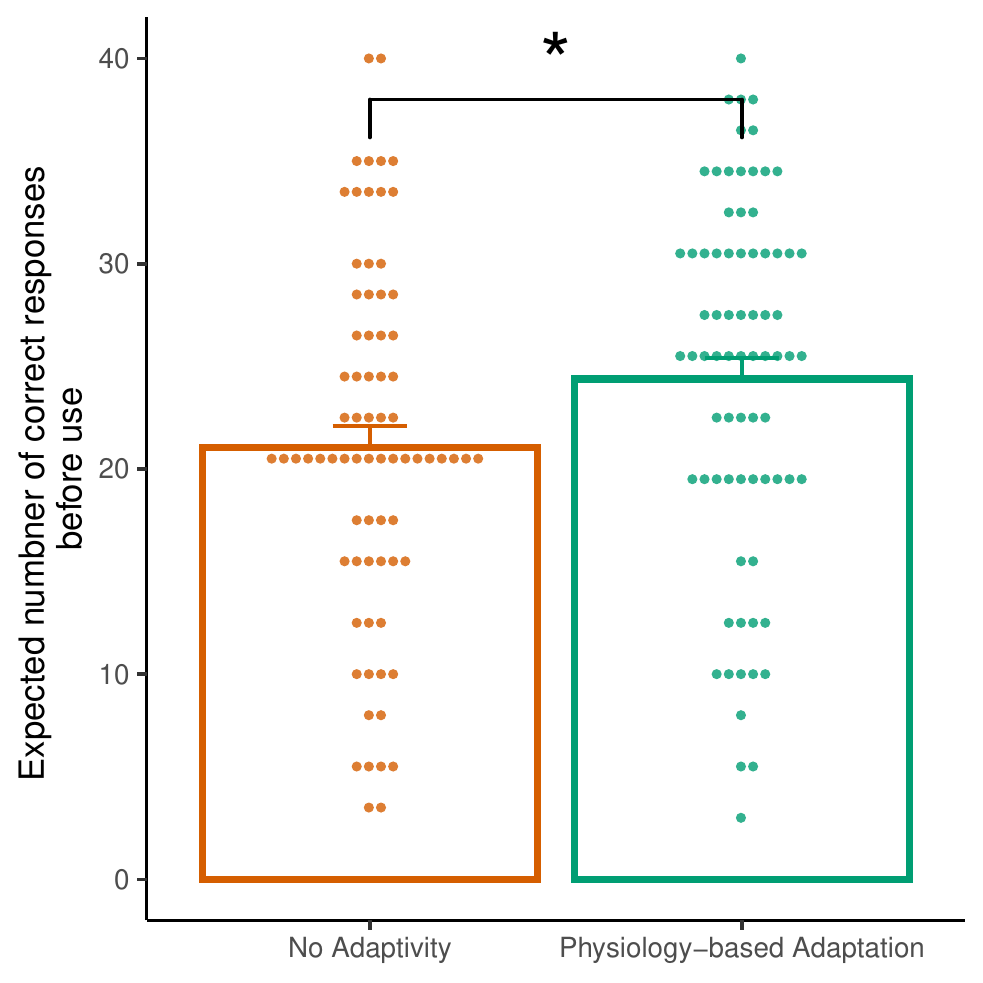}
     \caption{Averaged expected number of correct responses as a function of the \textit{Adaptivity}. Error bars denote +/- 1 standard error of the mean. Blue dots are individual data points.}
    \label{fig:expec_obj}
\end{figure*}

\subsubsection{Subjective Evaluation}
\mychange{We computed a rank-aligned repeated measures ANOVA} to analyze user performance expectations prior and post interaction (within-subjects factor) with the adaptivity, (within-subjects factor). There was a significant effect of the adaptivity, \mychange{\textit{F}(1, 222) = 74.70,} \textit{p} < .001. Participants believed that their performance would be superior when assigned the \textit{\nontransparent} (\textit{M} = 4.31, \textit{SD} =  1.02) placebo condition as compared to the control-condition, (\textit{M} = 2.93, \textit{SD} =  1.26). There was also a significant effect of \textit{Time point} (before vs. after UI interaction), \mychange{\textit{F}(1, 222) = 67.21, \textit{p} < .001,} indicating that after interaction expectations were readjusted. Interaction with the system diminished expectations at an average of about 1.5 points when comparing expectations before (\textit{M} =4.29, \textit{SD} =  0.95) and after system exploration (\textit{M} = 2.95, \textit{SD} =  1.35). The interaction effect of adaptivity and time was again non-significant, \mychange{\textit{F}(1, 222) = 0.04, \textit{p} = .837}. 

Still, we compared group means prior and post interaction and both were distinguishable, see Figure \ref{fig:expect_pre_group_s2}. Note that this lack of an interaction-effect \footnote{To verify if it is justified to conclude that expectation readjustment is uniform across the adaptivity, which amounts to rejecting the NULL-hypothesis, we carried out an additional \mychange{parametric} Bayesian analysis. We performed a repeated-measures Bayesian ANOVA \cite{wagenmakers2018bayesian} on user expectations for each time-point. The Bayes factor comparing the model with and without interaction effect suggests that the data were 4.99 : 1 in favor of the simple effect model, making it five times more likely to be true as compared to the alternative hypothesis of differing means for each adaptivity over time given the data. Thus, the data are in agreement with user expectation decrease being uniform.} implies that the readjustment of expectation was consistent across the adaptivity and thus the placebo effect persisted after interaction with the sham system. 

In line with subjective expectations on the Likert scale ratings, numeric estimates of performance also increased in response to our placebo-manipulation \textit{t}(74) = -2.73, \textit{p} = .008, \textit{d} = -0.32. Participants expected to solve about three more word puzzles when instructed that they would be receiving \textit{\nontransparent} (\textit{M} =24.39, \textit{SD} =  8.93) as compared to \textit{No Adaptivity} (\textit{M} =21.05, \textit{SD} =  8.90, see Figure~\ref{fig:expec_obj}).

\begin{figure*}
    \centering
    \subfloat[][]{
        \includegraphics[width=0.5\columnwidth]{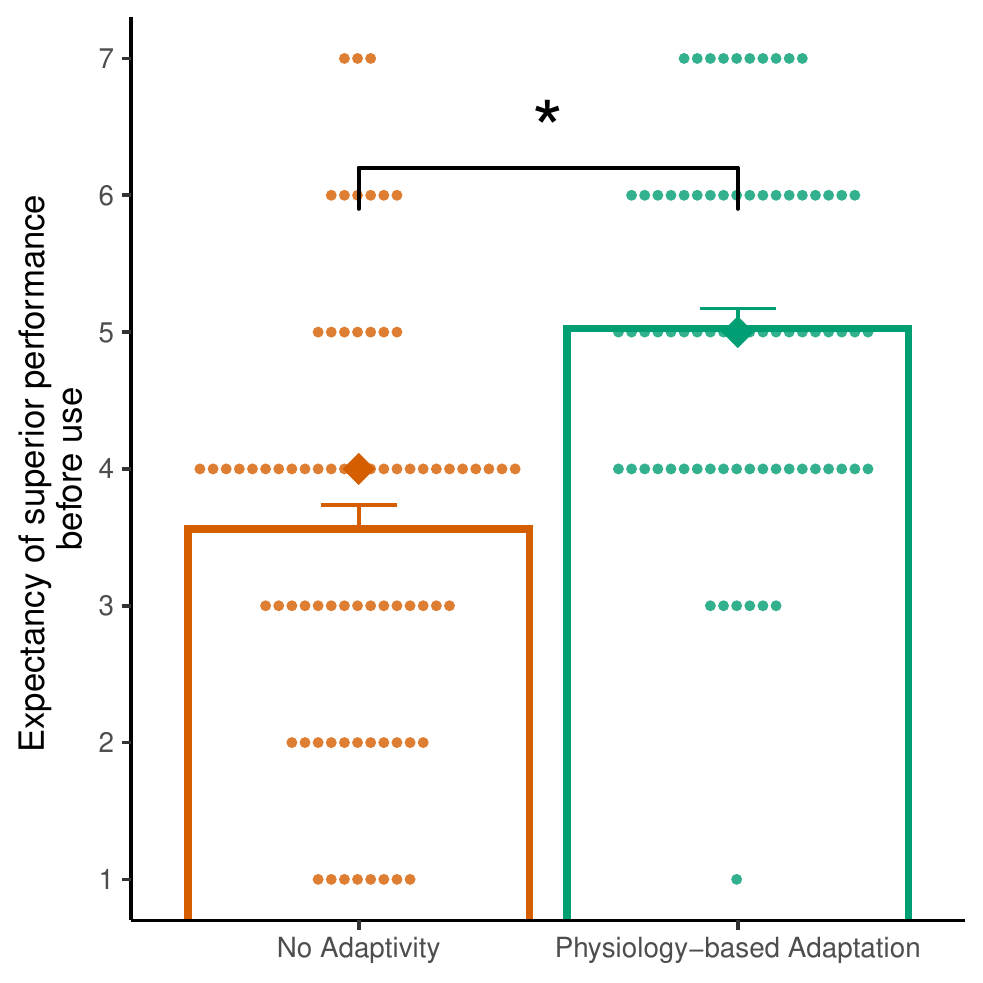}
        \label{fig:expect_pre_group_s2}}
    \subfloat[][]{
        \includegraphics[width=0.5\columnwidth]{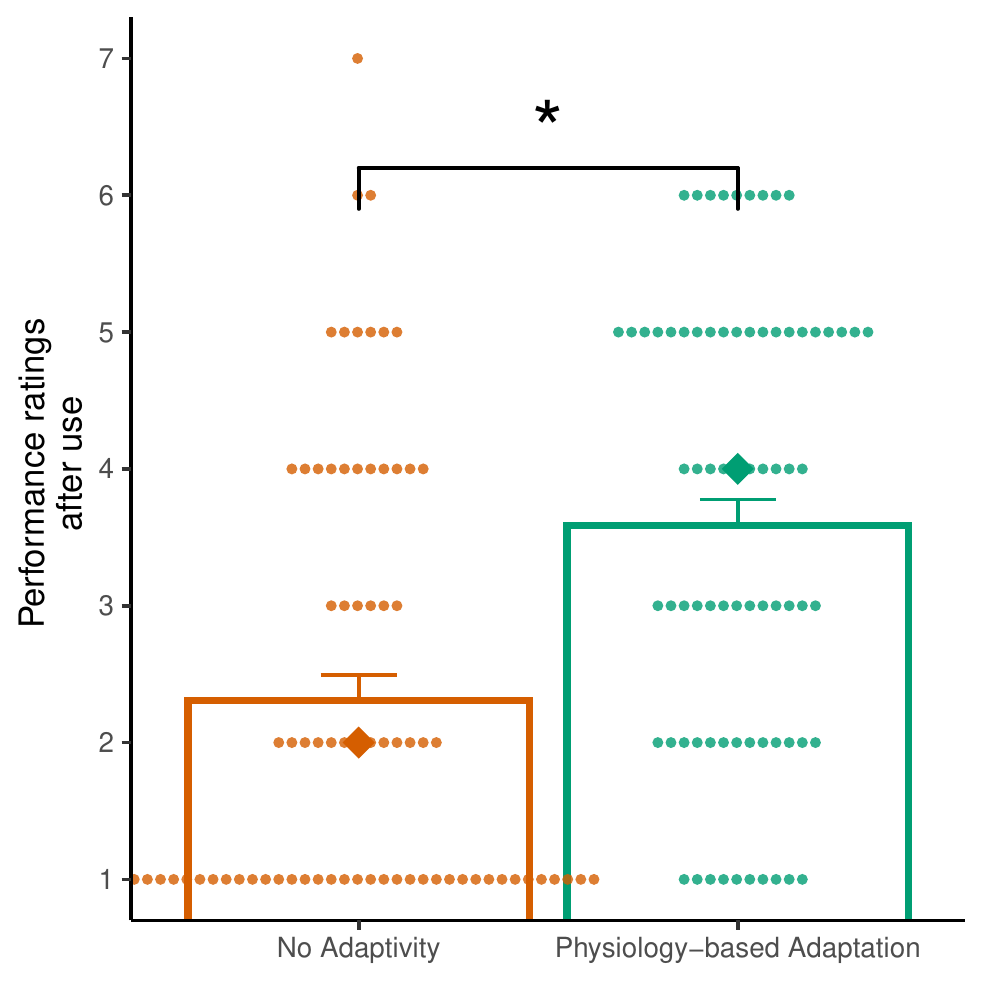}
        \label{fig:expect_post_group_s2}}
     \caption{\textbf{(a):} Averaged measured user expectations regarding the anticipated task performance of the assigned \textit{Adaptivity} \textbf{before} the experiment, \mychange{\textit{Z} = -5.59, \textit{p} < .001, \textit{r} = 0.64. \textbf{(b):} Averaged measured user expectations regarding the perceived usefulness of the assigned \textit{Adaptivity} \textbf{after} the experiment, \textit{Z} = -3.94, \textit{p} < .001, \textit{r} = 0.45}. The error bars depict the standard error of the mean. Brackets indicate significant differences.}
    \label{fig:expect_pre_group_s2}
\end{figure*}

\subsubsection{Number of Errors}
We found no difference in task performance between groups, \textit{t}(74) = -0.15, \textit{p} = .882, \textit{d} = -0.02\footnote{Using the Bayesian equivalent of a \textit{t}-test \cite{kruschke2013bayesian}, we estimate the the NULL-model with no differences between conditions is 7.7 times more likely, given the data, than the alternative model assuming differences in means}. Both the control-group (\textit{M} = 60.77 \%, \textit{SD} =  20.19) and the \textit{\nontransparent} (\textit{M} = 61.20\%, \textit{SD} =  19.63) group did not differ in terms of solving of the word puzzles. Again, we investigated whether individual differences in subjective performance expectations could explain differences in performance between condition, computing the difference in performance expectations between conditions and the difference between conditions in objective performance, i.e. percent correct. In line with the previous study, \mychange{we found a correlation of performance (percent correct) and expectation}, \textit{r}(73)=. 28, \textit{p}=.014. The larger the difference in expectations in favor of the \textit{\nontransparent} prior to the interaction the larger the difference in performance in favor of the \textit{\nontransparent}. 

\subsubsection{Workload}
Again, we found no significant effect on workload as operationalized by the NASA-TLX, between the two conditions, \textit{t}(74) = -1.81, \textit{p} = .074, \textit{d} = -0.21. Still on a descriptive level, the \textit{\nontransparent}-condition produced more workload (\textit{M} = 71.79, \textit{SD} =  14.94) as compared to the control-condition(\textit{M} = 66.83, \textit{SD} = 17.86). There were no significant differences between conditions on the subscales, all \textit{d} < .20.

In sum, we found in Experiment II that user expectations could be manipulated by a description of an adaptive system (H1). Participants expected superior performance when being assigned to the \textit{\nontransparent}-condition as compared to the \textit{No Adaptivity}-condition. As indicated by the number of errors when solving the word puzzles, user performance was not affected by our sham-treatment (H2); however, as in Experiment I, we found a substantial correlation between task performance and user expectations, which gives partial support to H2. Much like in Experiment I, we found no effect of assistance on workload during completion of our task (H3). 

\section{Discussion}

\mychange{Our study investigated if and how evaluation of AI  systems are affected by a placebo effect. A perceived improvement occurred after using the AI system. While we have not found any indications of a placebo effect on objective metrics of task performance, our study shows how system descriptions change performance expectancy and how these affect common usability metrics, e.g. performance ratings of the UTAUT, for different AI systems. The sustained belief of improvement after interaction constitutes the placebo effect.  Considering that describing system functionality is common in HCI studies of AI,  it is possible that some studies using subjective evaluations have fallen short on controlling for placebo effects in their study design. Our results should thus motivate future researchers to consider placebo control to ensure that system evaluations are not biased by user expectations.}

We attribute our findings to three critical factors. First, the \mychange{manipulating system descriptions} manifests subjectively in performance expectations of the \textit{\nontransparent} (H1). These expectations decrease slightly but persist after the use of the system \mychange{(placebo effect; H2)}. Second, there was no direct effect of manipulating performance expectations on task performance across both studies (H2). Notably, this is in line with the literature on placebo and cognitive enhancement. For instance, a sham sound treatment \cite{schwarz2015cognition} or an expectation manipulation of receiving Ritalin, a potent drug that increases focus~\cite{looby2011expectation}, only increased subjective expectations but not the objective performance metrics themselves. \mychange{This notion of placebo effects only manifesting subjectively is also supported by meta-analysis that compared subjective and objective placebo-responses across 130 medical studies \cite{powerless2001}}. However, we found a relationship between performance expectations and the percentage of correct answers: The more an individual participant believed in the superiority of the \textit{\nontransparent}, the larger the performance gain when solving word puzzles. Third, we found in Experiment I that participants do not validate the system's functionality. In Experiment I, both allegedly adaptive systems were indeed judged to be adaptive. Thus, the possibility to validate the adaptive algorithm does not alter the formation of performance expectations. \mychange{This is in line with the literature on automation-bias; the tendency to disregard contradictory information when being supported by automated decision making systems \cite{cummings2017automation}}. We did not find any effects on workload (H3).

\subsection{Theoretical Implications}
\subsubsection{Technology Acceptance}
Our data can be integrated into the current evaluation models of technology acceptance \cite{davis1985technology,venkatesh2012consumer,venkatesh2016unified} based on three points. First, we found that the mere presence of a system description, albeit with no functionality, influences system evaluation. Therefore, system usability and ease of use on a behavioral level are not the only critical factors to technological acceptance; subjective perception must also be considered. Second, in terms of performance expectation, beneficial subjective judgment persisted after system use in both of our experiments. One could explain this continuance of belief in terms of confirmation bias \cite{nickerson1998confirmation}), which describes the propensity of participants to gather evidence for confirming a hypothesis. Here, the confirmatory evidence of receiving an easy puzzle appearing after making an error, could have influenced the judgment more than non-confirmatory evidence, i.e., of receiving no adaptivity at all. Therefore, in line with theories on technology acceptance, we can conclude that expectations can persist even if participants are presented with counterfactual information. Our experimental design was not intended to test expectation confirmation effects, hence our system descriptions were not constructed to implement judgments regarding satisfaction. Instead, we focused on the performance of the interaction with  \mychange{adaptive AI} systems. Still, we observed decreases in performance expectations after interaction with the alleged adaptive systems. However, this decrease was largely uniform across all adaptivities. Third, we observed performance gains only in participants who expected these gains. Technology acceptance models such as the UTAUT~\cite{10.2307/41410412} emphasize the role of subjective evaluation, as participants' demographics, prior experience, and so on can mitigate both performance expectations and their influence over user behavior. 
In sum, our findings can emphasize the role of subjective expectations in theories of technological acceptance and call for close control for these when evaluating "superior" technology. 
\subsubsection{Placebo Theory and Classification}
In placebo research, there are two concurrent mechanisms for explaining placebo effects. Expectancy-oriented accounts suggest that placebo effects appear due to an increased belief in the efficacy of a treatment. Conditioned response accounts, on the other hand, describe the placebo effect on the level of previously obtained stimulus-response (e.g., between taking a pill and healing) associations. Our AI-based adaptive system was entirely novel to the subjects, thus we argue that no stimulus-response associations could form. Therefore, our results agree with the former expectancy-based account of placebo effects. 

The placebo literature in Medicine and Psychology emphasizes either physical artifacts (e.g. pills) or psychological treatment \cite{stewart2004placebo,de1996placebo}. \mychange{However, placebo effects impact upon the confidence that can be placed in allegedly functional power-ups in a range of fields, for example, influencing gaming immersion when employing a phantom AI~\cite{Denisova2019,denisova2015placebo}} or control modules \cite{vaccaro2018illusion}, or when cooperating with an assistive AI. Therefore, our literature survey and the present experiments can add a novel category to the placebo literature; namely, placebos introduced by digital artifacts. Our study is the first to investigate such a placebo in interaction with an alleged \mychange{adaptive AI}. Future research may investigate how digital placebos differ from physical or purely psychological placebo treatments about subjective and objective evaluation metrics.

\subsection{Controlling for Placebo Effects}
Adding digital placebos to the placebo literature would entail quantifying and experimentally controlling for placebo effects in \ac{HCI} studies, much like in psychological and medical studies.

\subsubsection{Control in Pharmacological Trials is not Applicable to HCI}
Indeed, the placebo effect can be so prominent in medicine that many pharmacological trials typically compare a novel medical treatment (e.g., a new medicine or vaccine) against a placebo-control group (see Figure \ref{fig:pcs}) that receives only the placebo. This is done to ensure that participants cannot tell if they are in the experimental or control group, hence controlling user expectations before the actual treatment. User studies in \ac{HCI} are confronted with larger obstacles in terms of placebo-control compared to pharmacological trials~\cite{geers2005goal}. 
\mychange{Participants in user studies can be aware of the novelty of the user interface they are assessing by knowing in which user group they are, i.e., assuming that the novel interface, that is different from a known one, can be better. To investigate the expectancy with the participants’ subjective perception and objective performance, we measure the expectancies before and after using each \textit{Adaptivity}.}

\subsubsection{Polling Expectations Prior and Post-Interaction}
\mychange{In this case, participants are unlikely to be naive or hold neutral expectations when presented with a novel and allegedly powerful technology \mychange{such as AI} that suggests improvement \mychange{in terms of human capabilities}. In other words, it is incorrect to assume an equivalence between control and experimental groups prior to interaction with \mychange{intelligent and non-intelligent systems}. Unfortunately, this casts doubt on the validity and robustness of user studies in that area that involve the interaction with intelligent technology. As far as we know, it is atypical to poll user expectations before and after a user study. Nonetheless, this should be a necessary consideration if a strong belief in the superiority of an intelligent system can generate favorable evaluations. \mychange{Thus, in line with a recent discourse on research practices \cite{Caraban2019}, HCI researchers must entertain the idea that many reported findings might be biased in favor of novel technologies. }}

\subsubsection{Adding Placebo-Control Conditions}
What are the practical implications of placebo effects for conducting user research? User studies should focus on normalizing and controlling the bias of user expectations. By introducing an active control group, or even a placebo-control to the study design, one can investigate whether the novel system exceeds the usability above and beyond the expectations. However, employing a placebo-control can be inefficient, unethical, or not possible, for example, when a placebo can be easily identified or when working with vulnerable patient samples requiring active treatment. In such cases, one can statistically control for the effect of expectations as presented in this study. This requires the quantification of expectations before the test itself with subsequent normalization of the bias afterward in the analysis.

\subsubsection{Objective Measures are Less Susceptible to Placebo Effects}
One could also try to resolve the issue by turning to so-called objective measures~\cite{rosenzweig1933experimental}, where the participant is blind to the measurement rationale employed (e.g. an implicit association test) or physiological measures where the response is hard to modulate deliberately (e.g. electroencephalography~\cite{millett2001hans}). We do not think that this is a promising approach for two reasons. First, we found effects not only on self-reports but also on a behavioral level, albeit no main effect but only a correlation. The participants' performance expectancy improved when they had the assistance of a superior technology (i.e. \textit{\nontransparent}). Second, physiological responses can also be altered by the participants' expectations~\cite{CHITTARO2014663}, hence explaining mixed results when researching implicit physiologically-adapting systems~\cite{strait2014reliability}. In line with placebo studies \cite{kirsch1999specifying, stewart2004placebo}, we recommend assessing the participant's expectations before use and compare them with the expectations after use. Correlating the UI superiority with the user's task performance can also serve as a pretest to anticipate the impact of expectations on the study results. One could also entertain the idea that qualitative methods may be superior when it comes to the bias of user expectations. A skilled interviewer may narrow down and identify the design aspects that might yield a favorable attitude towards the system or if a positive evaluation is only due to differential expectations. A qualitative interview before interaction with a UI could also normalize user expectations by explicitly addressing the role of novelty in evaluating a technological invention and thus making study participants aware of their potentially biased responses beforehand. Whether these strategies present a suitable solution to placebo-control needs to be investigated in future studies.

\subsection{Exploiting the Placebo Effect}

\subsubsection{Persistence of Effects}
Our results show a strong positive correlation between the description-induced subjective user expectations and the task performance during \textit{\nontransparent}. Here, we find that the narrative of the condition leads to a belief towards the utility of adaptivity, which reflects in improved performance compared to participants who do not believe in the superiority of the adaptivity. This shows that high user expectations indeed produce a placebo effect regarding the participant's perception (i.e., H1 and H2 are supported). This trend was clearly visible for the \textit{\nontransparent} condition, but not in the \textit{\transparent}. A potential reason for this is the participants' ability to self-validate their own provided input directly during the \textit{\transparent}. In contrast, \textit{\nontransparent} adds an encouraging non-transparent layer of improvement. Probing the users' beliefs after the experiment resulted in a lower score for the \textit{\nontransparent} compared to \textit{\transparent}, effectively reversing their belief in the superiority of the two adaptivities. This was revealed on the \ac{UTAUT} questionnaire, that showed a significant difference between \textit{\transparent} and \textit{\nontransparent} on the ``Performance Expectancy'' scale. This finding validates our assumption that high user expectations before interaction with an implicitly adapting UI (i.e. \textit{\nontransparent}) lower one's belief in superiority more after interaction, compared to an explicit UI adaptation (i.e., \textit{\transparent}). 

\subsubsection{Designing for Placebo}
\mychange{We suggest, in line with~\cite{denisova2015placebo,vaccaro2018illusion,garcia2021}}, that shaping the user's expectations before interaction with a novel technology can create better user experiences. Thus, communication in the system's abilities and functioning, even if highly inaccurate, can potentially increase usability. This hypothesis on communication in technological innovation certainly deserves further investigation regarding the constraints and applicability of placebo effects in a user evaluation. Especially when considering that the \textit{\nontransparent} condition was judged to be less adaptive after interaction as compared to the \textit{\transparent}, this suggests that dis-confirmed expectations can potentially lead to unfavorable evaluations. Still, if carefully framed, a narrative may gauge the perceived functioning and utility of a system. 

\subsubsection{FATE: Fairness, Accountability, Transparency, and Explainability}
\mychange{
We have not polled the perceived system complexity but assume that there are differences between the \textit{\transparent} and the \textit{\nontransparent} that could relate to differences in perceived transparency. While the \textit{\transparent} could be considered a rather trivial system of an error resulting in an easier item, the \textit{\nontransparent} could be considered a rather complex and less transparent system. For expectations to manifest in performance ratings post-interaction a semi-accurate mental model needs to be established, as only then can performance expectations be confirmed or disconfirmed \cite{thong2006effects,venkatesh2016unified,davis1985technology,garcia2021}. Revisiting our results from this perspective can explain why the \textit{\nontransparent} system in Experiment I could not produce an expectancy effect or a placebo effect. Participants were unable to probe the system before doing the task, thus increasing the perceived transparency of the AI. In contrast, Experiment II was designed as a within-subjects study with a standardized system exploration, doubled number of trials and familiarity with the task due to repeated measurements. Providing participants with more information to build their individual mental model of the relatively complex \textit{\nontransparent} system could have increased perceived transparency. Making the inner working of the AI accessible is a core feature in strengthening the user's belief in performance gains which should be tested in future studies. Considering our results on the placebo effect for the \textit{\transparent} \cite{garcia2021}, post-hoc explanations of system decisions could increase performance ratings.}

\mychange{Our studies can also add to the literature on trust in automated systems. Overtrust refers to the notion that subjective system capabilities exceed objective system capabilities \cite{kundinger2019over}. While these studies typically investigate trust with regard to functional systems, our studies show that non-functional systems can also provide a belief in performance capabilities. One could argue that this is negligible in practice or may even be desirable for technology adoption in private domains including video games \cite{Denisova2019}, but could be detrimental for safety-critical contexts such as a semi-autonomous vehicle. Indeed, the literature on automation bias can show that contradictory information is not integrated in subsequent interaction when using automated systems \cite{cummings2017automation,skitka2000accountability}. Hence, automation bias paired with increased expectations could foster an even stronger placebo effect of AI but also shift accountability from the user towards the autonomous system.
Aligned with this, we know that user perceived  fairness of AI is subjective \cite{araujo2020ai}. Manipulating perceived fairness by changing system descriptions could likewise be possible. 
}

\subsection{Limitations and Future Work}
\subsubsection{Assessing Qualitative Data}
\mychange{
Although we demonstrate the placebo effect induced by expectancy in two experiments, we have only found out a little about the factors that make up the placebo effect. Moving from a large-scale online study to a more controlled in-person study could allow for a closer examination of what factors contributed to participants' ratings of superior performance after interaction. We thus encourage future work that investigates the placebo effect in human-machine interaction to use mixed methods, thus supplementing quantitative data (e.g. physiological and behavioral data) with qualitative statements that can closely examine what made participants believe in the alleged system's functionality. }

\subsubsection{Providing Participants with Incentives}
\mychange{
About 30\% of the participants in each study had to be excluded from the analysis due to careless responding or lack of engagement in the study. While participants were paid for their participation, thoughtful responding was not incentivized. Garcia et al. \cite{garcia2021} have used a consensus-oriented financial incentive, that is, they gave participants a bonus if they chose the system that was preferred by the majority. Future studies that evaluate placebo effects in HCI should follow this approach to use such a consensus-oriented financial incentive in system performance ratings which could promote careful evaluation and building of a representative mental model. Whether this weakens or strengthens the placebo effects needs to be investigated. }

\subsubsection{Long-Term Effects}
Some limitations have to be taken into account. Our current results showed that our participants' positive expectations of the adaptivity systems decreased after experiencing them. However, we could not establish if they might be eradicated entirely or even turn negative. It would be necessary to perform studies on prolonged use to understand this better. A more thorough understanding of the placebo effect, the factors that mitigate it with system familiarity, and the time-scale of these interactions would help us better understand higher-level issues such as early technological adoption, acceptance, and disillusionment.

\subsubsection{Nocebo Effect}
In this study, the placebo was an alleged \mychange{adaptive AI} that should increase performance by selecting appropriate word puzzles and presenting them to the participant. Other studies in the domain of cognitive enhancement also use a so-called nocebo-condition, which is a sham treatment that shapes an expectation of impairing or even worsening task performance. In our case, this would add another condition to the experimental design in which an alleged challenging AI is used, i.e., one that suggests the most challenging word puzzles. This would allow researchers to closely see whether expectations of improvement and impairment similarly produce subjective performance expectations as well as gains or decreases in task performance. We are aware of one study that investigated biasing effects of AI descriptions. Describing art as being generated by AI prompted less favorable evaluations compared to descriptions of human-made artwork \cite{ragot2020}. Thus, adding a nocebo-condition may allow for closer inspection of expectation effects in HCI-studies. 



\subsubsection{Prior User Experience}
It is possible that our instructions on the alleged adaptivities could have directly contributed to the strength of the placebo effect. However, it is worth noting that this is an arbitrary decision that all HCI researchers have to make in introducing their systems to users. The current study is inconclusive for the specific factors that could have resulted in a larger placebo effect in the \textit{\nontransparent} condition. We believe that this is because participants had no access to verifying if the alleged system worked. An alternative explanation that is in line with placebo-conditioning accounts \cite{stewart2004placebo}, could be that the \textit{\nontransparent} was less familiar to our participants than \textit{\transparent} systems. As far as the subjective results indicate, our participants considered both alleged systems to be equally novel. 

\subsubsection{Variation in Novelty}
A valid criticism of our system descriptions could be that the \textit{\transparent} and the \textit{\nontransparent} condition differed with respect to novelty. This could have produced the difference between the conditions for the relation of task performance and expectations. We do not think this is a likely scenario as perceived novelty did not differ between the \mychange{adaptive conditions}. Both adaptive systems were considered to be equally novel. We did not control for individual differences (e.g. verbal fluency, stress) and how this could have impacted upon task performance across the test conditions. However, this study benefits from a large sample size, and participants were randomly assigned to the different adaptivities. Follow-up research could model individual variance in perceived novelty explicitly by performing a pre-study test for individual traits that could modulate the placebo effect and include it in regression analyses.

\subsubsection{Generalizability to other HCI Studies}
Finally, our study presents a simple interaction with technology that constrains the placebo effect to an  \textit{implicit} condition and to a set of participants who believed in the superiority of the assistant. Note that these characteristics are also relevant in medical and psychological placebo studies \cite{beecher1955powerful,enck2013placebo}. Whether other \ac{HCI} studies inherent characteristics such as heightened user expectations and \textit{implicit} technology, and how placebo effects can unfold in response to other study characteristics still needs to be investigated. Nevertheless, our study could strengthen the argument that employing a placebo-control group is necessary to properly infer the superiority of a novel UI~\cite{strait2014reliability}. Our current recommendations are limited to the placebo effect on task performance and user acceptance. However, we plan to evaluate alternative metrics for their susceptibility as well. For this, it would be necessary to extensively evaluate the metrics that are currently used to determine technological adoption and deployment. This could lead to a more comprehensive framework for developing analytical tools that control user expectations in user studies.

\section{Conclusion}
Users approach technology with certain expectations, and researchers themselves will describe their technology prior to testing in ways that could bias expectations. This work shows that subjective expectations and task performance concerning alleged technological innovation can be affected by the system description. We report a placebo effect; a belief of system functionality induced by a sham-treatment after interaction, in \ac{HCI} in the case of collaboration with an AI. Our study suggests that placebo effects can be present in \ac{HCI}. Thus, we believe that much like other fields of human-related research, such as Psychology, medicine, and sport sciences, \ac{HCI} studies should consider placebo-control. We provide the source code of the application, the collected data, and the analysis scripts to foster further research in this area\footnote{\url{https://doi.org/10.17605/OSF.IO/W4Q6J} - last access \lastaccess}.

\section*{Acknowledgments}
This publication has been partially funded by the research initiative ``Instant Teaming between Humans and Production Systems'' co-financed by tax funds of the Saxony State Ministry of Science and Art (SMWK3-7304/35/3-2021/4819) on the basis of the budget passed by the deputies of the Saxony state parliament. Parts of this research were supported by the European Union's Horizon 2020 Programme under ERCEA grant no. 683008 AMPLIFY.

\section*{Appendix}

\appendix

We present the used questionnaires in the following. When referring to Likert items, we use a seven-point Likert scale\footnote{1: Do not agree at all; 7: Absolutely agree}.

\section{Validation Questions}
We asked participants three validation questions in experiment I and experiment II after presenting the instructions and before the participants started to solve the word puzzles. We asked the validation questions to investigate if participants understood the instructions correctly. Participants who did not answer all three questions correctly were excluded from the analysis. 

\begin{enumerate}
    \item What are the two conditions present in this study with regard to the adaptation of the task difficulty?
    \begin{enumerate}
        \item No support through adaptation and AI-based analysis of the camera image.
        \item AI-based analysis of the camera image and adaptation by weather conditions.
        \item Adaptation by analysis of surroundings and adaptation by the current location.
    \end{enumerate}
    \item What is analyzed by the AI during the study?
    \begin{enumerate}
        \item Daylight.
        \item Pulse and emotional states through facial features using the camera image.
        \item Surrounding objects in the image.
    \end{enumerate}
    \item What are the measured metrics used for?
    \begin{enumerate}
        \item Evaluation of the Internet connection.
        \item Adaptation of the task difficulty.
        \item Selection of correct browser settings.
    \end{enumerate}
\end{enumerate}

\section{Customized Questions}
We asked the following custom questions using seven-point Likert items after each condition.

\begin{enumerate}
    \item The word puzzles adapted to my stress level.
    \item The word puzzles were easy to solve.
    \item I am using the same or a similar assistive system on a regular basis.
\end{enumerate}

\section{Adapted Questions}
We adapted the Likert scales from Denisova et al.~\cite{denisova2015placebo} to match the narrative of the task description. We used five out of six questions that were presented using seven-point Likert items. We omitted one question measuring the system intervention on a player's character.

\begin{enumerate}
    \item The word puzzles were generated according to my stress level.
    \item New word puzzles in the task appeared based on my stress level.
    \item The word puzzles were matched to my stress level.
    \item The behavior of the word puzzles changed when I was feeling bored or stressed.
    \item The word puzzles were not adapted sensibly to my stress level.
\end{enumerate}

\section{UTAUT}
 
We used the three UTAUT items \textit{Performance Expectancy}, \textit{Effort Expectancy}, and \textit{Hedonic Motivation} to measure the user expectancy and satisfaction \cite{venkatesh2012consumer}. We adapted the questions to meet the objectives of our study. The questions were asked after every condition (see Table~\ref{tab:appendix_utaut}).

\begin{table}
    \centering
    \caption{Adapted UTAUT questions asked after every condition.}
    \begin{tabular}{ll}
    \toprule
         \textbf{Performance Expectancy} &  \\
         & I find the assistant useful in my daily life.\\
         & Using the assistant increases my chances of achieving things that are important to me.\\
         & Using the assistant helps me accomplish things more quickly. \\
         & Using the assistant increases my productivity.\\
         \midrule
         \textbf{Effort Expectancy} & \\
         & Learning how to use the assistant was easy for me. \\
         & My interaction with the assistant is clear and understandable. \\
         & I find the assistant easy to use. \\ 
         &  It is easy for me to become skillful at using the assistant.\\
         \midrule
         \textbf{Hedonic Motivation} & \\
         & Using the assistant is fun. \\
         & Using the assistant is enjoyable. \\
         & Using the assistant is very entertaining. \\
         \bottomrule
    \end{tabular}
    \label{tab:appendix_utaut}
\end{table}

\section{Wells et al.}
We used all six question items used by Wells et al.~\cite{doi:10.1111/j.1540-5915.2010.00292.x}. We probed the participants after each condition using seven-point Likert items (see Table~\ref{tab:appendix_wells}).

\begin{table}
    \centering
    \caption{Questions about the effect of perceived novelty on the adoption of information technology innovations by Wells et al.}
    \begin{tabular}{l p{10.5cm}}
        \toprule
         \textbf{Personal Innovativeness} &  \\
         & If I heard about a new information technology, I would look for ways to experiment with it. \\
         & Among my peers, I am usually the first to try out new information technologies. \\
         & I like to experiment with new information technologies. \\
         \midrule
         \textbf{Novelty} & \\
         & I found using the assistant to be a novel experience. \\
         & Using the assistant is new and refreshing. \\
         & The assistant represents a neat and novel way of adapting the task difficulty.\\
         \midrule
         \textbf{Attitude} & \\
         & Overall, it is a good idea to use the assistant. \\
         & Overall, it is wise to use the assistant. \\
         & Overall, it is effective to use the assistant. \\
         \midrule
         \textbf{Overall Reward} & \\
         & On the whole, considering all sorts of factors combined, it is rewarding to sign up for and use the assistant.\\
         & Using the assistant will be rewarding. \\
         
         \midrule
         \textbf{Behavioral Intention} & \\
         & I intend to use the assistant to adapt the task difficulty. \\
         & My intention is to use the assistant instead of other non-adaptive systems. \\
         & It is my intention to use the assistant at home or work. \\
         \textbf{Overall Risk} & \\
         & On the whole, considering all sorts of factors combined, it is risky to sign up for and use the assistant. \\
         & Using the assistant would be risky. \\
         & Using the assistant would add great uncertainty to my interaction.\\
         \bottomrule
    \end{tabular}
    \label{tab:appendix_wells}
\end{table}

\section{NASA-TLX}
We used the NASA-TLX questionnaire to quantify the perceived task load for each adaptivity~\cite{HART1988139}. We calculated the raw NASA-TLX based on the six items ranging between one and twenty\footnote{1: Very Low; 20: Very High}.

\begin{enumerate}
    \item Mental Demand: How mentally demanding was the task?
    \item Physical Demand: How physically demanding was the task?
    \item Temporal Demand: How hurried or rushed was the pace of the task?
    \item Performance: How successful were you in accomplishing what you were asked to do?\footnote{1: Perfect; 20: Failure}
    \item Effort: How hard did you have to work to accomplish your level of performance?
    \item Frustration: How insecure, discouraged, irritated, stressed, and annoyed were you?
\end{enumerate}

\bibliographystyle{ACM-Reference-Format}
\bibliography{main}


\begin{thebibliography}{111}


\ifx \showCODEN    \undefined \def \showCODEN     #1{\unskip}     \fi
\ifx \showDOI      \undefined \def \showDOI       #1{#1}\fi
\ifx \showISBNx    \undefined \def \showISBNx     #1{\unskip}     \fi
\ifx \showISBNxiii \undefined \def \showISBNxiii  #1{\unskip}     \fi
\ifx \showISSN     \undefined \def \showISSN      #1{\unskip}     \fi
\ifx \showLCCN     \undefined \def \showLCCN      #1{\unskip}     \fi
\ifx \shownote     \undefined \def \shownote      #1{#1}          \fi
\ifx \showarticletitle \undefined \def \showarticletitle #1{#1}   \fi
\ifx \showURL      \undefined \def \showURL       {\relax}        \fi
\providecommand\bibfield[2]{#2}
\providecommand\bibinfo[2]{#2}
\providecommand\natexlab[1]{#1}
\providecommand\showeprint[2][]{arXiv:#2}

\bibitem[\protect\citeauthoryear{Ang and Mitchell}{Ang and Mitchell}{2017}]%
        {10.1145/3116595.3116623}
\bibfield{author}{\bibinfo{person}{Dennis Ang} {and} \bibinfo{person}{Alex
  Mitchell}.} \bibinfo{year}{2017}\natexlab{}.
\newblock \showarticletitle{Comparing Effects of Dynamic Difficulty Adjustment
  Systems on Video Game Experience}. In \bibinfo{booktitle}{\emph{Proceedings
  of the Annual Symposium on Computer-Human Interaction in Play}} (Amsterdam,
  The Netherlands) \emph{(\bibinfo{series}{CHI PLAY '17})}.
  \bibinfo{publisher}{Association for Computing Machinery},
  \bibinfo{address}{New York, NY, USA}, \bibinfo{pages}{317–327}.
\newblock
\showISBNx{9781450348980}
\urldef\tempurl%
\url{https://doi.org/10.1145/3116595.3116623}
\showDOI{\tempurl}


\bibitem[\protect\citeauthoryear{Araujo, Helberger, Kruikemeier, and
  De~Vreese}{Araujo et~al\mbox{.}}{2020}]%
        {araujo2020ai}
\bibfield{author}{\bibinfo{person}{Theo Araujo}, \bibinfo{person}{Natali
  Helberger}, \bibinfo{person}{Sanne Kruikemeier}, {and}
  \bibinfo{person}{Claes~H De~Vreese}.} \bibinfo{year}{2020}\natexlab{}.
\newblock \showarticletitle{In AI we trust? Perceptions about automated
  decision-making by artificial intelligence}.
\newblock \bibinfo{journal}{\emph{AI \& SOCIETY}} \bibinfo{volume}{35},
  \bibinfo{number}{3} (\bibinfo{year}{2020}), \bibinfo{pages}{611--623}.
\newblock
\urldef\tempurl%
\url{https://doi.org/10.1007/s00146-019-00931-w}
\showDOI{\tempurl}


\bibitem[\protect\citeauthoryear{Beecher}{Beecher}{1955}]%
        {beecher1955powerful}
\bibfield{author}{\bibinfo{person}{Henry~K. Beecher}.}
  \bibinfo{year}{1955}\natexlab{}.
\newblock \showarticletitle{{The powerful placebo}}.
\newblock \bibinfo{journal}{\emph{Journal of the American Medical Association}}
  \bibinfo{volume}{159}, \bibinfo{number}{17} (\bibinfo{date}{12}
  \bibinfo{year}{1955}), \bibinfo{pages}{1602--1606}.
\newblock
\showISSN{0002-9955}
\urldef\tempurl%
\url{https://doi.org/10.1001/jama.1955.02960340022006}
\showDOI{\tempurl}
\showeprint{https://jamanetwork.com/journals/jama/articlepdf/303530/jama\_159\_17\_006.pdf}


\bibitem[\protect\citeauthoryear{Bentley}{Bentley}{2000}]%
        {bentley2000biasing}
\bibfield{author}{\bibinfo{person}{T. Bentley}.}
  \bibinfo{year}{2000}\natexlab{}.
\newblock \showarticletitle{Biasing web site user evaluations: A study}. In
  \bibinfo{booktitle}{\emph{Proceedings of the Conference of the Computer Human
  Interaction Special Interest Group of the Ergonomics Society of Australia
  (OzCHI2000)}}. \bibinfo{pages}{130--134}.
\newblock


\bibitem[\protect\citeauthoryear{Bethge, Kosch, Grosse-Puppendahl, Chuang,
  Kari, Jagaciak, and Schmidt}{Bethge et~al\mbox{.}}{2021}]%
        {10.1145/3472749.3474775}
\bibfield{author}{\bibinfo{person}{David Bethge}, \bibinfo{person}{Thomas
  Kosch}, \bibinfo{person}{Tobias Grosse-Puppendahl}, \bibinfo{person}{Lewis~L.
  Chuang}, \bibinfo{person}{Mohamed Kari}, \bibinfo{person}{Alexander
  Jagaciak}, {and} \bibinfo{person}{Albrecht Schmidt}.}
  \bibinfo{year}{2021}\natexlab{}.
\newblock \bibinfo{booktitle}{\emph{VEmotion: Using Driving Context for
  Indirect Emotion Prediction in Real-Time}}.
\newblock \bibinfo{publisher}{Association for Computing Machinery},
  \bibinfo{address}{New York, NY, USA}, \bibinfo{pages}{638–651}.
\newblock
\showISBNx{9781450386357}
\urldef\tempurl%
\url{https://doi.org/10.1145/3472749.3474775}
\showURL{%
\tempurl}


\bibitem[\protect\citeauthoryear{Boot, Simons, Stothart, and Stutts~Berry}{Boot
  et~al\mbox{.}}{2013}]%
        {Boot2013Psychplac}
\bibfield{author}{\bibinfo{person}{Walter Boot}, \bibinfo{person}{Daniel
  Simons}, \bibinfo{person}{Cary Stothart}, {and} \bibinfo{person}{Cassie
  Stutts~Berry}.} \bibinfo{year}{2013}\natexlab{}.
\newblock \showarticletitle{The Pervasive Problem With Placebos in Psychology
  Why Active Control Groups Are Not Sufficient to Rule Out Placebo Effects}.
\newblock \bibinfo{journal}{\emph{Perspectives on Psychological Science}}
  \bibinfo{volume}{8} (\bibinfo{date}{07} \bibinfo{year}{2013}),
  \bibinfo{pages}{445--454}.
\newblock
\urldef\tempurl%
\url{https://doi.org/10.1177/1745691613491271}
\showDOI{\tempurl}


\bibitem[\protect\citeauthoryear{Briand, Falessi, Nejati, Sabetzadeh, and
  Yue}{Briand et~al\mbox{.}}{2014}]%
        {10.1145/2559978}
\bibfield{author}{\bibinfo{person}{Lionel Briand}, \bibinfo{person}{Davide
  Falessi}, \bibinfo{person}{Shiva Nejati}, \bibinfo{person}{Mehrdad
  Sabetzadeh}, {and} \bibinfo{person}{Tao Yue}.}
  \bibinfo{year}{2014}\natexlab{}.
\newblock \showarticletitle{Traceability and SysML Design Slices to Support
  Safety Inspections: A Controlled Experiment}.
\newblock \bibinfo{journal}{\emph{ACM Trans. Softw. Eng. Methodol.}}
  \bibinfo{volume}{23}, \bibinfo{number}{1}, Article \bibinfo{articleno}{9}
  (\bibinfo{date}{Feb.} \bibinfo{year}{2014}), \bibinfo{numpages}{43}~pages.
\newblock
\showISSN{1049-331X}
\urldef\tempurl%
\url{https://doi.org/10.1145/2559978}
\showDOI{\tempurl}


\bibitem[\protect\citeauthoryear{Browne}{Browne}{2016}]%
        {browne2016adaptive}
\bibfield{author}{\bibinfo{person}{Dermot Browne}.}
  \bibinfo{year}{2016}\natexlab{}.
\newblock \bibinfo{booktitle}{\emph{Adaptive user interfaces}}.
\newblock \bibinfo{publisher}{Elsevier}.
\newblock


\bibitem[\protect\citeauthoryear{Caraban, Karapanos, Gonçalves, and
  Campos}{Caraban et~al\mbox{.}}{2019}]%
        {Caraban2019}
\bibfield{author}{\bibinfo{person}{Ana Caraban}, \bibinfo{person}{Evangelos
  Karapanos}, \bibinfo{person}{Daniel Gonçalves}, {and} \bibinfo{person}{Pedro
  Campos}.} \bibinfo{year}{2019}\natexlab{}.
\newblock \showarticletitle{23 Ways to Nudge: A Review of Technology-Mediated
  Nudging in Human-Computer Interaction}. \bibinfo{pages}{1--15}.
\newblock
\urldef\tempurl%
\url{https://doi.org/10.1145/3290605.3300733}
\showDOI{\tempurl}


\bibitem[\protect\citeauthoryear{Chakraborti and Kambhampati}{Chakraborti and
  Kambhampati}{2019}]%
        {10.1145/3306618.3314281}
\bibfield{author}{\bibinfo{person}{Tathagata Chakraborti} {and}
  \bibinfo{person}{Subbarao Kambhampati}.} \bibinfo{year}{2019}\natexlab{}.
\newblock \showarticletitle{(When) Can AI Bots Lie?}. In
  \bibinfo{booktitle}{\emph{Proceedings of the 2019 AAAI/ACM Conference on AI,
  Ethics, and Society}} (Honolulu, HI, USA) \emph{(\bibinfo{series}{AIES
  '19})}. \bibinfo{publisher}{Association for Computing Machinery},
  \bibinfo{address}{New York, NY, USA}, \bibinfo{pages}{53–59}.
\newblock
\showISBNx{9781450363242}
\urldef\tempurl%
\url{https://doi.org/10.1145/3306618.3314281}
\showDOI{\tempurl}


\bibitem[\protect\citeauthoryear{Chittaro and Sioni}{Chittaro and
  Sioni}{2014}]%
        {CHITTARO2014663}
\bibfield{author}{\bibinfo{person}{Luca Chittaro} {and}
  \bibinfo{person}{Riccardo Sioni}.} \bibinfo{year}{2014}\natexlab{}.
\newblock \showarticletitle{Affective computing vs. affective placebo: Study of
  a biofeedback-controlled game for relaxation training}.
\newblock \bibinfo{journal}{\emph{International Journal of Human-Computer
  Studies}} \bibinfo{volume}{72}, \bibinfo{number}{8} (\bibinfo{year}{2014}),
  \bibinfo{pages}{663 -- 673}.
\newblock
\showISSN{1071-5819}
\urldef\tempurl%
\url{https://doi.org/10.1016/j.ijhcs.2014.01.007}
\showDOI{\tempurl}
\newblock
\shownote{Designing for emotional wellbeing.}


\bibitem[\protect\citeauthoryear{Correll}{Correll}{2020}]%
        {correll2020actually}
\bibfield{author}{\bibinfo{person}{Michael Correll}.}
  \bibinfo{year}{2020}\natexlab{}.
\newblock \bibinfo{title}{What Do We Actually Learn from Evaluations in the
  "Heroic Era" of Visualization?}
\newblock
\newblock
\showeprint[arxiv]{2008.11250}~[cs.HC]


\bibitem[\protect\citeauthoryear{Costa, Adams, Jung, Guimbreti\`{e}re, and
  Choudhury}{Costa et~al\mbox{.}}{2016}]%
        {costa2016}
\bibfield{author}{\bibinfo{person}{Jean Costa}, \bibinfo{person}{Alexander~T.
  Adams}, \bibinfo{person}{Malte~F. Jung}, \bibinfo{person}{Fran\c{c}ois
  Guimbreti\`{e}re}, {and} \bibinfo{person}{Tanzeem Choudhury}.}
  \bibinfo{year}{2016}\natexlab{}.
\newblock \showarticletitle{EmotionCheck: Leveraging Bodily Signals and False
  Feedback to Regulate Our Emotions}. In \bibinfo{booktitle}{\emph{Proceedings
  of the 2016 ACM International Joint Conference on Pervasive and Ubiquitous
  Computing}} (Heidelberg, Germany) \emph{(\bibinfo{series}{UbiComp '16})}.
  \bibinfo{publisher}{Association for Computing Machinery},
  \bibinfo{address}{New York, NY, USA}, \bibinfo{pages}{758–769}.
\newblock
\showISBNx{9781450344616}
\urldef\tempurl%
\url{https://doi.org/10.1145/2971648.2971752}
\showDOI{\tempurl}


\bibitem[\protect\citeauthoryear{Cummings}{Cummings}{2017}]%
        {cummings2017automation}
\bibfield{author}{\bibinfo{person}{Mary~L Cummings}.}
  \bibinfo{year}{2017}\natexlab{}.
\newblock \showarticletitle{Automation bias in intelligent time critical
  decision support systems}.
\newblock In \bibinfo{booktitle}{\emph{Decision Making in Aviation}}.
  \bibinfo{publisher}{Routledge}, \bibinfo{pages}{289--294}.
\newblock


\bibitem[\protect\citeauthoryear{Davis}{Davis}{1985}]%
        {davis1985technology}
\bibfield{author}{\bibinfo{person}{Fred~D. Davis}.}
  \bibinfo{year}{1985}\natexlab{}.
\newblock \emph{\bibinfo{title}{A technology acceptance model for empirically
  testing new end-user information systems: Theory and results}}.
\newblock \bibinfo{thesistype}{Ph.D. Dissertation}.
  \bibinfo{school}{Massachusetts Institute of Technology}.
\newblock


\bibitem[\protect\citeauthoryear{Davis}{Davis}{1989}]%
        {10.2307/249008}
\bibfield{author}{\bibinfo{person}{Fred~D. Davis}.}
  \bibinfo{year}{1989}\natexlab{}.
\newblock \showarticletitle{Perceived Usefulness, Perceived Ease of Use, and
  User Acceptance of Information Technology}.
\newblock \bibinfo{journal}{\emph{MIS Quarterly}} \bibinfo{volume}{13},
  \bibinfo{number}{3} (\bibinfo{year}{1989}), \bibinfo{pages}{319--340}.
\newblock
\showISSN{02767783}
\urldef\tempurl%
\url{http://www.jstor.org/stable/249008}
\showURL{%
\tempurl}


\bibitem[\protect\citeauthoryear{de~Jong, van Baast, Arntz, and
  Merckelbach}{de~Jong et~al\mbox{.}}{1996}]%
        {de1996placebo}
\bibfield{author}{\bibinfo{person}{Peter~J de Jong}, \bibinfo{person}{Robert
  van Baast}, \bibinfo{person}{Amoud Arntz}, {and} \bibinfo{person}{Harald
  Merckelbach}.} \bibinfo{year}{1996}\natexlab{}.
\newblock \showarticletitle{The placebo effect in pain reduction: the influence
  of conditioning experiences and response expectancies}.
\newblock \bibinfo{journal}{\emph{International journal of behavioral
  medicine}} \bibinfo{volume}{3}, \bibinfo{number}{1} (\bibinfo{year}{1996}),
  \bibinfo{pages}{14--29}.
\newblock
\urldef\tempurl%
\url{https://doi.org/10.1207/s15327558ijbm0301_2}
\showDOI{\tempurl}


\bibitem[\protect\citeauthoryear{Denisova and Cairns}{Denisova and
  Cairns}{2015}]%
        {denisova2015placebo}
\bibfield{author}{\bibinfo{person}{Alena Denisova} {and} \bibinfo{person}{Paul
  Cairns}.} \bibinfo{year}{2015}\natexlab{}.
\newblock \showarticletitle{The placebo effect in digital games: Phantom
  perception of adaptive artificial intelligence}. In
  \bibinfo{booktitle}{\emph{Proceedings of the 2015 annual symposium on
  computer-human interaction in play}}. \bibinfo{pages}{23--33}.
\newblock


\bibitem[\protect\citeauthoryear{Denisova and Cairns}{Denisova and
  Cairns}{2019}]%
        {DENISOVA201956}
\bibfield{author}{\bibinfo{person}{Alena Denisova} {and} \bibinfo{person}{Paul
  Cairns}.} \bibinfo{year}{2019}\natexlab{}.
\newblock \showarticletitle{Player experience and deceptive expectations of
  difficulty adaptation in digital games}.
\newblock \bibinfo{journal}{\emph{Entertainment Computing}}
  \bibinfo{volume}{29} (\bibinfo{year}{2019}), \bibinfo{pages}{56 -- 68}.
\newblock
\showISSN{1875-9521}
\urldef\tempurl%
\url{https://doi.org/10.1016/j.entcom.2018.12.001}
\showDOI{\tempurl}


\bibitem[\protect\citeauthoryear{Denisova and Cook}{Denisova and Cook}{2019}]%
        {Denisova2019}
\bibfield{author}{\bibinfo{person}{Alena Denisova} {and}
  \bibinfo{person}{Eliott Cook}.} \bibinfo{year}{2019}\natexlab{}.
\newblock \showarticletitle{Power-Ups in Digital Games: The Rewarding Effect of
  Phantom Game Elements on Player Experience}. In
  \bibinfo{booktitle}{\emph{Proceedings of the Annual Symposium on
  Computer-Human Interaction in Play}} (Barcelona, Spain)
  \emph{(\bibinfo{series}{CHI PLAY '19})}. \bibinfo{publisher}{Association for
  Computing Machinery}, \bibinfo{address}{New York, NY, USA},
  \bibinfo{pages}{161–168}.
\newblock
\showISBNx{9781450366885}
\urldef\tempurl%
\url{https://doi.org/10.1145/3311350.3347173}
\showDOI{\tempurl}


\bibitem[\protect\citeauthoryear{Dinda, Memik, Dick, Lin, Mallik, Gupta, and
  Rossoff}{Dinda et~al\mbox{.}}{2007}]%
        {10.1145/1281700.1281710}
\bibfield{author}{\bibinfo{person}{Peter~A. Dinda}, \bibinfo{person}{Gokhan
  Memik}, \bibinfo{person}{Robert~P. Dick}, \bibinfo{person}{Bin Lin},
  \bibinfo{person}{Arindam Mallik}, \bibinfo{person}{Ashish Gupta}, {and}
  \bibinfo{person}{Samuel Rossoff}.} \bibinfo{year}{2007}\natexlab{}.
\newblock \showarticletitle{The User in Experimental Computer Systems
  Research}. In \bibinfo{booktitle}{\emph{Proceedings of the 2007 Workshop on
  Experimental Computer Science}} (San Diego, California)
  \emph{(\bibinfo{series}{ExpCS '07})}. \bibinfo{publisher}{Association for
  Computing Machinery}, \bibinfo{address}{New York, NY, USA},
  \bibinfo{pages}{10–es}.
\newblock
\showISBNx{9781595937513}
\urldef\tempurl%
\url{https://doi.org/10.1145/1281700.1281710}
\showDOI{\tempurl}


\bibitem[\protect\citeauthoryear{Duarte and Carri\c{c}o}{Duarte and
  Carri\c{c}o}{2012}]%
        {Duarte12}
\bibfield{author}{\bibinfo{person}{Lu\'{\i}s Duarte} {and}
  \bibinfo{person}{Lu\'{\i}s Carri\c{c}o}.} \bibinfo{year}{2012}\natexlab{}.
\newblock \showarticletitle{"Blue Pill or Red Pill?": Placebo Effect and the
  Outcome on Physiological \& Player Performance Metrics}. In
  \bibinfo{booktitle}{\emph{Proceedings of the 4th International Conference on
  Fun and Games}} (Toulouse, France) \emph{(\bibinfo{series}{FnG '12})}.
  \bibinfo{publisher}{Association for Computing Machinery},
  \bibinfo{address}{New York, NY, USA}, \bibinfo{pages}{93–96}.
\newblock
\showISBNx{9781450315708}
\urldef\tempurl%
\url{https://doi.org/10.1145/2367616.2367627}
\showDOI{\tempurl}


\bibitem[\protect\citeauthoryear{Duarte and Carri{\c{c}}o}{Duarte and
  Carri{\c{c}}o}{2013}]%
        {duarte2013cake}
\bibfield{author}{\bibinfo{person}{Lu{\'\i}s Duarte} {and}
  \bibinfo{person}{Luis Carri{\c{c}}o}.} \bibinfo{year}{2013}\natexlab{}.
\newblock \showarticletitle{The cake can be a lie: placebos as persuasive
  videogame elements}.
\newblock In \bibinfo{booktitle}{\emph{CHI'13 Extended Abstracts on Human
  Factors in Computing Systems}}. \bibinfo{pages}{1113--1118}.
\newblock


\bibitem[\protect\citeauthoryear{Eiband, Buschek, Kremer, and Hussmann}{Eiband
  et~al\mbox{.}}{2019}]%
        {10.1145/3290607.3312787}
\bibfield{author}{\bibinfo{person}{Malin Eiband}, \bibinfo{person}{Daniel
  Buschek}, \bibinfo{person}{Alexander Kremer}, {and} \bibinfo{person}{Heinrich
  Hussmann}.} \bibinfo{year}{2019}\natexlab{}.
\newblock \showarticletitle{The Impact of Placebic Explanations on Trust in
  Intelligent Systems}. In \bibinfo{booktitle}{\emph{Extended Abstracts of the
  2019 CHI Conference on Human Factors in Computing Systems}} (Glasgow,
  Scotland Uk) \emph{(\bibinfo{series}{CHI EA '19})}.
  \bibinfo{publisher}{Association for Computing Machinery},
  \bibinfo{address}{New York, NY, USA}, \bibinfo{pages}{1–6}.
\newblock
\showISBNx{9781450359719}
\urldef\tempurl%
\url{https://doi.org/10.1145/3290607.3312787}
\showDOI{\tempurl}


\bibitem[\protect\citeauthoryear{Ekman and Friesen}{Ekman and Friesen}{2003}]%
        {ekman2003unmasking}
\bibfield{author}{\bibinfo{person}{Paul Ekman} {and} \bibinfo{person}{Wallace~V
  Friesen}.} \bibinfo{year}{2003}\natexlab{}.
\newblock \bibinfo{booktitle}{\emph{Unmasking the face: A guide to recognizing
  emotions from facial clues}}.
\newblock


\bibitem[\protect\citeauthoryear{Elkin, Kay, Higgins, and Wobbrock}{Elkin
  et~al\mbox{.}}{2021}]%
        {Elkin21}
\bibfield{author}{\bibinfo{person}{Lisa~A. Elkin}, \bibinfo{person}{Matthew
  Kay}, \bibinfo{person}{James~J. Higgins}, {and} \bibinfo{person}{Jacob~O.
  Wobbrock}.} \bibinfo{year}{2021}\natexlab{}.
\newblock \showarticletitle{An Aligned Rank Transform Procedure for Multifactor
  Contrast Tests}. In \bibinfo{booktitle}{\emph{The 34th Annual ACM Symposium
  on User Interface Software and Technology}} (Virtual Event, USA)
  \emph{(\bibinfo{series}{UIST '21})}. \bibinfo{publisher}{Association for
  Computing Machinery}, \bibinfo{address}{New York, NY, USA},
  \bibinfo{pages}{754–768}.
\newblock
\showISBNx{9781450386357}
\urldef\tempurl%
\url{https://doi.org/10.1145/3472749.3474784}
\showDOI{\tempurl}


\bibitem[\protect\citeauthoryear{Enck, Bingel, Schedlowski, and Rief}{Enck
  et~al\mbox{.}}{2013}]%
        {enck2013placebo}
\bibfield{author}{\bibinfo{person}{Paul Enck}, \bibinfo{person}{Ulrike Bingel},
  \bibinfo{person}{Manfred Schedlowski}, {and} \bibinfo{person}{Winfried
  Rief}.} \bibinfo{year}{2013}\natexlab{}.
\newblock \showarticletitle{The placebo response in medicine: minimize,
  maximize or personalize?}
\newblock \bibinfo{journal}{\emph{Nature reviews Drug discovery}}
  \bibinfo{volume}{12}, \bibinfo{number}{3} (\bibinfo{year}{2013}),
  \bibinfo{pages}{191--204}.
\newblock


\bibitem[\protect\citeauthoryear{Farahat}{Farahat}{2013}]%
        {farahat2013}
\bibfield{author}{\bibinfo{person}{Ayman Farahat}.}
  \bibinfo{year}{2013}\natexlab{}.
\newblock \showarticletitle{How effective is targeted advertising?}. In
  \bibinfo{booktitle}{\emph{2013 American Control Conference}}.
  \bibinfo{pages}{6014--6021}.
\newblock
\urldef\tempurl%
\url{https://doi.org/10.1109/ACC.2013.6580780}
\showDOI{\tempurl}


\bibitem[\protect\citeauthoryear{Finniss, Kaptchuk, Miller, and
  Benedetti}{Finniss et~al\mbox{.}}{2010}]%
        {finniss2010biological}
\bibfield{author}{\bibinfo{person}{Damien~G Finniss}, \bibinfo{person}{Ted~J
  Kaptchuk}, \bibinfo{person}{Franklin Miller}, {and} \bibinfo{person}{Fabrizio
  Benedetti}.} \bibinfo{year}{2010}\natexlab{}.
\newblock \showarticletitle{Biological, clinical, and ethical advances of
  placebo effects}.
\newblock \bibinfo{journal}{\emph{The Lancet}} \bibinfo{volume}{375},
  \bibinfo{number}{9715} (\bibinfo{year}{2010}), \bibinfo{pages}{686--695}.
\newblock


\bibitem[\protect\citeauthoryear{Flaten and Blumenthal}{Flaten and
  Blumenthal}{1999}]%
        {flaten1999caffeine}
\bibfield{author}{\bibinfo{person}{Magne~Arve Flaten} {and}
  \bibinfo{person}{Terry~D Blumenthal}.} \bibinfo{year}{1999}\natexlab{}.
\newblock \showarticletitle{Caffeine-associated stimuli elicit conditioned
  responses: an experimental model of the placebo effect}.
\newblock \bibinfo{journal}{\emph{Psychopharmacology}} \bibinfo{volume}{145},
  \bibinfo{number}{1} (\bibinfo{year}{1999}), \bibinfo{pages}{105--112}.
\newblock


\bibitem[\protect\citeauthoryear{Funk, Dingler, Cooper, and Schmidt}{Funk
  et~al\mbox{.}}{2015}]%
        {10.1145/2800835.2807942}
\bibfield{author}{\bibinfo{person}{Markus Funk}, \bibinfo{person}{Tilman
  Dingler}, \bibinfo{person}{Jennifer Cooper}, {and} \bibinfo{person}{Albrecht
  Schmidt}.} \bibinfo{year}{2015}\natexlab{}.
\newblock \showarticletitle{Stop Helping Me - I'm Bored! Why Assembly
  Assistance Needs to Be Adaptive}. In \bibinfo{booktitle}{\emph{Adjunct
  Proceedings of the 2015 ACM International Joint Conference on Pervasive and
  Ubiquitous Computing and Proceedings of the 2015 ACM International Symposium
  on Wearable Computers}} (Osaka, Japan)
  \emph{(\bibinfo{series}{UbiComp/ISWC'15 Adjunct})}.
  \bibinfo{publisher}{Association for Computing Machinery},
  \bibinfo{address}{New York, NY, USA}, \bibinfo{pages}{1269–1273}.
\newblock
\showISBNx{9781450335751}
\urldef\tempurl%
\url{https://doi.org/10.1145/2800835.2807942}
\showDOI{\tempurl}


\bibitem[\protect\citeauthoryear{Gajos, Czerwinski, Tan, and Weld}{Gajos
  et~al\mbox{.}}{2006}]%
        {10.1145/1133265.1133306}
\bibfield{author}{\bibinfo{person}{Krzysztof~Z. Gajos}, \bibinfo{person}{Mary
  Czerwinski}, \bibinfo{person}{Desney~S. Tan}, {and}
  \bibinfo{person}{Daniel~S. Weld}.} \bibinfo{year}{2006}\natexlab{}.
\newblock \showarticletitle{Exploring the Design Space for Adaptive Graphical
  User Interfaces}. In \bibinfo{booktitle}{\emph{Proceedings of the Working
  Conference on Advanced Visual Interfaces}} (Venezia, Italy)
  \emph{(\bibinfo{series}{AVI '06})}. \bibinfo{publisher}{Association for
  Computing Machinery}, \bibinfo{address}{New York, NY, USA},
  \bibinfo{pages}{201–208}.
\newblock
\showISBNx{1595933530}
\urldef\tempurl%
\url{https://doi.org/10.1145/1133265.1133306}
\showDOI{\tempurl}


\bibitem[\protect\citeauthoryear{García, Costanza, Verame, Nowacka, and
  Ramchurn}{García et~al\mbox{.}}{2021}]%
        {garcia2021}
\bibfield{author}{\bibinfo{person}{Pedro~García García},
  \bibinfo{person}{Enrico Costanza}, \bibinfo{person}{Jhim Verame},
  \bibinfo{person}{Diana Nowacka}, {and} \bibinfo{person}{Sarvapali~D.
  Ramchurn}.} \bibinfo{year}{2021}\natexlab{}.
\newblock \showarticletitle{Seeing (Movement) is Believing: The Effect of
  Motion on Perception of Automatic Systems Performance}.
\newblock \bibinfo{journal}{\emph{Human–Computer Interaction}}
  \bibinfo{volume}{36}, \bibinfo{number}{1} (\bibinfo{year}{2021}),
  \bibinfo{pages}{1--51}.
\newblock
\urldef\tempurl%
\url{https://doi.org/10.1080/07370024.2018.1453815}
\showDOI{\tempurl}
\showeprint{https://doi.org/10.1080/07370024.2018.1453815}


\bibitem[\protect\citeauthoryear{Geers, Weiland, Kosbab, Landry, and
  Helfer}{Geers et~al\mbox{.}}{2005}]%
        {geers2005goal}
\bibfield{author}{\bibinfo{person}{Andrew~L Geers}, \bibinfo{person}{Paul~E
  Weiland}, \bibinfo{person}{Kristin Kosbab}, \bibinfo{person}{Sarah~J Landry},
  {and} \bibinfo{person}{Suzanne~G Helfer}.} \bibinfo{year}{2005}\natexlab{}.
\newblock \showarticletitle{Goal activation, expectations, and the placebo
  effect.}
\newblock \bibinfo{journal}{\emph{Journal of personality and social
  psychology}} \bibinfo{volume}{89}, \bibinfo{number}{2}
  (\bibinfo{year}{2005}), \bibinfo{pages}{143}.
\newblock


\bibitem[\protect\citeauthoryear{Ghandeharioun and Picard}{Ghandeharioun and
  Picard}{2017}]%
        {10.1145/3027063.3053164}
\bibfield{author}{\bibinfo{person}{Asma Ghandeharioun} {and}
  \bibinfo{person}{Rosalind Picard}.} \bibinfo{year}{2017}\natexlab{}.
\newblock \bibinfo{booktitle}{\emph{BrightBeat: Effortlessly Influencing
  Breathing for Cultivating Calmness and Focus}}.
\newblock \bibinfo{publisher}{Association for Computing Machinery},
  \bibinfo{address}{New York, NY, USA}, \bibinfo{pages}{1624–1631}.
\newblock
\showISBNx{9781450346566}
\urldef\tempurl%
\url{https://doi.org/10.1145/3027063.3053164}
\showURL{%
\tempurl}


\bibitem[\protect\citeauthoryear{Gilhooly and Johnson}{Gilhooly and
  Johnson}{1978}]%
        {doi:10.1080/14640747808400654}
\bibfield{author}{\bibinfo{person}{K.~J. Gilhooly} {and} \bibinfo{person}{C.~E.
  Johnson}.} \bibinfo{year}{1978}\natexlab{}.
\newblock \showarticletitle{Effects of Solution Word Attributes on Anagram
  Difficulty: A Regression Analysis}.
\newblock \bibinfo{journal}{\emph{Quarterly Journal of Experimental
  Psychology}} \bibinfo{volume}{30}, \bibinfo{number}{1}
  (\bibinfo{year}{1978}), \bibinfo{pages}{57--70}.
\newblock
\urldef\tempurl%
\url{https://doi.org/10.1080/14640747808400654}
\showDOI{\tempurl}


\bibitem[\protect\citeauthoryear{Hart}{Hart}{2006}]%
        {doi:10.1177/154193120605000909}
\bibfield{author}{\bibinfo{person}{Sandra~G. Hart}.}
  \bibinfo{year}{2006}\natexlab{}.
\newblock \showarticletitle{Nasa-Task Load Index (NASA-TLX); 20 Years Later}.
\newblock \bibinfo{journal}{\emph{Proceedings of the Human Factors and
  Ergonomics Society Annual Meeting}} \bibinfo{volume}{50}, \bibinfo{number}{9}
  (\bibinfo{year}{2006}), \bibinfo{pages}{904--908}.
\newblock
\urldef\tempurl%
\url{https://doi.org/10.1177/154193120605000909}
\showDOI{\tempurl}


\bibitem[\protect\citeauthoryear{Hart and Staveland}{Hart and
  Staveland}{1988}]%
        {HART1988139}
\bibfield{author}{\bibinfo{person}{Sandra~G. Hart} {and}
  \bibinfo{person}{Lowell~E. Staveland}.} \bibinfo{year}{1988}\natexlab{}.
\newblock \showarticletitle{Development of NASA-TLX (Task Load Index): Results
  of Empirical and Theoretical Research}.
\newblock In \bibinfo{booktitle}{\emph{Human Mental Workload}},
  \bibfield{editor}{\bibinfo{person}{Peter~A. Hancock} {and}
  \bibinfo{person}{Najmedin Meshkati}} (Eds.). \bibinfo{series}{Advances in
  Psychology}, Vol.~\bibinfo{volume}{52}. \bibinfo{publisher}{North-Holland},
  \bibinfo{pages}{139 -- 183}.
\newblock
\showISSN{0166-4115}
\urldef\tempurl%
\url{https://doi.org/10.1016/S0166-4115(08)62386-9}
\showDOI{\tempurl}


\bibitem[\protect\citeauthoryear{Hartmann, De~Angeli, and Sutcliffe}{Hartmann
  et~al\mbox{.}}{2008}]%
        {hartmann2008framing}
\bibfield{author}{\bibinfo{person}{Jan Hartmann}, \bibinfo{person}{Antonella
  De~Angeli}, {and} \bibinfo{person}{Alistair Sutcliffe}.}
  \bibinfo{year}{2008}\natexlab{}.
\newblock \showarticletitle{Framing the user experience: information biases on
  website quality judgement}. In \bibinfo{booktitle}{\emph{Proceedings of the
  SIGCHI conference on human factors in computing systems}}.
  \bibinfo{pages}{855--864}.
\newblock


\bibitem[\protect\citeauthoryear{Helm, Swiergosz, Haeberle, Karnuta, Schaffer,
  Krebs, Spitzer, and Ramkumar}{Helm et~al\mbox{.}}{2020}]%
        {helm2020machine}
\bibfield{author}{\bibinfo{person}{Matthew~J. Helm}, \bibinfo{person}{Andrew~M.
  Swiergosz}, \bibinfo{person}{Heather~S. Haeberle}, \bibinfo{person}{Jaret~M.
  Karnuta}, \bibinfo{person}{Jonathan~L. Schaffer}, \bibinfo{person}{Viktor~E.
  Krebs}, \bibinfo{person}{Andrew~I. Spitzer}, {and} \bibinfo{person}{Prem~N.
  Ramkumar}.} \bibinfo{year}{2020}\natexlab{}.
\newblock \showarticletitle{Machine learning and artificial intelligence:
  Definitions, applications, and future directions}.
\newblock \bibinfo{journal}{\emph{Current reviews in musculoskeletal medicine}}
  \bibinfo{volume}{13}, \bibinfo{number}{1} (\bibinfo{year}{2020}),
  \bibinfo{pages}{69--76}.
\newblock


\bibitem[\protect\citeauthoryear{Hollis, Pekurovsky, Wu, and Whittaker}{Hollis
  et~al\mbox{.}}{2018}]%
        {hollis2018}
\bibfield{author}{\bibinfo{person}{Victoria Hollis}, \bibinfo{person}{Alon
  Pekurovsky}, \bibinfo{person}{Eunika Wu}, {and} \bibinfo{person}{Steve
  Whittaker}.} \bibinfo{year}{2018}\natexlab{}.
\newblock \showarticletitle{On Being Told How We Feel: How Algorithmic Sensor
  Feedback Influences Emotion Perception}.
\newblock \bibinfo{journal}{\emph{Proc. ACM Interact. Mob. Wearable Ubiquitous
  Technol.}} \bibinfo{volume}{2}, \bibinfo{number}{3}, Article
  \bibinfo{articleno}{114} (\bibinfo{date}{Sept.} \bibinfo{year}{2018}),
  \bibinfo{numpages}{31}~pages.
\newblock
\urldef\tempurl%
\url{https://doi.org/10.1145/3264924}
\showDOI{\tempurl}


\bibitem[\protect\citeauthoryear{Hornb{\ae}k and Law}{Hornb{\ae}k and
  Law}{2007}]%
        {hornbaek2007meta}
\bibfield{author}{\bibinfo{person}{Kasper Hornb{\ae}k} {and}
  \bibinfo{person}{Effie Lai-Chong Law}.} \bibinfo{year}{2007}\natexlab{}.
\newblock \showarticletitle{Meta-analysis of correlations among usability
  measures}. In \bibinfo{booktitle}{\emph{Proceedings of the SIGCHI conference
  on Human factors in computing systems}}. \bibinfo{pages}{617--626}.
\newblock


\bibitem[\protect\citeauthoryear{Hornbæk}{Hornbæk}{2006}]%
        {HORNBAEK200679}
\bibfield{author}{\bibinfo{person}{Kasper Hornbæk}.}
  \bibinfo{year}{2006}\natexlab{}.
\newblock \showarticletitle{Current practice in measuring usability: Challenges
  to usability studies and research}.
\newblock \bibinfo{journal}{\emph{International Journal of Human-Computer
  Studies}} \bibinfo{volume}{64}, \bibinfo{number}{2} (\bibinfo{year}{2006}),
  \bibinfo{pages}{79 -- 102}.
\newblock
\showISSN{1071-5819}
\urldef\tempurl%
\url{https://doi.org/10.1016/j.ijhcs.2005.06.002}
\showDOI{\tempurl}


\bibitem[\protect\citeauthoryear{Hróbjartsson and Gøtzsche}{Hróbjartsson and
  Gøtzsche}{2001}]%
        {powerless2001}
\bibfield{author}{\bibinfo{person}{Asbjørn Hróbjartsson} {and}
  \bibinfo{person}{Peter~C. Gøtzsche}.} \bibinfo{year}{2001}\natexlab{}.
\newblock \showarticletitle{Is the Placebo Powerless?}
\newblock \bibinfo{journal}{\emph{New England Journal of Medicine}}
  \bibinfo{volume}{344}, \bibinfo{number}{21} (\bibinfo{year}{2001}),
  \bibinfo{pages}{1594--1602}.
\newblock
\urldef\tempurl%
\url{https://doi.org/10.1056/NEJM200105243442106}
\showDOI{\tempurl}
\showeprint{https://doi.org/10.1056/NEJM200105243442106}
\newblock
\shownote{PMID: 11372012.}


\bibitem[\protect\citeauthoryear{Hudlicka and Mcneese}{Hudlicka and
  Mcneese}{2002}]%
        {hudlicka2002assessment}
\bibfield{author}{\bibinfo{person}{Eva Hudlicka} {and}
  \bibinfo{person}{Michael~D Mcneese}.} \bibinfo{year}{2002}\natexlab{}.
\newblock \showarticletitle{Assessment of user affective and belief states for
  interface adaptation: Application to an Air Force pilot task}.
\newblock \bibinfo{journal}{\emph{User Modeling and User-Adapted Interaction}}
  \bibinfo{volume}{12}, \bibinfo{number}{1} (\bibinfo{year}{2002}),
  \bibinfo{pages}{1--47}.
\newblock


\bibitem[\protect\citeauthoryear{Indrati, Minaji, Binastuti, and
  Raharjo}{Indrati et~al\mbox{.}}{2014}]%
        {indrati2014comparation}
\bibfield{author}{\bibinfo{person}{Aviarini Indrati}, \bibinfo{person}{Edi
  Minaji}, \bibinfo{person}{Sugiharti Binastuti}, {and}
  \bibinfo{person}{Philipus~Dwi Raharjo}.} \bibinfo{year}{2014}\natexlab{}.
\newblock \showarticletitle{Comparation of Model Unified Theory of Acceptance
  and Use Technology (UTAUT) And Technology Acceptance Model (TAM) for Internet
  Adoption of Credit Union Staff}. In \bibinfo{booktitle}{\emph{The First
  International Credit Union Conference on Social Micro}}.
\newblock


\bibitem[\protect\citeauthoryear{Jeon}{Jeon}{2012}]%
        {10.1145/2370216.2370455}
\bibfield{author}{\bibinfo{person}{Myounghoon Jeon}.}
  \bibinfo{year}{2012}\natexlab{}.
\newblock \showarticletitle{A Systematic Approach to Using Music for Mitigating
  Affective Effects on Driving Performance and Safety}. In
  \bibinfo{booktitle}{\emph{Proceedings of the 2012 ACM Conference on
  Ubiquitous Computing}} (Pittsburgh, Pennsylvania)
  \emph{(\bibinfo{series}{UbiComp '12})}. \bibinfo{publisher}{Association for
  Computing Machinery}, \bibinfo{address}{New York, NY, USA},
  \bibinfo{pages}{1127–1132}.
\newblock
\showISBNx{9781450312240}
\urldef\tempurl%
\url{https://doi.org/10.1145/2370216.2370455}
\showDOI{\tempurl}


\bibitem[\protect\citeauthoryear{Kaptchuk}{Kaptchuk}{1998}]%
        {kaptchuk1998powerful}
\bibfield{author}{\bibinfo{person}{Ted~J Kaptchuk}.}
  \bibinfo{year}{1998}\natexlab{}.
\newblock \showarticletitle{Powerful placebo: the dark side of the randomised
  controlled trial}.
\newblock \bibinfo{journal}{\emph{The lancet}} \bibinfo{volume}{351},
  \bibinfo{number}{9117} (\bibinfo{year}{1998}), \bibinfo{pages}{1722--1725}.
\newblock
\urldef\tempurl%
\url{https://doi.org/10.1016/S0140-6736(97)10111-8}
\showDOI{\tempurl}


\bibitem[\protect\citeauthoryear{Kay and Wobbrock}{Kay and Wobbrock}{2016}]%
        {kay2016package}
\bibfield{author}{\bibinfo{person}{Matthew Kay} {and} \bibinfo{person}{Jacob~O
  Wobbrock}.} \bibinfo{year}{2016}\natexlab{}.
\newblock \showarticletitle{Package ‘ARTool’}.
\newblock \bibinfo{journal}{\emph{CRAN Repository}} (\bibinfo{year}{2016}),
  \bibinfo{pages}{1--13}.
\newblock


\bibitem[\protect\citeauthoryear{Kirsch}{Kirsch}{1999}]%
        {kirsch1999specifying}
\bibfield{author}{\bibinfo{person}{Irving Kirsch}.}
  \bibinfo{year}{1999}\natexlab{}.
\newblock \showarticletitle{Specifying Nonspecifics: Psychological}.
\newblock \bibinfo{journal}{\emph{The placebo effect: An interdisciplinary
  exploration}}  \bibinfo{volume}{8} (\bibinfo{year}{1999}),
  \bibinfo{pages}{166}.
\newblock


\bibitem[\protect\citeauthoryear{Kosch, Funk, Schmidt, and Chuang}{Kosch
  et~al\mbox{.}}{2018a}]%
        {10.1145/3229093}
\bibfield{author}{\bibinfo{person}{Thomas Kosch}, \bibinfo{person}{Markus
  Funk}, \bibinfo{person}{Albrecht Schmidt}, {and} \bibinfo{person}{Lewis~L.
  Chuang}.} \bibinfo{year}{2018}\natexlab{a}.
\newblock \showarticletitle{Identifying Cognitive Assistance with Mobile
  Electroencephalography: A Case Study with In-Situ Projections for Manual
  Assembly}.
\newblock \bibinfo{journal}{\emph{Proc. ACM Hum.-Comput. Interact.}}
  \bibinfo{volume}{2}, \bibinfo{number}{EICS}, Article \bibinfo{articleno}{11}
  (\bibinfo{date}{jun} \bibinfo{year}{2018}), \bibinfo{numpages}{20}~pages.
\newblock
\urldef\tempurl%
\url{https://doi.org/10.1145/3229093}
\showDOI{\tempurl}


\bibitem[\protect\citeauthoryear{Kosch, Hassib, Reutter, and Alt}{Kosch
  et~al\mbox{.}}{2020a}]%
        {10.1145/3399715.3399928}
\bibfield{author}{\bibinfo{person}{Thomas Kosch}, \bibinfo{person}{Mariam
  Hassib}, \bibinfo{person}{Robin Reutter}, {and} \bibinfo{person}{Florian
  Alt}.} \bibinfo{year}{2020}\natexlab{a}.
\newblock \showarticletitle{Emotions on the Go: Mobile Emotion Assessment in
  Real-Time Using Facial Expressions}. In \bibinfo{booktitle}{\emph{Proceedings
  of the International Conference on Advanced Visual Interfaces}} (Salerno,
  Italy) \emph{(\bibinfo{series}{AVI '20})}. \bibinfo{publisher}{Association
  for Computing Machinery}, \bibinfo{address}{New York, NY, USA}, Article
  \bibinfo{articleno}{18}, \bibinfo{numpages}{9}~pages.
\newblock
\showISBNx{9781450375351}
\urldef\tempurl%
\url{https://doi.org/10.1145/3399715.3399928}
\showDOI{\tempurl}


\bibitem[\protect\citeauthoryear{Kosch, Hassib, Wozniak, Buschek, and
  Alt}{Kosch et~al\mbox{.}}{2018b}]%
        {10.1145/3173574.3174010}
\bibfield{author}{\bibinfo{person}{Thomas Kosch}, \bibinfo{person}{Mariam
  Hassib}, \bibinfo{person}{Pawe\l{}~W. Wozniak}, \bibinfo{person}{Daniel
  Buschek}, {and} \bibinfo{person}{Florian Alt}.}
  \bibinfo{year}{2018}\natexlab{b}.
\newblock \showarticletitle{Your Eyes Tell: Leveraging Smooth Pursuit for
  Assessing Cognitive Workload}. In \bibinfo{booktitle}{\emph{Proceedings of
  the 2018 CHI Conference on Human Factors in Computing Systems}} (Montreal QC,
  Canada) \emph{(\bibinfo{series}{CHI '18})}. \bibinfo{publisher}{Association
  for Computing Machinery}, \bibinfo{address}{New York, NY, USA},
  \bibinfo{pages}{1–13}.
\newblock
\showISBNx{9781450356206}
\urldef\tempurl%
\url{https://doi.org/10.1145/3173574.3174010}
\showDOI{\tempurl}


\bibitem[\protect\citeauthoryear{Kosch, Karolus, Ha, and Schmidt}{Kosch
  et~al\mbox{.}}{2019}]%
        {kosch2019your}
\bibfield{author}{\bibinfo{person}{Thomas Kosch}, \bibinfo{person}{Jakob
  Karolus}, \bibinfo{person}{Havy Ha}, {and} \bibinfo{person}{Albrecht
  Schmidt}.} \bibinfo{year}{2019}\natexlab{}.
\newblock \showarticletitle{Your Skin Resists: Exploring Electrodermal Activity
  As Workload Indicator During Manual Assembly}. In
  \bibinfo{booktitle}{\emph{Proceedings of the ACM SIGCHI Symposium on
  Engineering Interactive Computing Systems}} (Valencia, Spain)
  \emph{(\bibinfo{series}{EICS '19})}. \bibinfo{publisher}{ACM},
  \bibinfo{address}{New York, NY, USA}, Article \bibinfo{articleno}{8},
  \bibinfo{numpages}{5}~pages.
\newblock
\showISBNx{978-1-4503-6745-5}
\urldef\tempurl%
\url{https://doi.org/10.1145/3319499.3328230}
\showDOI{\tempurl}


\bibitem[\protect\citeauthoryear{Kosch, Schmidt, Thanheiser, and Chuang}{Kosch
  et~al\mbox{.}}{2020b}]%
        {koschrsvp}
\bibfield{author}{\bibinfo{person}{Thomas Kosch}, \bibinfo{person}{Albrecht
  Schmidt}, \bibinfo{person}{Simon Thanheiser}, {and} \bibinfo{person}{Lewis~L.
  Chuang}.} \bibinfo{year}{2020}\natexlab{b}.
\newblock \showarticletitle{One Does Not Simply RSVP: Mental Workload to Select
  Speed Reading Parameters Using Electroencephalography}. In
  \bibinfo{booktitle}{\emph{Proceedings of the 2020 CHI Conference on Human
  Factors in Computing Systems}} (Honolulu, HI, USA)
  \emph{(\bibinfo{series}{CHI '20})}. \bibinfo{publisher}{Association for
  Computing Machinery}, \bibinfo{address}{New York, NY, USA},
  \bibinfo{pages}{1–13}.
\newblock
\showISBNx{9781450367080}
\urldef\tempurl%
\url{https://doi.org/10.1145/3313831.3376766}
\showDOI{\tempurl}


\bibitem[\protect\citeauthoryear{Kosmyna, Tarpin-Bernard, and Rivet}{Kosmyna
  et~al\mbox{.}}{2015}]%
        {10.1145/2808228}
\bibfield{author}{\bibinfo{person}{Nataliya Kosmyna}, \bibinfo{person}{Franck
  Tarpin-Bernard}, {and} \bibinfo{person}{Bertrand Rivet}.}
  \bibinfo{year}{2015}\natexlab{}.
\newblock \showarticletitle{Conceptual Priming for In-Game BCI Training}.
\newblock \bibinfo{journal}{\emph{ACM Trans. Comput.-Hum. Interact.}}
  \bibinfo{volume}{22}, \bibinfo{number}{5}, Article \bibinfo{articleno}{26}
  (\bibinfo{date}{oct} \bibinfo{year}{2015}), \bibinfo{numpages}{25}~pages.
\newblock
\showISSN{1073-0516}
\urldef\tempurl%
\url{https://doi.org/10.1145/2808228}
\showDOI{\tempurl}


\bibitem[\protect\citeauthoryear{Kruschke}{Kruschke}{2013}]%
        {kruschke2013bayesian}
\bibfield{author}{\bibinfo{person}{John~K. Kruschke}.}
  \bibinfo{year}{2013}\natexlab{}.
\newblock \showarticletitle{Bayesian estimation supersedes the t test.}
\newblock \bibinfo{journal}{\emph{Journal of Experimental Psychology: General}}
  \bibinfo{volume}{142}, \bibinfo{number}{2} (\bibinfo{year}{2013}),
  \bibinfo{pages}{573}.
\newblock


\bibitem[\protect\citeauthoryear{Kundinger, Wintersberger, and
  Riener}{Kundinger et~al\mbox{.}}{2019}]%
        {kundinger2019over}
\bibfield{author}{\bibinfo{person}{Thomas Kundinger}, \bibinfo{person}{Philipp
  Wintersberger}, {and} \bibinfo{person}{Andreas Riener}.}
  \bibinfo{year}{2019}\natexlab{}.
\newblock \showarticletitle{(Over) Trust in Automated Driving: The Sleeping
  Pill of Tomorrow?}. In \bibinfo{booktitle}{\emph{Extended Abstracts of the
  2019 CHI Conference on Human Factors in Computing Systems}}.
  \bibinfo{pages}{1--6}.
\newblock


\bibitem[\protect\citeauthoryear{Kwak, Holtkamp, and Kim}{Kwak
  et~al\mbox{.}}{2019}]%
        {kwak2019measuring}
\bibfield{author}{\bibinfo{person}{Dong-Heon Kwak}, \bibinfo{person}{Philipp
  Holtkamp}, {and} \bibinfo{person}{Sung~S Kim}.}
  \bibinfo{year}{2019}\natexlab{}.
\newblock \showarticletitle{Measuring and controlling social desirability bias:
  Applications in information systems research}.
\newblock \bibinfo{journal}{\emph{Journal of the Association for Information
  Systems}} \bibinfo{volume}{20}, \bibinfo{number}{4} (\bibinfo{year}{2019}),
  \bibinfo{pages}{5}.
\newblock


\bibitem[\protect\citeauthoryear{Lasagna, Mosteller, von Felsinger, and
  Beecher}{Lasagna et~al\mbox{.}}{1954}]%
        {lasagna1954study}
\bibfield{author}{\bibinfo{person}{Louis Lasagna}, \bibinfo{person}{Frederick
  Mosteller}, \bibinfo{person}{John~M von Felsinger}, {and}
  \bibinfo{person}{Henry~K Beecher}.} \bibinfo{year}{1954}\natexlab{}.
\newblock \showarticletitle{A study of the placebo response}.
\newblock \bibinfo{journal}{\emph{The American journal of medicine}}
  \bibinfo{volume}{16}, \bibinfo{number}{6} (\bibinfo{year}{1954}),
  \bibinfo{pages}{770--779}.
\newblock


\bibitem[\protect\citeauthoryear{Lee, Kim, Kim, and Kang}{Lee
  et~al\mbox{.}}{2014}]%
        {10.1145/2668056.2668058}
\bibfield{author}{\bibinfo{person}{Eunjung Lee}, \bibinfo{person}{Gyu-Wan Kim},
  \bibinfo{person}{Byung-Soo Kim}, {and} \bibinfo{person}{Mi-Ae Kang}.}
  \bibinfo{year}{2014}\natexlab{}.
\newblock \showarticletitle{A Design Platform for Emotion-Aware User
  Interfaces}. In \bibinfo{booktitle}{\emph{Proceedings of the 2014 Workshop on
  Emotion Representation and Modelling in Human-Computer-Interaction-Systems}}
  (Istanbul, Turkey) \emph{(\bibinfo{series}{ERM4HCI '14})}.
  \bibinfo{publisher}{Association for Computing Machinery},
  \bibinfo{address}{New York, NY, USA}, \bibinfo{pages}{19–24}.
\newblock
\showISBNx{9781450301244}
\urldef\tempurl%
\url{https://doi.org/10.1145/2668056.2668058}
\showDOI{\tempurl}


\bibitem[\protect\citeauthoryear{{Lewandowska}, {Rumiński}, {Kocejko}, and
  {Nowak}}{{Lewandowska} et~al\mbox{.}}{2011}]%
        {6078233}
\bibfield{author}{\bibinfo{person}{Magdalena {Lewandowska}},
  \bibinfo{person}{Jacek {Rumiński}}, \bibinfo{person}{Tomasz {Kocejko}},
  {and} \bibinfo{person}{Jędrzej {Nowak}}.} \bibinfo{year}{2011}\natexlab{}.
\newblock \showarticletitle{Measuring pulse rate with a webcam — A
  non-contact method for evaluating cardiac activity}. In
  \bibinfo{booktitle}{\emph{2011 Federated Conference on Computer Science and
  Information Systems (FedCSIS)}}. \bibinfo{pages}{405--410}.
\newblock


\bibitem[\protect\citeauthoryear{Li, Kumar, Lasecki, and Hilliges}{Li
  et~al\mbox{.}}{2020}]%
        {li2020}
\bibfield{author}{\bibinfo{person}{Yang Li}, \bibinfo{person}{Ranjitha Kumar},
  \bibinfo{person}{Walter~S. Lasecki}, {and} \bibinfo{person}{Otmar Hilliges}.}
  \bibinfo{year}{2020}\natexlab{}.
\newblock \showarticletitle{Artificial Intelligence for HCI: A Modern
  Approach}. In \bibinfo{booktitle}{\emph{Extended Abstracts of the 2020 CHI
  Conference on Human Factors in Computing Systems}} (Honolulu, HI, USA)
  \emph{(\bibinfo{series}{CHI EA '20})}. \bibinfo{publisher}{Association for
  Computing Machinery}, \bibinfo{address}{New York, NY, USA},
  \bibinfo{pages}{1–8}.
\newblock
\showISBNx{9781450368193}
\urldef\tempurl%
\url{https://doi.org/10.1145/3334480.3375147}
\showDOI{\tempurl}


\bibitem[\protect\citeauthoryear{Lin, Guo, Chen, Yao, and Ying}{Lin
  et~al\mbox{.}}{2020}]%
        {lin2020}
\bibfield{author}{\bibinfo{person}{Yuyu Lin}, \bibinfo{person}{Jiahao Guo},
  \bibinfo{person}{Yang Chen}, \bibinfo{person}{Cheng Yao}, {and}
  \bibinfo{person}{Fangtian Ying}.} \bibinfo{year}{2020}\natexlab{}.
\newblock \showarticletitle{It Is Your Turn: Collaborative Ideation With a
  Co-Creative Robot through Sketch}. In \bibinfo{booktitle}{\emph{Proceedings
  of the 2020 CHI Conference on Human Factors in Computing Systems}} (Honolulu,
  HI, USA) \emph{(\bibinfo{series}{CHI '20})}. \bibinfo{publisher}{Association
  for Computing Machinery}, \bibinfo{address}{New York, NY, USA},
  \bibinfo{pages}{1–14}.
\newblock
\showISBNx{9781450367080}
\urldef\tempurl%
\url{https://doi.org/10.1145/3313831.3376258}
\showDOI{\tempurl}


\bibitem[\protect\citeauthoryear{Liu, Gummadi, Krishnamurthy, and Mislove}{Liu
  et~al\mbox{.}}{2011}]%
        {Liu2011FB}
\bibfield{author}{\bibinfo{person}{Yabing Liu}, \bibinfo{person}{Krishna~P.
  Gummadi}, \bibinfo{person}{Balachander Krishnamurthy}, {and}
  \bibinfo{person}{Alan Mislove}.} \bibinfo{year}{2011}\natexlab{}.
\newblock \showarticletitle{Analyzing Facebook privacy settings: User
  expectations vs. reality}.
\newblock \bibinfo{journal}{\emph{Proceedings of the ACM SIGCOMM Internet
  Measurement Conference, IMC}} (\bibinfo{date}{11} \bibinfo{year}{2011}).
\newblock
\urldef\tempurl%
\url{https://doi.org/10.1145/2068816.2068823}
\showDOI{\tempurl}


\bibitem[\protect\citeauthoryear{Looby and Earleywine}{Looby and
  Earleywine}{2011}]%
        {looby2011expectation}
\bibfield{author}{\bibinfo{person}{Alison Looby} {and} \bibinfo{person}{Mitch
  Earleywine}.} \bibinfo{year}{2011}\natexlab{}.
\newblock \showarticletitle{Expectation to receive methylphenidate enhances
  subjective arousal but not cognitive performance.}
\newblock \bibinfo{journal}{\emph{Experimental and clinical
  psychopharmacology}} \bibinfo{volume}{19}, \bibinfo{number}{6}
  (\bibinfo{year}{2011}), \bibinfo{pages}{433}.
\newblock


\bibitem[\protect\citeauthoryear{Luque-Casado, Perales, Cárdenas, and
  Sanabria}{Luque-Casado et~al\mbox{.}}{2016}]%
        {LUQUECASADO201683}
\bibfield{author}{\bibinfo{person}{Antonio Luque-Casado},
  \bibinfo{person}{José~C. Perales}, \bibinfo{person}{David Cárdenas}, {and}
  \bibinfo{person}{Daniel Sanabria}.} \bibinfo{year}{2016}\natexlab{}.
\newblock \showarticletitle{Heart rate variability and cognitive processing:
  The autonomic response to task demands}.
\newblock \bibinfo{journal}{\emph{Biological Psychology}}
  \bibinfo{volume}{113} (\bibinfo{year}{2016}), \bibinfo{pages}{83 -- 90}.
\newblock
\showISSN{0301-0511}
\urldef\tempurl%
\url{https://doi.org/10.1016/j.biopsycho.2015.11.013}
\showDOI{\tempurl}


\bibitem[\protect\citeauthoryear{Matthews, Woodall, and Allen}{Matthews
  et~al\mbox{.}}{1993}]%
        {doi:10.1161/01.HYP.22.4.479}
\bibfield{author}{\bibinfo{person}{Karen~A. Matthews},
  \bibinfo{person}{Karen~L. Woodall}, {and} \bibinfo{person}{Michael~T.
  Allen}.} \bibinfo{year}{1993}\natexlab{}.
\newblock \showarticletitle{Cardiovascular reactivity to stress predicts future
  blood pressure status.}
\newblock \bibinfo{journal}{\emph{Hypertension}} \bibinfo{volume}{22},
  \bibinfo{number}{4} (\bibinfo{year}{1993}), \bibinfo{pages}{479--485}.
\newblock
\urldef\tempurl%
\url{https://doi.org/10.1161/01.HYP.22.4.479}
\showDOI{\tempurl}
\showeprint{https://www.ahajournals.org/doi/pdf/10.1161/01.HYP.22.4.479}


\bibitem[\protect\citeauthoryear{Michalco, Simonsen, and Hornbaek}{Michalco
  et~al\mbox{.}}{2015}]%
        {doi:10.1080/10447318.2015.1065696}
\bibfield{author}{\bibinfo{person}{Jaroslav Michalco},
  \bibinfo{person}{Jakob~Grue Simonsen}, {and} \bibinfo{person}{Kasper
  Hornbaek}.} \bibinfo{year}{2015}\natexlab{}.
\newblock \showarticletitle{An Exploration of the Relation Between Expectations
  and User Experience}.
\newblock \bibinfo{journal}{\emph{International Journal of Human-Computer
  Interaction}} \bibinfo{volume}{31}, \bibinfo{number}{9}
  (\bibinfo{year}{2015}), \bibinfo{pages}{603--617}.
\newblock
\urldef\tempurl%
\url{https://doi.org/10.1080/10447318.2015.1065696}
\showDOI{\tempurl}
\showeprint{https://doi.org/10.1080/10447318.2015.1065696}


\bibitem[\protect\citeauthoryear{Millett}{Millett}{2001}]%
        {millett2001hans}
\bibfield{author}{\bibinfo{person}{David Millett}.}
  \bibinfo{year}{2001}\natexlab{}.
\newblock \showarticletitle{Hans Berger: From psychic energy to the EEG}.
\newblock \bibinfo{journal}{\emph{Perspectives in biology and medicine}}
  \bibinfo{volume}{44}, \bibinfo{number}{4} (\bibinfo{year}{2001}),
  \bibinfo{pages}{522--542}.
\newblock


\bibitem[\protect\citeauthoryear{Miroglio, Zeber, Kaye, and Weiss}{Miroglio
  et~al\mbox{.}}{2018}]%
        {10.1145/3178876.3186162}
\bibfield{author}{\bibinfo{person}{Ben Miroglio}, \bibinfo{person}{David
  Zeber}, \bibinfo{person}{Jofish Kaye}, {and} \bibinfo{person}{Rebecca
  Weiss}.} \bibinfo{year}{2018}\natexlab{}.
\newblock \bibinfo{booktitle}{\emph{The Effect of Ad Blocking on User
  Engagement with the Web}}.
\newblock \bibinfo{publisher}{International World Wide Web Conferences Steering
  Committee}, \bibinfo{address}{Republic and Canton of Geneva, CHE},
  \bibinfo{pages}{813–821}.
\newblock
\showISBNx{9781450356398}
\urldef\tempurl%
\url{https://doi.org/10.1145/3178876.3186162}
\showURL{%
\tempurl}


\bibitem[\protect\citeauthoryear{Montgomery and Kirsch}{Montgomery and
  Kirsch}{1996}]%
        {montgomery1996mechanisms}
\bibfield{author}{\bibinfo{person}{Guy Montgomery} {and}
  \bibinfo{person}{Irving Kirsch}.} \bibinfo{year}{1996}\natexlab{}.
\newblock \showarticletitle{Mechanisms of placebo pain reduction: an empirical
  investigation}.
\newblock \bibinfo{journal}{\emph{Psychological science}} \bibinfo{volume}{7},
  \bibinfo{number}{3} (\bibinfo{year}{1996}), \bibinfo{pages}{174--176}.
\newblock


\bibitem[\protect\citeauthoryear{Morey, Rouder, Jamil, and Morey}{Morey
  et~al\mbox{.}}{2015}]%
        {morey2015package}
\bibfield{author}{\bibinfo{person}{Richard~D Morey}, \bibinfo{person}{Jeffrey~N
  Rouder}, \bibinfo{person}{Tahira Jamil}, {and} \bibinfo{person}{Maintainer
  Richard~D Morey}.} \bibinfo{year}{2015}\natexlab{}.
\newblock \showarticletitle{Package ‘bayesfactor’}.
\newblock \bibinfo{journal}{\emph{URLh
  http://cran/r-projectorg/web/packages/BayesFactor/BayesFactor pdf i (accessed
  1006 15)}} (\bibinfo{year}{2015}).
\newblock


\bibitem[\protect\citeauthoryear{Narayan, TeBlunthuis, Hale, Hill, and
  Shaw}{Narayan et~al\mbox{.}}{2019}]%
        {10.1145/3359203}
\bibfield{author}{\bibinfo{person}{Sneha Narayan}, \bibinfo{person}{Nathan
  TeBlunthuis}, \bibinfo{person}{Wm~Salt Hale}, \bibinfo{person}{Benjamin~Mako
  Hill}, {and} \bibinfo{person}{Aaron Shaw}.} \bibinfo{year}{2019}\natexlab{}.
\newblock \showarticletitle{All Talk: How Increasing Interpersonal
  Communication on Wikis May Not Enhance Productivity}.
\newblock \bibinfo{journal}{\emph{Proc. ACM Hum.-Comput. Interact.}}
  \bibinfo{volume}{3}, \bibinfo{number}{CSCW}, Article \bibinfo{articleno}{101}
  (\bibinfo{date}{Nov.} \bibinfo{year}{2019}), \bibinfo{numpages}{19}~pages.
\newblock
\urldef\tempurl%
\url{https://doi.org/10.1145/3359203}
\showDOI{\tempurl}


\bibitem[\protect\citeauthoryear{Natesan, Walker, and Clark}{Natesan
  et~al\mbox{.}}{2016}]%
        {natesan2016cognitive}
\bibfield{author}{\bibinfo{person}{Divya Natesan}, \bibinfo{person}{Morgan
  Walker}, {and} \bibinfo{person}{Shannon Clark}.}
  \bibinfo{year}{2016}\natexlab{}.
\newblock \showarticletitle{Cognitive bias in usability testing}. In
  \bibinfo{booktitle}{\emph{Proceedings of the International Symposium on Human
  Factors and Ergonomics in Health Care}}, Vol.~\bibinfo{volume}{5}. SAGE
  Publications Sage CA: Los Angeles, CA, \bibinfo{pages}{86--88}.
\newblock


\bibitem[\protect\citeauthoryear{Nickerson}{Nickerson}{1998}]%
        {nickerson1998confirmation}
\bibfield{author}{\bibinfo{person}{Raymond~S Nickerson}.}
  \bibinfo{year}{1998}\natexlab{}.
\newblock \showarticletitle{Confirmation bias: A ubiquitous phenomenon in many
  guises}.
\newblock \bibinfo{journal}{\emph{Review of general psychology}}
  \bibinfo{volume}{2}, \bibinfo{number}{2} (\bibinfo{year}{1998}),
  \bibinfo{pages}{175--220}.
\newblock


\bibitem[\protect\citeauthoryear{Ojanen et~al\mbox{.}}{Ojanen
  et~al\mbox{.}}{1994}]%
        {ojanen1994can}
\bibfield{author}{\bibinfo{person}{M Ojanen} {et~al\mbox{.}}}
  \bibinfo{year}{1994}\natexlab{}.
\newblock \showarticletitle{Can the true effects of exercise on psychological
  variables be separated from placebo effects?}
\newblock \bibinfo{journal}{\emph{International Journal of Sport Psychology}}
  \bibinfo{volume}{25}, \bibinfo{number}{1} (\bibinfo{year}{1994}),
  \bibinfo{pages}{63--80}.
\newblock


\bibitem[\protect\citeauthoryear{Oliver}{Oliver}{1977}]%
        {oliver1977effect}
\bibfield{author}{\bibinfo{person}{Richard~L Oliver}.}
  \bibinfo{year}{1977}\natexlab{}.
\newblock \showarticletitle{Effect of expectation and disconfirmation on
  postexposure product evaluations: An alternative interpretation.}
\newblock \bibinfo{journal}{\emph{Journal of applied psychology}}
  \bibinfo{volume}{62}, \bibinfo{number}{4} (\bibinfo{year}{1977}),
  \bibinfo{pages}{480}.
\newblock


\bibitem[\protect\citeauthoryear{Ottenbreit-Leftwich, Glazewski, Jeon,
  Hmelo-Silver, Mott, Lee, and Lester}{Ottenbreit-Leftwich
  et~al\mbox{.}}{2021}]%
        {ottenbreit2021}
\bibfield{author}{\bibinfo{person}{Anne Ottenbreit-Leftwich},
  \bibinfo{person}{Krista Glazewski}, \bibinfo{person}{Minji Jeon},
  \bibinfo{person}{Cindy Hmelo-Silver}, \bibinfo{person}{Bradford Mott},
  \bibinfo{person}{Seung Lee}, {and} \bibinfo{person}{James Lester}.}
  \bibinfo{year}{2021}\natexlab{}.
\newblock \bibinfo{booktitle}{\emph{How Do Elementary Students Conceptualize
  Artificial Intelligence?}}
\newblock \bibinfo{publisher}{Association for Computing Machinery},
  \bibinfo{address}{New York, NY, USA}, \bibinfo{pages}{1261}.
\newblock
\showISBNx{9781450380621}
\urldef\tempurl%
\url{https://doi.org/10.1145/3408877.3439642}
\showURL{%
\tempurl}


\bibitem[\protect\citeauthoryear{Paepcke and Takayama}{Paepcke and
  Takayama}{2010}]%
        {Paepcke2010}
\bibfield{author}{\bibinfo{person}{Steffi Paepcke} {and} \bibinfo{person}{Leila
  Takayama}.} \bibinfo{year}{2010}\natexlab{}.
\newblock \showarticletitle{Judging a Bot by Its Cover: An Experiment on
  Expectation Setting for Personal Robots}. In
  \bibinfo{booktitle}{\emph{Proceedings of the 5th ACM/IEEE International
  Conference on Human-Robot Interaction}} (Osaka, Japan)
  \emph{(\bibinfo{series}{HRI '10})}. \bibinfo{publisher}{IEEE Press},
  \bibinfo{pages}{45–52}.
\newblock
\showISBNx{9781424448937}


\bibitem[\protect\citeauthoryear{Palanisamy, Muthusamy, and Sazali}{Palanisamy
  et~al\mbox{.}}{2013}]%
        {doi:10.1142/S0219519413500383}
\bibfield{author}{\bibinfo{person}{Karthikeyan Palanisamy},
  \bibinfo{person}{Murguappan Muthusamy}, {and} \bibinfo{person}{Yaacob
  Sazali}.} \bibinfo{year}{2013}\natexlab{}.
\newblock \showarticletitle{Detection of Human Stress Using Short-Term ECG and
  HRV Signals}.
\newblock \bibinfo{journal}{\emph{Journal of Mechanics in Medicine and
  Biology}} \bibinfo{volume}{13}, \bibinfo{number}{02} (\bibinfo{year}{2013}),
  \bibinfo{pages}{1350038}.
\newblock
\urldef\tempurl%
\url{https://doi.org/10.1142/S0219519413500383}
\showDOI{\tempurl}
\showeprint{https://doi.org/10.1142/S0219519413500383}


\bibitem[\protect\citeauthoryear{Papou\v{s}ek and Pel\'{a}nek}{Papou\v{s}ek and
  Pel\'{a}nek}{2017}]%
        {10.1145/3099023.3099080}
\bibfield{author}{\bibinfo{person}{Jan Papou\v{s}ek} {and}
  \bibinfo{person}{Radek Pel\'{a}nek}.} \bibinfo{year}{2017}\natexlab{}.
\newblock \showarticletitle{Should We Give Learners Control Over Item
  Difficulty?}. In \bibinfo{booktitle}{\emph{Adjunct Publication of the 25th
  Conference on User Modeling, Adaptation and Personalization}} (Bratislava,
  Slovakia) \emph{(\bibinfo{series}{UMAP '17})}.
  \bibinfo{publisher}{Association for Computing Machinery},
  \bibinfo{address}{New York, NY, USA}, \bibinfo{pages}{299–303}.
\newblock
\showISBNx{9781450350679}
\urldef\tempurl%
\url{https://doi.org/10.1145/3099023.3099080}
\showDOI{\tempurl}


\bibitem[\protect\citeauthoryear{Piech, Sahami, Huang, and Guibas}{Piech
  et~al\mbox{.}}{2015}]%
        {10.1145/2724660.2724668}
\bibfield{author}{\bibinfo{person}{Chris Piech}, \bibinfo{person}{Mehran
  Sahami}, \bibinfo{person}{Jonathan Huang}, {and} \bibinfo{person}{Leonidas
  Guibas}.} \bibinfo{year}{2015}\natexlab{}.
\newblock \showarticletitle{Autonomously Generating Hints by Inferring Problem
  Solving Policies}. In \bibinfo{booktitle}{\emph{Proceedings of the Second
  (2015) ACM Conference on Learning @ Scale}} (Vancouver, BC, Canada)
  \emph{(\bibinfo{series}{L@S '15})}. \bibinfo{publisher}{Association for
  Computing Machinery}, \bibinfo{address}{New York, NY, USA},
  \bibinfo{pages}{195–204}.
\newblock
\showISBNx{9781450334112}
\urldef\tempurl%
\url{https://doi.org/10.1145/2724660.2724668}
\showDOI{\tempurl}


\bibitem[\protect\citeauthoryear{Prakash, Hussain, and Schirda}{Prakash
  et~al\mbox{.}}{2015}]%
        {prakash2015role}
\bibfield{author}{\bibinfo{person}{Ruchika~Shaurya Prakash},
  \bibinfo{person}{Mariam~A Hussain}, {and} \bibinfo{person}{Brittney
  Schirda}.} \bibinfo{year}{2015}\natexlab{}.
\newblock \showarticletitle{The role of emotion regulation and cognitive
  control in the association between mindfulness disposition and stress.}
\newblock \bibinfo{journal}{\emph{Psychology and aging}} \bibinfo{volume}{30},
  \bibinfo{number}{1} (\bibinfo{year}{2015}), \bibinfo{pages}{160}.
\newblock
\urldef\tempurl%
\url{https://doi.org/10.1037/a0038544}
\showDOI{\tempurl}


\bibitem[\protect\citeauthoryear{Ragot, Martin, and Cojean}{Ragot
  et~al\mbox{.}}{2020}]%
        {ragot2020}
\bibfield{author}{\bibinfo{person}{Martin Ragot}, \bibinfo{person}{Nicolas
  Martin}, {and} \bibinfo{person}{Salom\'{e} Cojean}.}
  \bibinfo{year}{2020}\natexlab{}.
\newblock \showarticletitle{AI-Generated vs. Human Artworks. A Perception Bias
  Towards Artificial Intelligence?}. In \bibinfo{booktitle}{\emph{Extended
  Abstracts of the 2020 CHI Conference on Human Factors in Computing Systems}}
  (Honolulu, HI, USA) \emph{(\bibinfo{series}{CHI EA '20})}.
  \bibinfo{publisher}{Association for Computing Machinery},
  \bibinfo{address}{New York, NY, USA}, \bibinfo{pages}{1–10}.
\newblock
\showISBNx{9781450368193}
\urldef\tempurl%
\url{https://doi.org/10.1145/3334480.3382892}
\showDOI{\tempurl}


\bibitem[\protect\citeauthoryear{Reiley}{Reiley}{2011}]%
        {10.1145/2020408.2020535}
\bibfield{author}{\bibinfo{person}{David Reiley}.}
  \bibinfo{year}{2011}\natexlab{}.
\newblock \showarticletitle{"Which Half Is Wasted?": Controlled Experiments to
  Measure Online-Advertising Effectiveness}. In
  \bibinfo{booktitle}{\emph{Proceedings of the 17th ACM SIGKDD International
  Conference on Knowledge Discovery and Data Mining}} (San Diego, California,
  USA) \emph{(\bibinfo{series}{KDD '11})}. \bibinfo{publisher}{Association for
  Computing Machinery}, \bibinfo{address}{New York, NY, USA},
  \bibinfo{pages}{777}.
\newblock
\showISBNx{9781450308137}
\urldef\tempurl%
\url{https://doi.org/10.1145/2020408.2020535}
\showDOI{\tempurl}


\bibitem[\protect\citeauthoryear{Rosenzweig}{Rosenzweig}{1933}]%
        {rosenzweig1933experimental}
\bibfield{author}{\bibinfo{person}{Saul Rosenzweig}.}
  \bibinfo{year}{1933}\natexlab{}.
\newblock \showarticletitle{The experimental situation as a psychological
  problem.}
\newblock \bibinfo{journal}{\emph{Psychological Review}} \bibinfo{volume}{40},
  \bibinfo{number}{4} (\bibinfo{year}{1933}), \bibinfo{pages}{337}.
\newblock


\bibitem[\protect\citeauthoryear{Rutten and Geerts}{Rutten and Geerts}{2020}]%
        {rutten2020better}
\bibfield{author}{\bibinfo{person}{Isa Rutten} {and} \bibinfo{person}{David
  Geerts}.} \bibinfo{year}{2020}\natexlab{}.
\newblock \showarticletitle{Better Because It's New: The Impact of Perceived
  Novelty on the Added Value of Mid-Air Haptic Feedback}. In
  \bibinfo{booktitle}{\emph{Proceedings of the 2020 CHI Conference on Human
  Factors in Computing Systems}}. \bibinfo{pages}{1--13}.
\newblock


\bibitem[\protect\citeauthoryear{Schwarz and B{\"u}chel}{Schwarz and
  B{\"u}chel}{2015}]%
        {schwarz2015cognition}
\bibfield{author}{\bibinfo{person}{Katharina~A Schwarz} {and}
  \bibinfo{person}{Christian B{\"u}chel}.} \bibinfo{year}{2015}\natexlab{}.
\newblock \showarticletitle{Cognition and the placebo effect--dissociating
  subjective perception and actual performance}.
\newblock \bibinfo{journal}{\emph{PloS one}} \bibinfo{volume}{10},
  \bibinfo{number}{7} (\bibinfo{year}{2015}), \bibinfo{pages}{e0130492}.
\newblock


\bibitem[\protect\citeauthoryear{Shankar, Louis, Dascalu, Hayes, and
  Houmanfar}{Shankar et~al\mbox{.}}{2007}]%
        {10.1145/1216295.1216357}
\bibfield{author}{\bibinfo{person}{Anil Shankar}, \bibinfo{person}{Sushil~J.
  Louis}, \bibinfo{person}{Sergiu Dascalu}, \bibinfo{person}{Linda~J. Hayes},
  {and} \bibinfo{person}{Ramona Houmanfar}.} \bibinfo{year}{2007}\natexlab{}.
\newblock \showarticletitle{User-Context for Adaptive User Interfaces}. In
  \bibinfo{booktitle}{\emph{Proceedings of the 12th International Conference on
  Intelligent User Interfaces}} (Honolulu, Hawaii, USA)
  \emph{(\bibinfo{series}{IUI '07})}. \bibinfo{publisher}{Association for
  Computing Machinery}, \bibinfo{address}{New York, NY, USA},
  \bibinfo{pages}{321–324}.
\newblock
\showISBNx{1595934812}
\urldef\tempurl%
\url{https://doi.org/10.1145/1216295.1216357}
\showDOI{\tempurl}


\bibitem[\protect\citeauthoryear{Simmonds and Zikos}{Simmonds and
  Zikos}{2014}]%
        {10.1145/2674396.2674435}
\bibfield{author}{\bibinfo{person}{Maureen~J. Simmonds} {and}
  \bibinfo{person}{Dimitrios Zikos}.} \bibinfo{year}{2014}\natexlab{}.
\newblock \showarticletitle{Computer Games to Decrease Pain and Improve Mood
  and Movement}. In \bibinfo{booktitle}{\emph{Proceedings of the 7th
  International Conference on PErvasive Technologies Related to Assistive
  Environments}} (Rhodes, Greece) \emph{(\bibinfo{series}{PETRA '14})}.
  \bibinfo{publisher}{Association for Computing Machinery},
  \bibinfo{address}{New York, NY, USA}, Article \bibinfo{articleno}{56},
  \bibinfo{numpages}{4}~pages.
\newblock
\showISBNx{9781450327466}
\urldef\tempurl%
\url{https://doi.org/10.1145/2674396.2674435}
\showDOI{\tempurl}


\bibitem[\protect\citeauthoryear{Skitka, Mosier, and Burdick}{Skitka
  et~al\mbox{.}}{2000}]%
        {skitka2000accountability}
\bibfield{author}{\bibinfo{person}{Linda~J Skitka}, \bibinfo{person}{Kathleen
  Mosier}, {and} \bibinfo{person}{Mark~D Burdick}.}
  \bibinfo{year}{2000}\natexlab{}.
\newblock \showarticletitle{Accountability and automation bias}.
\newblock \bibinfo{journal}{\emph{International Journal of Human-Computer
  Studies}} \bibinfo{volume}{52}, \bibinfo{number}{4} (\bibinfo{year}{2000}),
  \bibinfo{pages}{701--717}.
\newblock


\bibitem[\protect\citeauthoryear{Spiel, Bertel, and Kayali}{Spiel
  et~al\mbox{.}}{2017}]%
        {10.1145/3025453.3025721}
\bibfield{author}{\bibinfo{person}{Katta Spiel}, \bibinfo{person}{Sven Bertel},
  {and} \bibinfo{person}{Fares Kayali}.} \bibinfo{year}{2017}\natexlab{}.
\newblock \showarticletitle{"Not Another Z Piece!": Adaptive Difficulty in
  TETRIS}. In \bibinfo{booktitle}{\emph{Proceedings of the 2017 CHI Conference
  on Human Factors in Computing Systems}} (Denver, Colorado, USA)
  \emph{(\bibinfo{series}{CHI '17})}. \bibinfo{publisher}{Association for
  Computing Machinery}, \bibinfo{address}{New York, NY, USA},
  \bibinfo{pages}{5126–5131}.
\newblock
\showISBNx{9781450346559}
\urldef\tempurl%
\url{https://doi.org/10.1145/3025453.3025721}
\showDOI{\tempurl}


\bibitem[\protect\citeauthoryear{Springer and Whittaker}{Springer and
  Whittaker}{2019}]%
        {10.1145/3301275.3302322}
\bibfield{author}{\bibinfo{person}{Aaron Springer} {and} \bibinfo{person}{Steve
  Whittaker}.} \bibinfo{year}{2019}\natexlab{}.
\newblock \showarticletitle{Progressive Disclosure: Empirically Motivated
  Approaches to Designing Effective Transparency}. In
  \bibinfo{booktitle}{\emph{Proceedings of the 24th International Conference on
  Intelligent User Interfaces}} (Marina del Ray, California)
  \emph{(\bibinfo{series}{IUI '19})}. \bibinfo{publisher}{Association for
  Computing Machinery}, \bibinfo{address}{New York, NY, USA},
  \bibinfo{pages}{107–120}.
\newblock
\showISBNx{9781450362726}
\urldef\tempurl%
\url{https://doi.org/10.1145/3301275.3302322}
\showDOI{\tempurl}


\bibitem[\protect\citeauthoryear{Stewart-Williams and Podd}{Stewart-Williams
  and Podd}{2004}]%
        {stewart2004placebo}
\bibfield{author}{\bibinfo{person}{Steve Stewart-Williams} {and}
  \bibinfo{person}{John Podd}.} \bibinfo{year}{2004}\natexlab{}.
\newblock \showarticletitle{The placebo effect: dissolving the expectancy
  versus conditioning debate.}
\newblock \bibinfo{journal}{\emph{Psychological bulletin}}
  \bibinfo{volume}{130}, \bibinfo{number}{2} (\bibinfo{year}{2004}),
  \bibinfo{pages}{324}.
\newblock


\bibitem[\protect\citeauthoryear{Strait, Canning, and Scheutz}{Strait
  et~al\mbox{.}}{2014}]%
        {strait2014reliability}
\bibfield{author}{\bibinfo{person}{Megan Strait}, \bibinfo{person}{Cody
  Canning}, {and} \bibinfo{person}{Matthias Scheutz}.}
  \bibinfo{year}{2014}\natexlab{}.
\newblock \showarticletitle{Reliability of NIRS-Based BCIs: a
  placebo-controlled replication and reanalysis of Brainput}.
\newblock In \bibinfo{booktitle}{\emph{CHI'14 Extended Abstracts on Human
  Factors in Computing Systems}}. \bibinfo{pages}{619--630}.
\newblock


\bibitem[\protect\citeauthoryear{Tanveer, Zhao, Chen, Tiet, and Hoque}{Tanveer
  et~al\mbox{.}}{2016}]%
        {10.1145/2856767.2856785}
\bibfield{author}{\bibinfo{person}{Iftekhar~M. Tanveer}, \bibinfo{person}{Ru
  Zhao}, \bibinfo{person}{Kezhen Chen}, \bibinfo{person}{Zoe Tiet}, {and}
  \bibinfo{person}{Mohammed~Ehsan Hoque}.} \bibinfo{year}{2016}\natexlab{}.
\newblock \showarticletitle{AutoManner: An Automated Interface for Making
  Public Speakers Aware of Their Mannerisms}. In
  \bibinfo{booktitle}{\emph{Proceedings of the 21st International Conference on
  Intelligent User Interfaces}} (Sonoma, California, USA)
  \emph{(\bibinfo{series}{IUI '16})}. \bibinfo{publisher}{Association for
  Computing Machinery}, \bibinfo{address}{New York, NY, USA},
  \bibinfo{pages}{385–396}.
\newblock
\showISBNx{9781450341370}
\urldef\tempurl%
\url{https://doi.org/10.1145/2856767.2856785}
\showDOI{\tempurl}


\bibitem[\protect\citeauthoryear{Thong, Hong, and Tam}{Thong
  et~al\mbox{.}}{2006}]%
        {thong2006effects}
\bibfield{author}{\bibinfo{person}{James~YL Thong}, \bibinfo{person}{Se-Joon
  Hong}, {and} \bibinfo{person}{Kar~Yan Tam}.} \bibinfo{year}{2006}\natexlab{}.
\newblock \showarticletitle{The effects of post-adoption beliefs on the
  expectation-confirmation model for information technology continuance}.
\newblock \bibinfo{journal}{\emph{International Journal of Human-computer
  studies}} \bibinfo{volume}{64}, \bibinfo{number}{9} (\bibinfo{year}{2006}),
  \bibinfo{pages}{799--810}.
\newblock


\bibitem[\protect\citeauthoryear{Trewin, Marques, and Guerreiro}{Trewin
  et~al\mbox{.}}{2015}]%
        {trewin2015usage}
\bibfield{author}{\bibinfo{person}{Shari Trewin}, \bibinfo{person}{Diogo
  Marques}, {and} \bibinfo{person}{Tiago Guerreiro}.}
  \bibinfo{year}{2015}\natexlab{}.
\newblock \showarticletitle{Usage of subjective scales in accessibility
  research}. In \bibinfo{booktitle}{\emph{Proceedings of the 17th International
  ACM SIGACCESS Conference on Computers \& Accessibility}}.
  \bibinfo{pages}{59--67}.
\newblock


\bibitem[\protect\citeauthoryear{Vaccaro, Huang, Eslami, Sandvig, Hamilton, and
  Karahalios}{Vaccaro et~al\mbox{.}}{2018}]%
        {vaccaro2018illusion}
\bibfield{author}{\bibinfo{person}{Kristen Vaccaro}, \bibinfo{person}{Dylan
  Huang}, \bibinfo{person}{Motahhare Eslami}, \bibinfo{person}{Christian
  Sandvig}, \bibinfo{person}{Kevin Hamilton}, {and} \bibinfo{person}{Karrie
  Karahalios}.} \bibinfo{year}{2018}\natexlab{}.
\newblock \showarticletitle{The illusion of control: Placebo effects of control
  settings}. In \bibinfo{booktitle}{\emph{Proceedings of the 2018 CHI
  Conference on Human Factors in Computing Systems}}. \bibinfo{pages}{1--13}.
\newblock


\bibitem[\protect\citeauthoryear{Venkatesh and Davis}{Venkatesh and
  Davis}{2000}]%
        {doi:10.1287/mnsc.46.2.186.11926}
\bibfield{author}{\bibinfo{person}{Viswanath Venkatesh} {and}
  \bibinfo{person}{Fred~D. Davis}.} \bibinfo{year}{2000}\natexlab{}.
\newblock \showarticletitle{A Theoretical Extension of the Technology
  Acceptance Model: Four Longitudinal Field Studies}.
\newblock \bibinfo{journal}{\emph{Management Science}} \bibinfo{volume}{46},
  \bibinfo{number}{2} (\bibinfo{year}{2000}), \bibinfo{pages}{186--204}.
\newblock
\urldef\tempurl%
\url{https://doi.org/10.1287/mnsc.46.2.186.11926}
\showDOI{\tempurl}
\showeprint{https://doi.org/10.1287/mnsc.46.2.186.11926}


\bibitem[\protect\citeauthoryear{Venkatesh, Thong, and Xu}{Venkatesh
  et~al\mbox{.}}{2012a}]%
        {venkatesh2012consumer}
\bibfield{author}{\bibinfo{person}{Viswanath Venkatesh},
  \bibinfo{person}{James~YL Thong}, {and} \bibinfo{person}{Xin Xu}.}
  \bibinfo{year}{2012}\natexlab{a}.
\newblock \showarticletitle{Consumer acceptance and use of information
  technology: extending the unified theory of acceptance and use of
  technology}.
\newblock \bibinfo{journal}{\emph{MIS quarterly}} (\bibinfo{year}{2012}),
  \bibinfo{pages}{157--178}.
\newblock


\bibitem[\protect\citeauthoryear{Venkatesh, Thong, and Xu}{Venkatesh
  et~al\mbox{.}}{2016}]%
        {venkatesh2016unified}
\bibfield{author}{\bibinfo{person}{Viswanath Venkatesh},
  \bibinfo{person}{James~YL Thong}, {and} \bibinfo{person}{Xin Xu}.}
  \bibinfo{year}{2016}\natexlab{}.
\newblock \showarticletitle{Unified theory of acceptance and use of technology:
  A synthesis and the road ahead}.
\newblock \bibinfo{journal}{\emph{Journal of the association for Information
  Systems}} \bibinfo{volume}{17}, \bibinfo{number}{5} (\bibinfo{year}{2016}),
  \bibinfo{pages}{328--376}.
\newblock


\bibitem[\protect\citeauthoryear{Venkatesh, Thong, and Xu}{Venkatesh
  et~al\mbox{.}}{2012b}]%
        {10.2307/41410412}
\bibfield{author}{\bibinfo{person}{Viswanath Venkatesh}, \bibinfo{person}{James
  Y.~L. Thong}, {and} \bibinfo{person}{Xin Xu}.}
  \bibinfo{year}{2012}\natexlab{b}.
\newblock \showarticletitle{Consumer Acceptance and Use of Information
  Technology: Extending the Unified Theory of Acceptance and Use of
  Technology}.
\newblock \bibinfo{journal}{\emph{MIS Quarterly}} \bibinfo{volume}{36},
  \bibinfo{number}{1} (\bibinfo{year}{2012}), \bibinfo{pages}{157--178}.
\newblock
\showISSN{02767783}
\urldef\tempurl%
\url{http://www.jstor.org/stable/41410412}
\showURL{%
\tempurl}


\bibitem[\protect\citeauthoryear{Voghoei, Tonekaboni, Yazdansepas, Soleymani,
  Farahani, and Arabnia}{Voghoei et~al\mbox{.}}{2020}]%
        {10.1145/3374135.3385274}
\bibfield{author}{\bibinfo{person}{Sahar Voghoei},
  \bibinfo{person}{Navid~Hashemi Tonekaboni}, \bibinfo{person}{Delaram
  Yazdansepas}, \bibinfo{person}{Saber Soleymani}, \bibinfo{person}{Abolfazl
  Farahani}, {and} \bibinfo{person}{Hamid~R. Arabnia}.}
  \bibinfo{year}{2020}\natexlab{}.
\newblock \showarticletitle{Personalized Feedback Emails: A Case Study on
  Online Introductory Computer Science Courses}. In
  \bibinfo{booktitle}{\emph{Proceedings of the 2020 ACM Southeast Conference}}
  (Tampa, FL, USA) \emph{(\bibinfo{series}{ACM SE '20})}.
  \bibinfo{publisher}{Association for Computing Machinery},
  \bibinfo{address}{New York, NY, USA}, \bibinfo{pages}{18–25}.
\newblock
\showISBNx{9781450371056}
\urldef\tempurl%
\url{https://doi.org/10.1145/3374135.3385274}
\showDOI{\tempurl}


\bibitem[\protect\citeauthoryear{Wagenmakers, Love, Marsman, Jamil, Ly,
  Verhagen, Selker, Gronau, Dropmann, Boutin, et~al\mbox{.}}{Wagenmakers
  et~al\mbox{.}}{2018}]%
        {wagenmakers2018bayesian}
\bibfield{author}{\bibinfo{person}{Eric-Jan Wagenmakers},
  \bibinfo{person}{Jonathon Love}, \bibinfo{person}{Maarten Marsman},
  \bibinfo{person}{Tahira Jamil}, \bibinfo{person}{Alexander Ly},
  \bibinfo{person}{Josine Verhagen}, \bibinfo{person}{Ravi Selker},
  \bibinfo{person}{Quentin~F Gronau}, \bibinfo{person}{Damian Dropmann},
  \bibinfo{person}{Bruno Boutin}, {et~al\mbox{.}}}
  \bibinfo{year}{2018}\natexlab{}.
\newblock \showarticletitle{Bayesian inference for psychology. Part II: Example
  applications with JASP}.
\newblock \bibinfo{journal}{\emph{Psychonomic bulletin \& review}}
  \bibinfo{volume}{25}, \bibinfo{number}{1} (\bibinfo{year}{2018}),
  \bibinfo{pages}{58--76}.
\newblock


\bibitem[\protect\citeauthoryear{Wehbe, Lank, and Nacke}{Wehbe
  et~al\mbox{.}}{2017}]%
        {10.1145/3064663.3064712}
\bibfield{author}{\bibinfo{person}{Rina~R. Wehbe}, \bibinfo{person}{Edward
  Lank}, {and} \bibinfo{person}{Lennart~E. Nacke}.}
  \bibinfo{year}{2017}\natexlab{}.
\newblock \showarticletitle{Left Them 4 Dead: Perception of Humans versus
  Non-Player Character Teammates in Cooperative Gameplay}. In
  \bibinfo{booktitle}{\emph{Proceedings of the 2017 Conference on Designing
  Interactive Systems}} (Edinburgh, United Kingdom) \emph{(\bibinfo{series}{DIS
  '17})}. \bibinfo{publisher}{Association for Computing Machinery},
  \bibinfo{address}{New York, NY, USA}, \bibinfo{pages}{403–415}.
\newblock
\showISBNx{9781450349222}
\urldef\tempurl%
\url{https://doi.org/10.1145/3064663.3064712}
\showDOI{\tempurl}


\bibitem[\protect\citeauthoryear{Wells, Campbell, Valacich, and
  Featherman}{Wells et~al\mbox{.}}{2010a}]%
        {wells2010effect}
\bibfield{author}{\bibinfo{person}{John~D Wells}, \bibinfo{person}{Damon~E
  Campbell}, \bibinfo{person}{Joseph~S Valacich}, {and}
  \bibinfo{person}{Mauricio Featherman}.} \bibinfo{year}{2010}\natexlab{a}.
\newblock \showarticletitle{The effect of perceived novelty on the adoption of
  information technology innovations: a risk/reward perspective}.
\newblock \bibinfo{journal}{\emph{Decision Sciences}} \bibinfo{volume}{41},
  \bibinfo{number}{4} (\bibinfo{year}{2010}), \bibinfo{pages}{813--843}.
\newblock


\bibitem[\protect\citeauthoryear{Wells, Campbell, Valacich, and
  Featherman}{Wells et~al\mbox{.}}{2010b}]%
        {doi:10.1111/j.1540-5915.2010.00292.x}
\bibfield{author}{\bibinfo{person}{John~D. Wells}, \bibinfo{person}{Damon~E.
  Campbell}, \bibinfo{person}{Joseph~S. Valacich}, {and}
  \bibinfo{person}{Mauricio Featherman}.} \bibinfo{year}{2010}\natexlab{b}.
\newblock \showarticletitle{The Effect of Perceived Novelty on the Adoption of
  Information Technology Innovations: A Risk/Reward Perspective}.
\newblock \bibinfo{journal}{\emph{Decision Sciences}} \bibinfo{volume}{41},
  \bibinfo{number}{4} (\bibinfo{year}{2010}), \bibinfo{pages}{813--843}.
\newblock
\urldef\tempurl%
\url{https://doi.org/10.1111/j.1540-5915.2010.00292.x}
\showDOI{\tempurl}
\showeprint{https://onlinelibrary.wiley.com/doi/pdf/10.1111/j.1540-5915.2010.00292.x}


\bibitem[\protect\citeauthoryear{Wiegand, Eiband, Haubelt, and
  Hussmann}{Wiegand et~al\mbox{.}}{2020}]%
        {10.1145/3379503.3403554}
\bibfield{author}{\bibinfo{person}{Gesa Wiegand}, \bibinfo{person}{Malin
  Eiband}, \bibinfo{person}{Maximilian Haubelt}, {and}
  \bibinfo{person}{Heinrich Hussmann}.} \bibinfo{year}{2020}\natexlab{}.
\newblock \showarticletitle{“I’d like an Explanation for That!”Exploring
  Reactions to Unexpected Autonomous Driving}. In
  \bibinfo{booktitle}{\emph{22nd International Conference on Human-Computer
  Interaction with Mobile Devices and Services}} (Oldenburg, Germany)
  \emph{(\bibinfo{series}{MobileHCI '20})}. \bibinfo{publisher}{Association for
  Computing Machinery}, \bibinfo{address}{New York, NY, USA}, Article
  \bibinfo{articleno}{36}, \bibinfo{numpages}{11}~pages.
\newblock
\showISBNx{9781450375160}
\urldef\tempurl%
\url{https://doi.org/10.1145/3379503.3403554}
\showDOI{\tempurl}


\bibitem[\protect\citeauthoryear{Wilde and Andersen}{Wilde and
  Andersen}{2010}]%
        {10.1145/1952222.1952262}
\bibfield{author}{\bibinfo{person}{Danielle Wilde} {and}
  \bibinfo{person}{Kristina Andersen}.} \bibinfo{year}{2010}\natexlab{}.
\newblock \showarticletitle{Part Science Part Magic: Analysing the OWL
  Outcomes}. In \bibinfo{booktitle}{\emph{Proceedings of the 22nd Conference of
  the Computer-Human Interaction Special Interest Group of Australia on
  Computer-Human Interaction}} (Brisbane, Australia)
  \emph{(\bibinfo{series}{OZCHI '10})}. \bibinfo{publisher}{Association for
  Computing Machinery}, \bibinfo{address}{New York, NY, USA},
  \bibinfo{pages}{188–191}.
\newblock
\showISBNx{9781450305020}
\urldef\tempurl%
\url{https://doi.org/10.1145/1952222.1952262}
\showDOI{\tempurl}


\end{thebibliography}

\end{document}